# Micromagnetic Simulation of Three-dimensional Nanoarchitectures

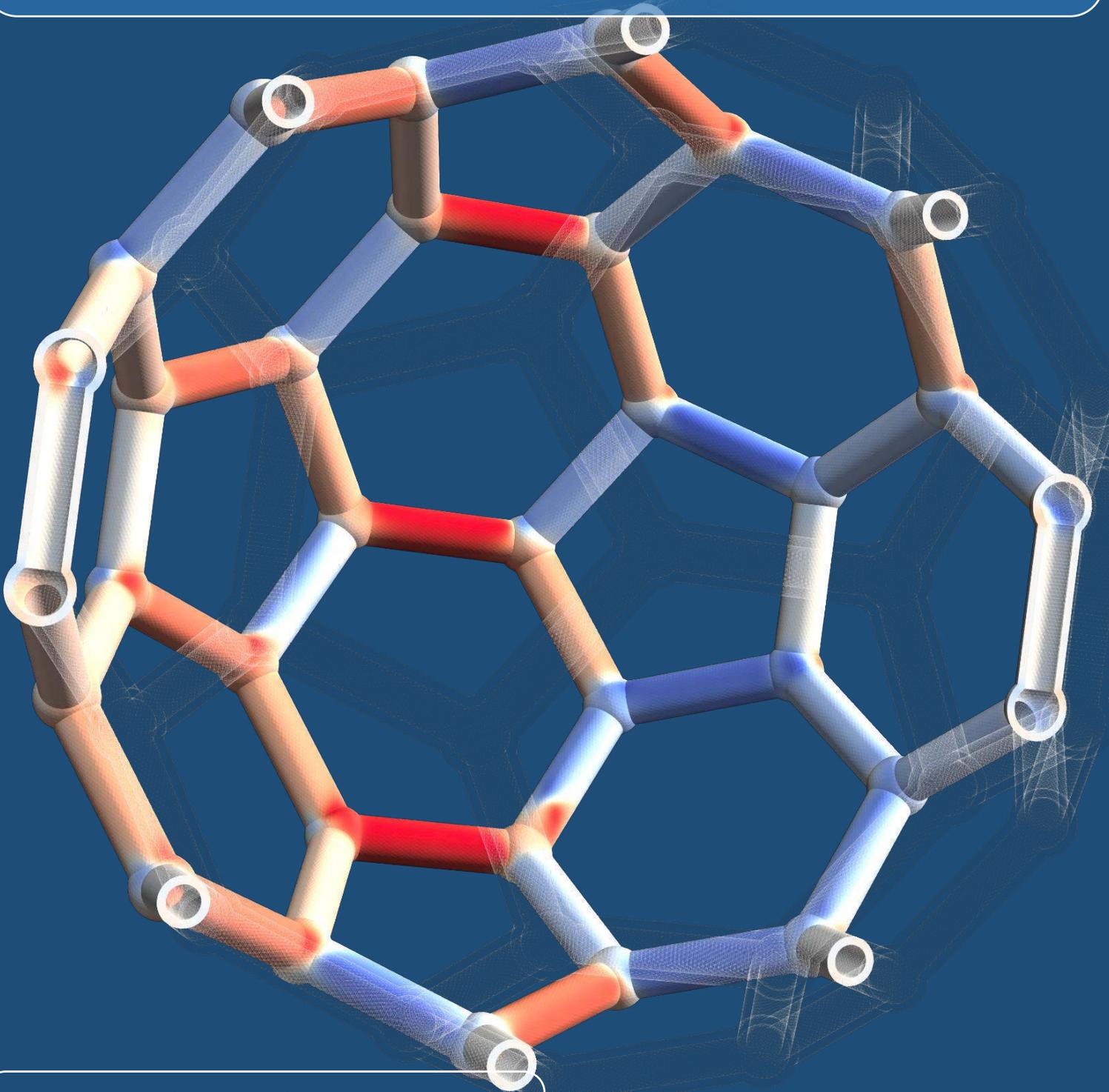

Ph.D. thesis
Rajgowrav Cheenikundil
Raj.cheenik@gmail.com

# THÈSE présentée par :

## Rajgowrav CHEENIKUNDIL

soutenue le : **16 décembre 2021**

pour obtenir le grade de : **Docteur de l'université de Strasbourg**

Discipline/ Spécialité : Physique / Physique de la Matière Condensée

---

## Simulations micromagnétiques de nano-architectures tridimensionnelles

## Micromagnetic simulation of three-dimensional nano-architectures

---

**THÈSE dirigée par :**

**M HERTEL Riccardo**  Directeur de Recherche CNRS, Institut de Physique et Chimie des Matériaux de Strasbourg

**RAPPORTEURS :**

**M GRUNDLER Dirk**  Professeur, École Polytechnique Fédérale de Lausanne (CH)
**Mme BUDA-PREJBEANU Liliana**  Professeure, SPINTEC, Université Grenoble Alpes / CNRS / CEA

---

**AUTRES MEMBRES DU JURY :**

**M LADAK Sam**  Professeur, Cardiff University (UK)
**M WEINMANN Dietmar**  Directeur de Recherche CNRS, Institut de Physique et Chimie des Matériaux de Strasbourg
**M LACOUR Daniel**  Chargé de recherche CNRS, Institut Jean Lamour, Nancy

# Déclaration sur l'honneur
## *Declaration of Honour*

J'affirme être informé que le plagiat est une faute grave susceptible de mener à des sanctions administratives et disciplinaires pouvant aller jusqu'au renvoi de l'Université de Strasbourg et passible de poursuites devant les tribunaux de la République Française.

Je suis conscient(e) que l'absence de citation claire et transparente d'une source empruntée à un tiers (texte, idée, raisonnement ou autre création) est constitutive de plagiat.

Au vu de ce qui précède, **j'atteste sur l'honneur que le travail décrit dans mon manuscrit de thèse est un travail original et que je n'ai pas eu recours au plagiat ou à toute autre forme de fraude.**

*I affirm that I am aware that plagiarism is a serious misconduct that may lead to administrative and disciplinary sanctions up to dismissal from the University of Strasbourg and liable to prosecution in the courts of the French Republic.*

*I am aware that the absence of a clear and transparent citation of a source borrowed from a third party (text, idea, reasoning or other creation) is constitutive of plagiarism.*

*In view of the foregoing, **I hereby certify that the work described in my thesis manuscript is original work and that I have not resorted to plagiarism or any other form of fraud.***

**Nom : Prénom : Rajgowrav CHEENIKUNDIL**

**Ecole doctorale : ED182**

**Laboratoire : Institut de physique et chimie des matériaux de Strasbourg**

**Date : 08/02/2022**

**Signature :**

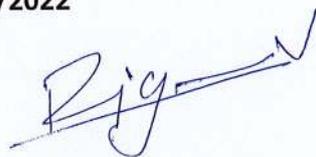

# Acknowledgements

The three years of my life in which I did this PhD is an important part of my life and I would like to express my heartfelt gratitude to all the people who helped me and supported me throughput this time. First of all I would like to express my gratitude to the thesis supervisor Riccardo Hertel for accepting me as his student and the excellent mentorship offered. I consider having received the opportunity to work with him a great honour and moving from India to a completely foreign environment was made much easier with his support and help. Riccardo was a patient supervisor throughout the three years and he always found time and interest for discussions and was patient enough to explain things in detail and to make corrections whenever necessary. From his guidance I learnt a lot: about magnetism and computation in particular and a wide range of other things.

I would also like to thank Salia Cherifi-Hertel for the discussions and valuable suggestions at various important points of the thesis.

I would like to express my thankfulness to the jury; Daniel Lacour, Dietmar Weinmann, Dirk Grundler, Liliana Buda-Prejbeanu and Sam Ladak for accepting the request to be a part of the jury, carefully reading the thesis and for the valuable discussions during the defense and also for the careful corrections.

I express my appreciation for the experimental collaboration with the group of David Schmool regarding the study of magnonic modes in Cobalt nanodots.

The studies in this thesis was carried out at the Institut de Physique et de Chimie des Matériaux de Strasbourg (IPCMS-CNRS). I would like to express my sincere thanks to our institute director Pierre Rabu, the Doctroral School (ED-182) director Aziz Dinia, the IPCMS secretary Catherine Bonnin, the IPCMS librarian Beatrice Masson, the head of department (DMONS) Yves Henry, the department secretary Veronique Wernher and all the other people in the IPCMS for hosting me for these years and all the academic and legislative services provided.

I would like to express my gratitude towards my colleagues and friends in IPCMS and in Strasbourg who helped me and supported me throughout these years: Aleena, Anatoli, Ankita, Aravind, Dheeksha, Gyandeep, Jinu,

i

Jose, Mathiew, Matias, Mehrdad, Mohammed Soliman, Nivedita, Sambith, Satakshi, Sharmin, Suvidya, and Swapneel. I'm grateful to have such nice colleagues and the PhD was made fun working with these people.

I would like to express my gratitude to Julien for proof reading the French abstract.

I would like to thank Senoy Thomas, my former professor for introducing me to the field of micromagnetism. I would like to express my love and gratitude towards my friends and my family for all the support.

Finally I would like to express my gratitude and love towards all my friends whom I made in Strasbourg, who made my life here memorable and amusing.

# Abstract

Le nanomagnétisme est la sous-branche du magnétisme qui traite de l'étude des phénomènes magnétiques à l'échelle du nanomètre. L'un des principaux moteurs de l'étude des systèmes de plus petite taille est le désir d'obtenir un moyen de manipuler les propriétés magnétiques telles que la structure de l'aimantation et ses propriétés à haute fréquence par confinement géométrique. Des exemples de cette capacité à contrôler la structure de l'aimantation par confinement géométrique incluent l'observation d'états aimantés axialement dans des géométries allongées et l'observation de tourbillons magnétiques dans des structures planes minces. L'existence de ces structures de spin particulières dans les particules de faible dimension leur attribue des propriétés statiques et dynamiques distinctes par rapport à l'état massif.

Dans le passé, les effets de ce modelage géométrique étaient principalement limités aux structures bidimensionnelles. Récemment, une nouvelle tendance a émergé dans le nanomagnétisme pour étendre ces effets à la troisième dimension et, par conséquent, le nanomagnétisme tridimensionnel est apparu comme une nouvelle frontière de la recherche. Des progrès considérables ont été réalisés dans les techniques de mesure et de fabrication pour répondre à cette tendance. Les progrès des techniques d'imagerie par rayons X et par électrons ont facilité la visualisation des structures magnétiques tridimensionnelles avec une précision nanométrique, même si elles sont enfouies profondément dans le matériau.

Parallèlement, les progrès récents dans les techniques de fabrication telles que le dépôt induit par faisceau d'électrons focalisé (FEBID), également connu sous le nom de nanoimpression 3D, et d'autres techniques de nanofabrication telles que la lithographie à deux photons (TPL) en combinaison avec diverses autres techniques de dépôt ont permis de fabriquer des matériaux ferromagnétiques avec des géométries de forme arbitraire a l'échelle nanométrique. Par conséquent, les nanoarchitectures tridimensionnelles sont apparues comme une nouvelle catégorie de nanostructures. L'accès à un degré de liberté supplémentaire dans ces structures pourrait faire apparaître des phénomènes physiques nouveaux et intéressants qui sont absents dans les structures bidimensionnelles. Parallèlement aux techniques de fabrication et de mesure de pointe, les simulations par éléments finis constituent un excellent outil théorique pour étudier les propriétés fondamentales de ces types de structures. Dans cette thèse, nous avons étudié la structure magnétique statique et les propriétés haute fréquence de différents types d'architectures 3D telles que les structures fractales 3D de type Sierpinski, les nanoarchitectures Buckyball, les nanoarchitectures de type diamant et les nanoarchitectures cubiques en utilisant des simulations micromagnétiques par éléments finis.

La thèse est divisée en huit chapitres. Dans le premier chapitre, le contexte général et la motivation scientifique de l'étude sont présentés. Dans le deuxième chapitre, la théorie de base du micromagnétisme est introduite, y compris les différents termes énergétiques pertinents pour les simulations numériques. Le micromagnétisme est la théorie du continu du ferromagnétism qui décrit les phénomènes magnétiques à une échelle de longueur mésoscopique. La principale quantité d'intérêt dans la théorie du micromagnétisme est l'aimantation $\boldsymbol{M}(\boldsymbol{r}, t)$ qui est définie comme la densité des moments magnétiques par unité de volume. L'hypothèse centrale de la théorie du micromagnétisme est que la magnitude du vecteur $\boldsymbol{M}(\boldsymbol{r}, t)$ reste constante tout au long du processus et que seule sa direction peut varier. La théorie micromagnétique décrit l'énergie totale du système comme la somme de plusieurs termes énergétiques différents, tels que l'énergie d'échange, l'énergie de démagnétisation, l'énergie d'anisotropie cristalline et l'énergie de Zeeman. L'énergie d'échange est la contribution de la force fondamentale qui favorise un alignement parallèle des moments magnétiques voisins dans les matériaux ferromagnétiques. Dans le modèle continu du micromagnétisme, l'expression de l'énergie d'échange prend la forme d'une fonction impliquant les gradients spatiaux de l'aimantation, qui est minimale dans le cas d'une distribution uniforme de l'aimantation, de sorte que toute inhomogénéité dans la structure de l'aimantation entraîne une augmentation de l'énergie d'échange. L'énergie de démagnétisation est due à l'interaction de l'aimantation avec le champ magnétostatique, c'est-à-dire le champ magnétique créé par l'ensemble des moments magnétiques de l'échantillon. Cette contribution énergétique favorise la formation de structures de fermeture du flux, comme les tourbillons, qui minimisent les charges magnétostatiques et réduisent ainsi le champ dipolaire. La magnétostatique est également à l'origine du phénomène d'anisotropie de forme, qui décrit la tendance des nanoparticules allongées à aligner l'aimantation le long de son axe. L'énergie d'anisotropie magnéto-cristalline résulte du couplage des moments de spin et d'orbite (couplage spin-orbite) et de l'interaction des ions et du champ cristallin. Le cas le plus simple de l'énergie d'anisotropie magnétocristalline est l'anisotropie uniaxiale dans laquelle il existe une direction d'aimantation préférée et toute déviation de l'aimantation par rapport à cette direction est énergiquement pénalisée. Enfin, l'énergie Zeeman décrit l'interaction de l'aimantation avec un champ externe. Ce terme énergétique favorise l'alignement de l'aimantation parallèlement au champ externe. L'énergie totale est obtenue comme la somme de ces différents termes énergétiques. À chacun de ces termes énergétiques, on peut attribuer un champ effectif décrivant la force de l'interaction. Par conséquent, le champ effectif total peut être décomposé en champs effectifs individuels pour chaque terme énergétique.

Les simulations numériques micromagnétiques impliquent une subdivision de la géométrie de la particule en un grand nombre de cellules de discrétisation. En plus de ces cellules, la subdivision implique un ensemble de points (nœuds) auxquels l'aimantation $\boldsymbol{M}(\boldsymbol{r}, t)$ est calculée et interpolée dans chaque cellule. Pour le cas statique, une version discrétisée d'une structure de l'aimantation d'équilibre peut être obtenue en intégrant numériquement les densités d'énergie micromagnétique sur tout le volume et en minimisant l'énergie totale du système qui en résulte. La dynamique de $\boldsymbol{M}(\boldsymbol{r}, t)$ est régie par l'équation de Landau-Lifshitz-Gilbert (LLG), l'équation de mouvement de l'aimantation, qui décrit une dynamique amortie-précessionnelle de $\boldsymbol{M}(\boldsymbol{r}, t)$ par rapport au champ effectif local.

$$\frac{d\boldsymbol{M}}{dt} = \gamma_L \left[\boldsymbol{M} \times \boldsymbol{H}_{\mathrm{eff}}\right] - \frac{\alpha_L}{M_s} \left[\boldsymbol{M} \times \left[\boldsymbol{M} \times \boldsymbol{H}_{\mathrm{eff}}\right]\right] \tag{1}$$

Numériquement, l'évolution temporelle du vecteur d'aimantation peut être obtenue en intégrant l'équation LLG dans le temps. Toutes les simulations abordées dans cette thèse sont réalisées avec le logiciel de simulation micromagnétique par éléments finis tetmag, qui a été développé à l'IPCMS par

Riccardo Hertel. Le logiciel tetmag prend en entrée le modèle à éléments finis de la géométrie, les paramètres du matériau ferromagnétique et les conditions externes telles que les champs ou les courants appliqués, et produit en sortie le champ vectoriel de magnétisation à chaque point en fonction du temps. Le logiciel tetmag utilise des techniques mathématiques avancées telles que les schémas de compression matricielle hiérarchique, qui permettent de simuler de grandes géométries tout en utilisant des ressources numériques modestes. Il met également en œuvre un calcul massivement parallèle en utilisant l'accélération GPU, ce qui permet d'atteindre des vitesses de simulation rapides, en particulier dans le cas de problèmes à grande échelle impliquant des millions de cellules de discrétisation. Une structure d'aimantation d'équilibre statique dans la géométrie concernée est obtenue en simulant l'évolution dissipative de l'aimantation dans le temps jusqu'à ce qu'un minimum énergétique soit atteint. Pour étudier les effets dynamiques tels que les modes de précession à petit angle et la propagation des ondes de spin, les structures d'aimantation détendues sont excitées avec une perturbation appropriée telle qu'une impulsion de champ gaussienne et la relaxation oscillatoire de l'aimantation (un phénomène appelé "ringdown" magnétique) depuis cet état excité jusqu'au minimum énergétique est enregistrée sous forme de série chronologique. La dynamique de l'aimantation dans un tel processus de ringdown contient généralement plusieurs signaux haute fréquence différents superposés, et il est généralement impossible d'identifier les composantes de fréquence par de simples inspections visuelles. La partie initiale du doctorat a consisté à développer un code de transformation de Fourier en Python qui peut identifier et extraire les modes de fréquence individuels d'une telle superposition.

Avant de les utiliser pour étudier la dynamique haute fréquence de nanoarchitectures tridimensionnelles, ces techniques ont été appliquées sur un système plus traditionnel : un ensemble de nanodisques cylindriques. Cette étude a été réalisée en collaboration avec un groupe expérimental de l'Université Versailles. Les collègues expérimentateurs ont fabriqué un réseau bidimensionnel de nanodisques en cobalt et ont mesuré les modes haute fréquence de la structure en fonction du champ appliqué au moyen d'un dispositif

VNA-FMR. Dans les données expérimentales, on a pu identifier deux modes de fréquence proéminents. En effectuant des simulations par éléments finis, nous avons pu reproduire cette oscillation à deux fréquences proches et identifier le profil spatial relatif à ces modes. L'excitation de fréquence plus basse est un mode localisé sur le bord, dans lequel l'oscillation de l'aimantation aux deux extrémités opposées du nanocylindre est synchronisée. Le deuxième mode, de fréquence plus élevée, a pu être identifié comme une oscillation plus complexe de l'aimantation au centre du nanocylindre. En effectuant des simulations de ce type à différentes intensités de champ, nous avons pu reproduire numériquement les fréquences des deux modes en fonction du champ obtenues dans les expériences de FMR. La fréquence des deux modes augmente presque linéairement avec l'intensité du champ externe appliqué. En raison des limitations des techniques de mesure, il y avait une incertitude concernant la valeur des paramètres du matériau, comme l'aimantation à saturation. En effectuant un grand nombre de simulations et en faisant correspondre la fréquence de ces modes aux données expérimentales, nous avons pu obtenir indirectement ces paramètres matériels à partir de nos simulations micromagnétiques.

Les premières structures artificielles tridimensionnelles que nous avons étudiées sont des géométries 3D fractales de tétraèdres du type Sierpinksi. Ces structures se composent de quatre tétraèdres disposés de manière récursive aux sommets d'un tétraèdre plus grand. Des itérations ou étapes supérieures des fractales ont été générées en divisant les tétraèdres individuels en unités plus petites. Ces structures fractales peuvent être considérées comme une version tridimensionnelle des triangles de Sierpinski bidimensionnels plus connus. Dans des études précédentes rapportées dans la littérature, de telles nanostructures fractales ferromagnétiques bidimensionnelles ont été étudiées expérimentalement et à l'aide de simulations micromagnétiques aux différences finies. Il a été démontré que leur structure auto-similaire et l'existence de différentes échelles de longueur ont un effet considérable sur leurs propriétés, notamment en ce qui concerne leur réponse en fréquence. Dans nos études, nous avons maintenu la taille globale de la fractale constante à 512 nm et simulé cinq itérations de la fractale. Ainsi, le premier étage de la fractale était composé de quatre tétraèdres, chacun de longueur de côté 256 nm, et le cinquième étage était composé de 1024 tétraèdres de longueur de côté 16 nm. En raison de leur disposition géométrique particulière, tous les étages des fractales ont la même surface, tandis que le volume total diminue exponentiellement avec chaque étage. Le modèle par éléments finis du premier étage était composé de près de 500 000 cellules d'éléments finis (qui sont aussi des tétraèdres, bien que de forme irrégulière) et l'étage final contenait près de 700 000 cellules. Pour la simulation, nous avons utilisé des paramètres de matériau correspondant à ceux du Permalloy avec une aimantation à saturation de $M_s = 8 \times 10^5 \, \mathrm{A \, m^{-1}}$, une rigidité d'échange de $A = 1.3 \times 10^{-11} \, \mathrm{J \, m^{-1}}$, et une anisotropie magnétocristalline nulle. Ces paramètres donnent une longueur d'échange de 5.6 nm. Toutes les mailles des éléments finis avaient une taille de maille inférieure à cette longueur intrinsèque. Les états d'aimantation relaxés qui se développent dans ces géométries à champ nul sont obtenus en saturant la structure le long d'un des sommets, puis en relaxant le système à champ externe nul.

Le premier stade de la fractale contient des tétraèdres individuels relativement grands par rapport à la longueur d'échange du matériau. Dans ce cas, la structure de l'aimantation rémanente à champ nul, obtenue après la relaxation de la saturation, donne une structure magnétique dans laquelle chaque tétraèdre est dans un état de vortex magnétique, dont le plan de circulation se trouve le long de la face perpendiculaire à la direction de saturation.

En se déplaçant davantage vers l'intérieur du tétraèdre, la distribution de l'aimantation converge pour former une structure en forme de "A". Dans la deuxième étape, même si les tétraèdres sont deux fois moins grands qu'avant, la structure tourbillonnaire reste la même, sauf dans trois des tétraèdres. Dans ces trois tétraèdres, au lieu de former des états de vortex individuels, l'aimantation s'arrange dans une configuration en forme de $\lambda$, et l'aimantation dans les tétraèdres forme collectivement un vortex déconnecté. Une structure de fermeture de flux partielle est ainsi obtenue en alignant l'aimantation des tétraèdres voisins, de sorte que l'aimantation s'enroule pour former une boucle macroscopique. À partir de la troisième étape, lorsque la taille des tétraèdres est encore réduite, les unités individuelles de tétraèdres sont trop petites pour entretenir un vortex et elles forment à la place des structures de vortex déconnectées s'étendant sur plusieurs tétraèdres.

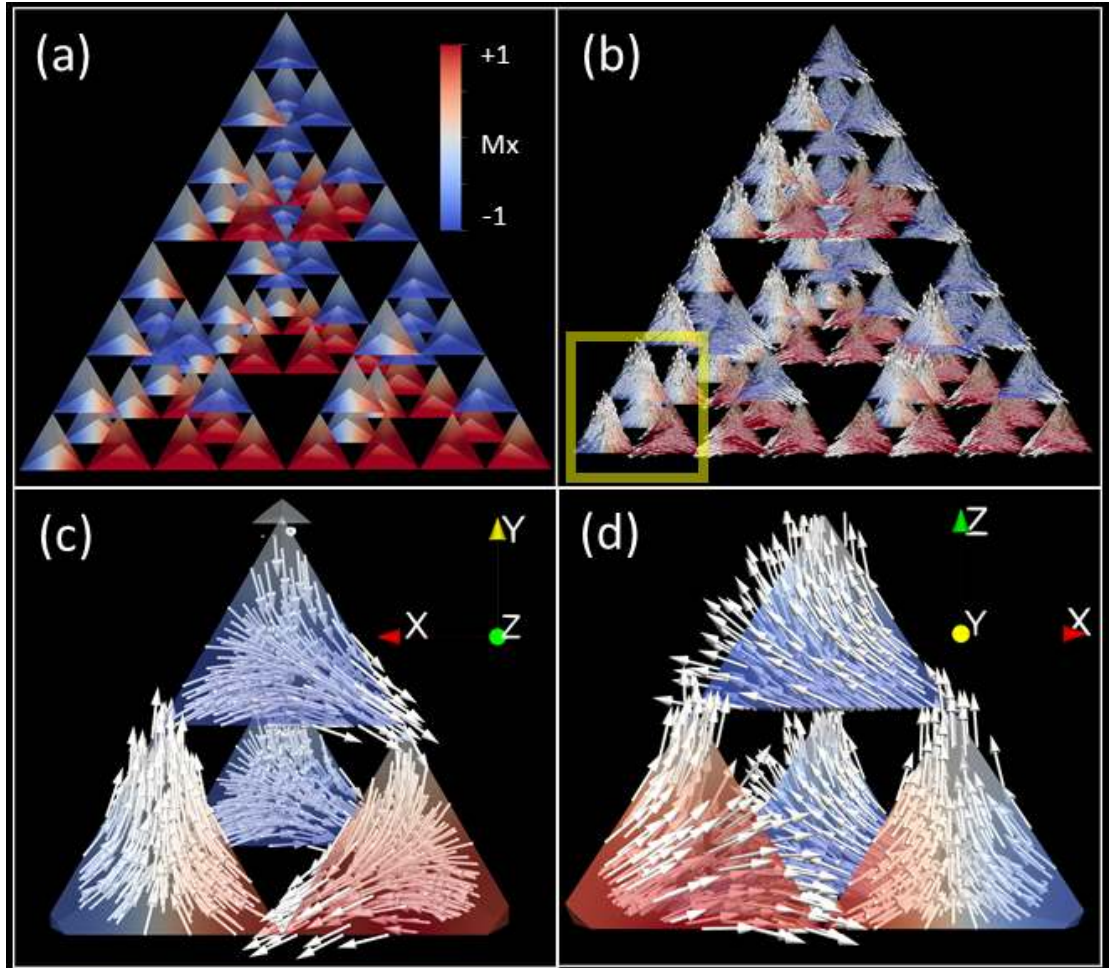

Figure 1: La structure de l'aimantation du troisième stade de la fractale : Tous les tétraèdres individuels sont aimantés dans une structure semblable à une forme "Λ". (c) Vue agrandie de l'amas formant une structure tourbillonnaire déconnectée. (d) Une perspective différente montrant la structure en forme de "A" de l'amas.

Les propriétés hystérétiques de ces structures sont simulées en appliquant un champ de 500 mT puis en réduisant l'intensité du champ à zéro par étapes de champ suffisamment petites. Nous observons des différences considérables dans la boucle M-H à chaque étape de la fractale. Dans les premiers et deuxièmes stades, la structure est dans un état de saturation magnétique à un champ de 500 mT. Lorsque le champ est progressivement réduit, l'aimantation moyenne diminue presque linéairement. À partir de la troisième étape, nous observons un comportement différent dans lequel l'aimantation reste presque constante jusqu'à une intensité de champ spécifique, au-delà de laquelle un saut abrupt se produit. Ce point de saut abrupt est caractérisé par la nucléation des structures tourbillonnaires déconnectées susmentionnées.

Pour calculer les modes magnétiques haute fréquence des différents états d'équilibre relaxés, les structures magnétiques ont été excités par différentes méthodes à l'aide d'une perturbation externe. Le processus oscillatoire "ringdown" magnétique se déroulant après l'application de cette petite perturbation a été enregistré et analysé. La réponse en fréquence des cinq étapes est affichée sur la figure 2. Dans la première étape, nous pouvons identifier deux modes importants : un mode basse fréquence à 700 MHz et un mode haute fréquence à 2.7 GHz. Ces modes correspondent à la gyration synchrone des noyaux de vortex. Dans la deuxième étape, nous pouvons identifier deux régions principales : plusieurs modes basse fréquence dans la plage entre 1,1 et 1,9 GHz, qui peuvent être attribués à l'activité des tourbillons dans les tétraèdres, et des modes comparativement plus faibles autour de 4,7 GHz, qui sont localisés au niveau des trois tétraèdres formant un tourbillon déconnecté. À partir de la troisième étape, au lieu d'avoir plusieurs pics nets bien définis, nous obtenons une large bande de modes intermixtes. La largeur de cette bande augmente au fur et à mesure que nous passons du troisième stade au quatrième stade, même si la fréquence médiane se déplace vers une valeur inférieure. Au cinquième stade, alors que la taille des unités tétraédriques est encore réduite vers la limite du domaine unique, nous observons une déviation de cette excitation à large bande. En étudiant les oscillations magnétiques haute fréquence dépendantes du champ à différents stades, nous avons observé que la fréquence des modes augmente linéairement avec l'intensité du champ appliqué.

La prochaine catégorie de structures tridimensionnelles que nous avons étudiée est celle des nanoarchitectures artificielles de type Buckyball. Ces

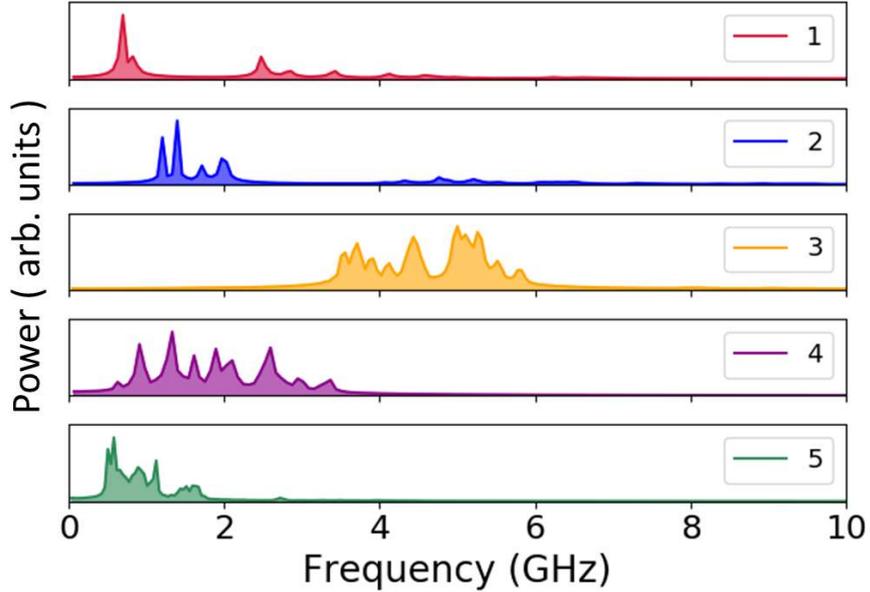

Figure 2: Spectres de Fourier des différentes étapes à champ zéro.

structures sont composées de 90 nanofils cylindriques interconnectés à 60 sommets sphériques afin de former un réseau fermé d'hexagones et de pentagones. Ces structures peuvent être fabriquées de manière à ce que les nanofils soient suffisamment fins pour être dans un état de magnétisation axiale. Par conséquent, ces géométries peuvent être considérées comme des systèmes modèles pour étudier la transition entre les systèmes de verre de spin artificiels bidimensionnels et tridimensionnels. Nous avons étudié la structure de l'aimantation statique dans ces géométries et leurs propriétés haute fréquence. Pour étudier la dépendance de leur comportement en fonction de la taille, nous avons gardé certains paramètres géométriques constants : Nous avons fixé le rapport entre la longueur du nanofil ($L$), le rayon du nanofil ($R$) et le rayon de la sphère ($S$) à un rapport constant de 25:3:4 et fait varier la longueur du nanofil de 50 à 250 nm. Pour simuler le système magnétique, nous avons utilisé des paramètres de matériau correspondant aux valeurs typiques du cobalt déposé par FEBID, avec une constante de rigidité d'échange $A = 1.5 \times 10^{-11}\,\mathrm{J\,m^{-1}}$ et une aimantation à saturation $\mu_0 M_s = 1.2\,\mathrm{T}$, où $\mu_0$ est la perméabilité du vide. Nous avons supposé que l'anisotropie magnéto-cristalline est nulle. Ces paramètres du matériau

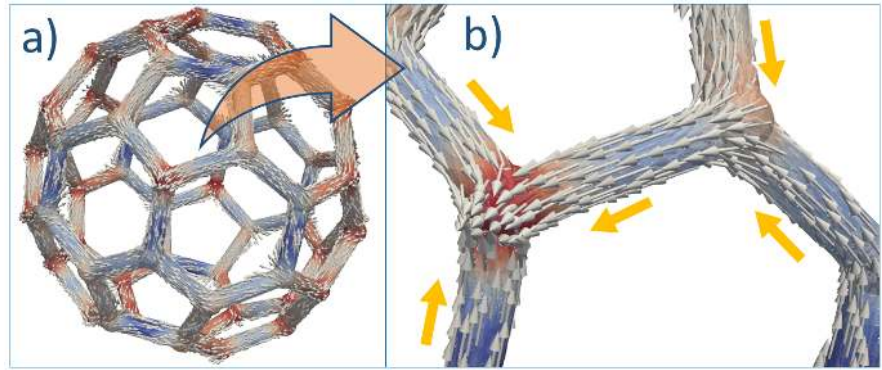

Figure 3: (a) Structure d'aimantation relaxée formée dans le Buckyball à champ nul (b) Vue agrandie de l'arrangement de l'aimantation aux sommets

.

donnent une longueur d'échange de 5.1 nm. Par conséquent, nous avons veillé à ce que tous nos modèles à éléments finis aient une longueur de maille maximale inférieure à cette valeur pour éviter les erreurs de discrétisation. De plus, pour une approximation géométrique lisse de la courbure des nanofils, la longueur de maille maximale a été maintenue en dessous de la moitié du rayon du nanofil. Les plus petites structures avaient donc une longueur de maille maximale de 3 nm, et les plus grandes utilisaient des cellules dont la taille pouvait atteindre 5 nm.

À champ nul, tous les nanofils de la structure Buckyball sont aimantés le long de leur axe. On peut donc attribuer un moment dipolaire de type Ising à chaque nanofil. Une situation micromagnétiquement intéressante se présente aux sommets, où trois de ces nanofils se rencontrent. En fonction de la direction d'aimantation relative de chaque nanofil se rencontrant à un sommet, quatre types différents d'arrangements de sommets frustrés peuvent se développer :

1. État deux entrées / une sortie : dans lequel l'aimantation de deux nanofils est orientée vers le sommet et un autre elle est orientée à l'opposé du sommet $(+1q)$

2. État une entrée / deux sorties $(-1q)$

3. État de trois entrées $(+3q)$

4. État de trois sorties $(-3q)$

Parmi ces quatre configurations possibles, les deux premières et les deux dernières sont équivalentes en raison de la symétrie d'inversion temporelle. Si nous attribuons une charge de $+1q$ à l'extrémité d'un nanofil avec un moment dirigé vers le sommet et une charge de $-1q$ à l'extrémité d'un moment de type Ising quittant le sommet, une valeur de charge peut être attribuée à chaque configuration de sommet comme indiqué ci-dessus. Cette valeur de charge est une mesure indirecte de la densité de charge du volume magnétostatique formée à ces points de sommet. Comme les deux premières configurations conduisent à une charge de $+1q$ et $-1q$, nous désignons ces états comme les "charges simples" et les deux dernières comme les "charges triples". Parmi ces configurations possibles, les états énergétiquement optimaux obéissant à la règle de la glace (*ice rule*) sont les configurations à charge unique, et les états à charge triple représentent des charges de défaut avec des propriétés de type monopole. En raison de leur structure micromagnétique, les charges triples ont une densité d'énergie d'échange et une énergie magnétostatique plus élevées. En exploitant la géométrie tridimensionnelle de la structure Buckyball, nous avons montré qu'il est possible de contrôler la formation et la suppression de ces défauts de charge triple au moyen d'un champ magnétique externe. Pour le démontrer, la boucle d'hystérésis de la Buckyball est simulée en appliquant un champ externe de $500\,\mathrm{mT}$ le long d'une diagonale reliant deux sommets opposés de la Buckyball, puis en réduisant progressivement le champ à $-500\,\mathrm{mT}$ par petites étapes. L'état rémanent obtenu lors de la relaxation à champ nul à partir de la saturation contient une paire de charges triples de signe opposé sur deux sommets situés sur les côtés opposés de la structure Buckyball. Même si les défauts de charge triple sont énergétiquement défavorables, ils restent stables une fois qu'ils sont formés. Cependant, si un champ de force croissante est appliqué dans la direction opposée, les charges triples finissent par se dissoudre, ce qui donne un arrangement magnétique contenant uniquement des sommets à charge simple. Ce processus est irréversible, c'est-à-dire que si l'on revient au champ zéro à partir de ce point, les charges triples dissoutes ne sont pas générées à nouveau. Ainsi, en faisant varier le champ externe, nous pouvons obtenir deux états stables qualitativement différents au champ zéro : un "état de charge triple" qui contient une paire de défauts de charge triple et un "état de charge simple" qui est exempt de toute charge de défaut. En répétant le processus pour des Buckyballs de différentes tailles, nous avons pu observer un comportement similaire dans toutes ces structures. La courbe M-H devient plus lisse lorsque la taille des Buckyballs augmente et, en même temps, la rémanence et le champ coercitif diminuent. De plus, la structure micromagnétique des sommets triplement chargés évolue d'une structure de paroi de domaine à trois têtes dans la plus petite Buckyball à une structure de tourbillon tridimensionnel dans le cas de la plus grande Buckyball.

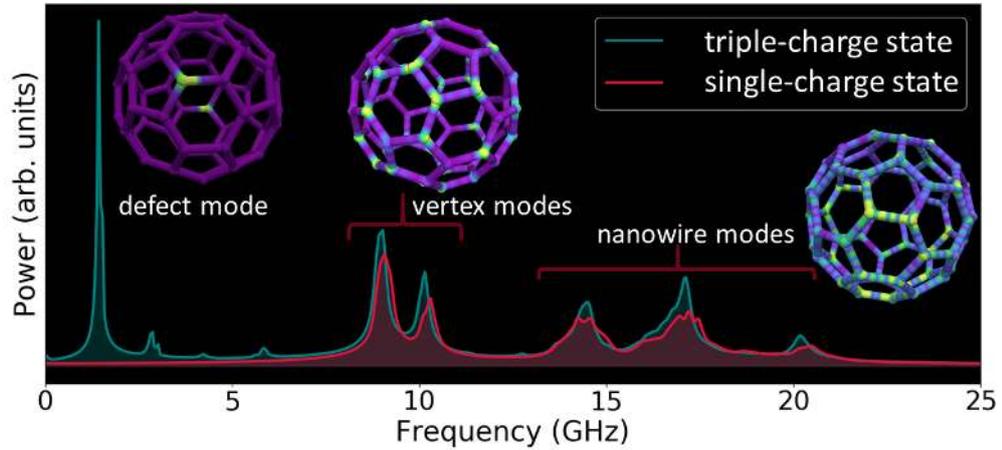

Figure 4: Spectre d'excitation magnétique à haute fréquence de l'état de charge simple et triple, avec le profil spatial correspondant des modes.

Le spectre d'excitation magnétique haute fréquence des différents états relaxés est simulé en excitant ces états avec une petite perturbation et en enregistrant les modes de précession à petit angle qui en découlent. Il y a cinq modes importants dans le spectre de fréquence de l'état "charge unique" : les deux premiers modes de plus basse fréquence sont localisés aux points de sommet, où trois nanofils se rencontrent. Les trois modes de fréquence plus élevée sont les modes des nanofils correspondant à la formation d'ondes stationnaires de différents ordres au sein des nanofils reliant ces sommets. Pour connaître l'effet de la formation de défauts à triple charge au sein de ces structures, la réponse en fréquence de l'état à triple charge est comparée à celle de l'état à simple charge. Nous observons que, dans les deux cas, tous les modes précédents restent les mêmes. Cependant, la présence de charges triples introduit un mode basse fréquence intense à environ 1.4 GHz qui est localisé aux sommets de la charge triple. Pour étudier la dépendance de la taille des propriétés haute fréquence, la procédure est répétée pour les Buckyballs de toutes tailles et les résultats sont comparés. La fréquence d'oscillation des sommets à charge simple et triple diminue linéairement avec l'augmentation de la longueur du nanofil. De plus, nous constatons que la fréquence des modes de défauts à triple charge peut être contrôlée au moyen d'un champ externe, ce qui donne une dépendance presque linéaire au champ.

Parallèlement aux techniques de nanofabrication directe telles que FEBID, ces types de nanoarchitectures peuvent également être fabriqués en gravant la géométrie requise sur une résine non magnétique, puis en recouvrant l'échantillon d'un matériau ferromagnétique. Les structures fabriquées de cette manière sont magnétiquement creuses. Elles peuvent donc être modélisées comme un ensemble de nanotubes magnétiques creux interconnectés, dont les points d'intersection sont des coquilles sphériques creuses. Par rapport aux modèles solides discutés précédemment, les structures creuses ont une surface interne supplémentaire qui peut leur conférer des propriétés additionnelles. Pour étudier de tels systèmes et vérifier la validité générale de nos observations, nous avons étendu une enquête similaire aux Buckyballs creuses. Nous avons obtenu, dans l'ensemble, des résultats similaires à ceux déjà trouvés dans le cas des modèles solides. En particulier, la configuration magnétique est très similaire, avec des nanotubes aimantés le long de l'axe et quatre types différents de configurations de sommets. Les Buckyballs creuses présentaient également des modes haute fréquence similaires, avec les sommets à triple charge oscillant à une fréquence beaucoup plus basse que les sommets à simple charge. Cependant, par rapport au cas solide, nous avons constaté que la fréquence d'oscillation des modes à triple charge est décalée vers des valeurs plus élevées, tandis que celle des sommets à charge unique est décalée vers une fréquence plus basse. De plus, les modes correspondant à l'oscillation des nanofils étaient presque entièrement supprimés dans les structures creuses.

Même si les Buckyballs sont des nanostructures tridimensionnelles, elles conservent certaines caractéristiques bidimensionnelles car le réseau de nanofils peut être considéré comme une surface bidimensionnelle enroulée sur la surface d'une sphère. En revanche, les prochains types de structures que nous avons étudiés sont des réseaux tridimensionnels à part entière dans le sens où ils s'étendent également dans toutes les directions. Ces types de structures sont formés par un réseau périodique de nanofils interconnectés, de sorte que leurs points de rencontre correspondent aux sites atomiques des cristaux naturels. Ces types de structures peuvent être considérés comme une extension des systèmes artificiels de verre de spin de deux à trois dimensions. Ces structures ont une importance scientifique car elles peuvent potentiellement combiner les caractéristiques des systèmes de verre de spin artificielle à celles du magnétisme tridimensionnel. Les propriétés et le comportement de ces types de structures peuvent être accordés en contrôlant leur configuration et, par conséquent, ces structures peuvent être classées comme une nouvelle sous-catégorie de métamatériaux magnétiques. De façon analogue à la façon dont les photons et les phonons peuvent être contrôlés dans les cristaux photoniques et acoustiques, ces cristaux magnoniques ont le potentiel de modifier artificiellement la propagation des magnons, l'unité fondamentale de l'excitation magnétique. Cette propriété donne également à ces structures des applications potentielles dans les dispositifs logiques basés sur les magnons. Bien qu'il s'agisse d'une catégorie de nanostructures prometteuse, on en sait peu sur leurs propriétés fondamentales. Dans cette thèse, nous avons étudié deux types de cristaux magnoniques artificiels de ce type.

Le premier type de structures de ce type que nous avons étudié sont des treillis artificiels de type diamant, qui sont formés par un réseau tridimensionnel interconnecté de nanofils, de sorte que leurs points de sommet sont situés à des positions correspondant aux sites atomiques d'un cristal de diamant naturel. Quatre nanofils se rencontrent à chaque sommet dans une disposition tétragonale. Le réseau que nous avons simulé a une taille totale de $450 \times 450 \times 450 \, \text{nm}^3$ et est composé de nanofils de longueur $70 \, \text{nm}$ et d'épaisseur $7 \, \text{nm}$. Le treillis entier contient 202 nanofils interconnectés à 83 points de sommet. Le modèle par éléments finis a une taille de cellule maximale de 4 nm et contient plus de 700 000 éléments tétraédriques. Nous avons étudié leurs structures d'aimantation statique, leurs propriétés hystérétiques et les modes d'oscillation haute fréquence apparaissant dans différents micro-états.

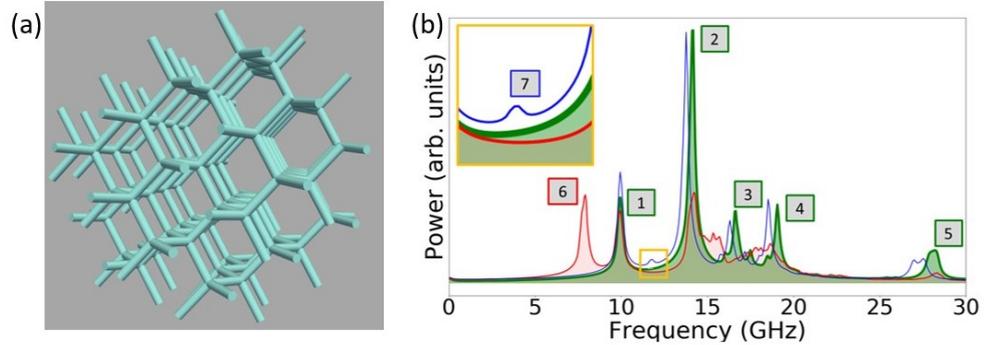

Figure 5: (a) Modèle par éléments finis du réseau de diamants utilisé pour la simulation (b). Spectre d'excitation haute fréquence des différents états ; vert : état de charge nulle, rouge : état désordonné, bleu : état de défaut géométrique.

Comme dans le cas des Buckyballs, les nanofils sont suffisamment fins pour qu'à champ nul, les branches individuelles aient tendance à être aimantées le long de leur axe. Par conséquent, comme dans le cas précédent, nous pouvons attribuer un moment dipolaire magnétique de type Ising à chaque nanofil. Puisque quatre nanofils se rencontrent à chaque sommet, il existe cinq dispositions possibles des sommets, à savoir :

1. deux entrées / deux sorties ($0q$)

2. trois entrées / une sortie ($+2q$)

3. une entrée / trois sorties ($-2q$)

4. quatre entrées ($+4q$)

5. quatre sorties ($-4q$)

Parmi ces quatre arrangements, le deuxième et le troisième, ainsi que le quatrième et le cinquième sont respectivement équivalents en raison de la symétrie d'inversion temporelle. L'état énergétiquement optimal, qui obéit à la règle de la glace ("ice rule"), est le premier arrangement dans lequel le flux magnétique vers et à l'opposé du sommet est parfaitement équilibré, ce qui conduit à une configuration dans laquelle la charge volumique nette au sommet est nulle. Les deuxième et troisième arrangements sont les charges doubles de type défaut avec des propriétés de type monopole. Contrairement aux sommets à charge nulle, ces types de sommets développent une charge magnétostatique nette non nulle. Étant donné que les charges magnétiques peuvent agir comme des sources et des puits du champ magnétique $\boldsymbol{H}$, la présence de telles charges de défaut donne à ces structures des propriétés supplémentaires. Les configurations 4 et 5 n'ont pas été observées dans un état stable dans aucune de nos simulations, car ces arrangements semblent être énergétiquement instables. En plus des sommets, nous pouvons également attribuer une charge de $+1q$ ou $-1q$ aux extrémités libres à la surface du treillis.

Les états d'aimantation relaxés formés dans ces structures sont sensibles à leur histoire magnétique. Ainsi, nous pouvons générer divers états qualitativement différents en saturant le réseau le long de différentes directions, puis en relaxant jusqu'au champ zéro. L'application d'un champ suffisamment fort pour saturer l'échantillon dans la direction [1,0,0], de sorte que le champ ne soit pas parallèle à l'un des nanofils, puis la relaxation jusqu'au champ zéro, donne un état ordonné à charge nulle qui ne contient que des sommets obéissant à la règle de la glace avec une charge nette nulle. En revanche, en saturant le réseau dans la direction [1,1,1] de façon à ce que le champ soit aligné parallèlement à l'une des directions d'orientation des nanofils, et en relaxant le système jusqu'au champ zéro, on obtient un état de charge défectueuse qui ne contient que des sommets chargés de type $+2q$ et $-2q$. En relaxant la structure à un état stable à champ nul à partir d'une configuration initiale aléatoire, on obtient un état désordonné, dans lequel les sommets à charge nulle et les sommets à charge double sont répartis de façon aléatoire. Le renversement de l'aimantation dans ces structures en fonction du champ est étudiée en simulant leur boucle d'hystérèse. Nous observons que le renversement de l'aimantation implique une commutation magnétique abrupte des nanofils, ce qui entraîne des sauts brusques dans la courbe M-H.

Les propriétés magnoniques du réseau ont été étudiées en excitant les différents états relaxés et en enregistrant le "ringdown" magnétique qui s'ensuit. Le spectre haute fréquence de l'état sans défaut contient trois types de modes. Premièrement, les modes de surface, qui sont formés par l'activité des extrémités libres avec des charges $+1q$ et $-1q$ à la surface du treillis. Deuxièmement, les modes de sommet qui sont dus à l'oscillation des sommets à charge nulle à l'intérieur du réseau, et enfin les modes de nanofils à plus haute fréquence, qui sont causés par des oscillations de type onde stationnaire au sein des nanofils. Pour étudier l'effet de la formation des charges de défaut sur le spectre magnonique, les modes haute fréquence de l'état désordonné (qui contient des charges $+2q$ et $-2q$, ainsi que des charges nulles) sont comparés à ceux de l'état à charge nulle. Nous avons observé que tous les modes précédents sont restés presque les mêmes, tandis qu'un mode basse fréquence supplémentaire et intense est apparu, qui est localisé aux sommets de la double charge. Il est intéressant de noter que les sommets à double charge oscillent à une fréquence considérablement plus basse que les sommets à charge nulle, bien que l'oscillation se produise à des emplacements géométriques similaires. Nous avons en outre constaté qu'il n'y a pas de différence de fréquence ou d'intensité d'oscillation entre les sommets chargés $+2q$ et $-2q$. Nous avons également observé une baisse de l'intensité de l'oscillation des sommets à charge nulle par rapport à l'état de référence. Cela est dû à la diminution du nombre de sommets à charge nulle dans l'état désordonné, car certains sommets acquièrent une configuration à double charge. Alors que les modes du nanofil dans l'état de référence apparaissent comme trois pics bien définis, dans l'état désordonné, ils sont mélangés et essentiellement indiscernables les uns des autres. Cela peut être attribué à l'effet des charges $+2q$ et $-2q$, qui agissent comme des sources de champs magnétiques locaux qui modifient la fréquence d'oscillation des nanofils correspondants.

L'évolution du spectre de fréquence du réseau en fonction d'un champ magnétique externe a été étudiée en appliquant le champ dans deux directions, puis en observant les changements résultants. Nous avons observé que la réponse à un champ externe dépend de l'orientation du champ ex-terne par rapport à la direction de saturation du treillis. Lorsque le champ est appliqué parallèlement à la direction de saturation, la fréquence de tous les modes augmente linéairement avec l'intensité du champ. En revanche, lorsque le champ est appliqué dans la direction opposée, la fréquence des modes diminue. Lorsque le champ externe est appliqué perpendiculairement à la direction de saturation, les modes de surface et des nanofils se divisent en deux et quatre sous-branches, en fonction de leur orientation par rapport au champ externe.

Après avoir établi comment la présence de défauts magnétiques affecte le spectre magnonique, nous avons étudié comment les propriétés magnoniques peuvent être manipulées par la formation de défauts géométriques. Pour cela, nous avons introduit un défaut structurel dans le réseau en retirant un nanofil individuel à l'intérieur du réseau, créant ainsi une paire de sommets avec un numéro de coordination de trois. Comme dans le cas des états obéissant à la règle de la glace dans la structure buckyball, y résultent deux types de configurations stables dans ces sommets défectueux :

1. deux entrées / une sortie
2. une entrée / deux sorties

Pour comprendre l'effet de la formation de défauts géométriques sur les propriétés magnoniques, les excitations haute fréquence qui se développent dans un réseau avec un défaut géométrique sont comparées à celles d'un réseau sans défaut en état de charge nulle. La présence d'un défaut structurel introduit un pic supplémentaire dans le spectre de puissance, localisé uniquement sur les sites des défauts. L'intensité de ce nouveau pic est comparativement faible, car il provient de seulement deux sommets sur les 83 sommets totaux. L'observation d'un pic d'absorption supplémentaire est comparable aux phénomènes de modification des spectres optiques des cristaux naturels par la présence d'impuretés et de défauts.

Jusqu'à présent, nous avons étudié les modes statiques dans le sens où les oscillations sont localisées à des positions géométriques spécifiques, sans évoluer dans le temps. Cependant, en analysant l'oscillation filtrée par une analyse de Fourier des sommets dans l'état de charge zéro, nous avons pu identifier une corrélation de phase à longue portée. Cela suggère l'existence d'une onde de spin macroscopique se propageant du cœur du réseau vers le centre, par le biais d'une corrélation spatiale et temporelle de l'oscillation des modes des sommets. Comme nous l'avons déjà dit, les propriétés magnoniques de ces structures peuvent être contrôlées en manipulant leur structure magnétique et aussi en introduisant des défauts structurels. Si une onde de spin macroscopique se propageant peut être réalisée dans une telle structure, cela pourrait ouvrir de nouvelles perspectives pour des applications potentielles en magnonique. Pour étudier cette possibilité, nous avons modélisé une version allongée du réseau et, au lieu d'exciter la structure entière, nous avons appliqué une impulsion de champ localisée à l'une des extrémités de la structure et nous avons analysé les oscillations magnétiques qui en résultent. Contrairement à ce que nous attendions, nous n'avons pas pu observer de propagation significative dans le sens longitudinal lors de nos premières tentatives. Cela suggère que le couplage magnéto-dipolaire entre les sommets n'est pas assez fort pour un transfert d'énergie efficace et pour une propagation macroscopique des ondes de spin qui en résulte. Cependant, il existe un large éventail de paramètres à faire varier, tels que la longueur et l'épaisseur des nanofils, les propriétés du matériau *etc.*, qui pourraient être abordés dans de futures études.

Dans la section précédente, nous avons étudié la formation de différentes configurations de sommets dans un cristal magnonique artificiel (AMC) de type diamant ainsi que leur effet sur la signature magnonique du réseau. Pour démontrer la robustesse de ces observations, nous étendons notre enquête à un autre type d'AMC avec une structure de réseau cubique. La structure de treillis cubique est formée par l'interconnexion de nanofils dans une disposition cubique simple, de sorte que six fils se rencontrent à chaque sommet. Notre structure modèle est composée de nanofils cylindriques de 7,0 nm de rayon et 70,0 nm de longueur, disposés en un treillis de $7 \times 6 \times 5$. Comme six nanofils se rencontrent à chaque sommet, au lieu des quatre du réseau de type diamant, un plus grand nombre de configurations de sommets sont possibles, comme indiqué ci-dessous :

1. Trois entrées / trois sorties ($0q$) (État respectant la règle de la glace)

2. Quatre entrées / deux sorties ($+2q$)

3. Deux entrées / quatre sorties ($-2q$)

4. Cinq entrées / une sortie ($+4q$)

5. Une entrée / cinq sorties ($-4q$)

6. Six entrées ($+6q$)

7. Six-sorties ($-6q$)

Parmi ces sept configurations possibles, les deux dernières n'ont pas été observées en tant que structures stables, ce qui suggère qu'elles sont énergétiquement instables. Dans le cas du réseau de type diamant, puisque les quatre nanofils sont disposés de manière tétragonale, les trois voisins d'un nanofil sont équivalents. En revanche, les cinq voisins du réseau cubique ne sont pas équivalents. Quatre des cinq voisins sont disposés normalement au nanofil, et le cinquième fil est disposé à l'opposé. À cause de cette disposition géométrique particulière, il existe différents "variants" de configurations de sommets ayant la même charge.

Par exemple, il existe deux types différents d'états "deux entrants/deux sortants", avec une structure micromagnétique différente et, par conséquent, des propriétés dynamiques différentes. Comme dans le cas précédent, il est possible de générer des états qualitativement différents en saturant la structure le long de différentes directions, et ensuite relaxant le système au champ zéro. Nous avons pu générer trois états différents : un état de référence qui ne contient que des sommets à charge nulle de type un obéissant à la règle de la glace et un autre état qui contient des sommets à charge double ainsi que les deux types de charges nulles et un troisième état qui contient des sommets à charge quadruple, ainsi que des sommets à charge double et nulle. Le spectre haute fréquence des différents états statiques a été simulé et comparé. Le réseau cubique présente des caractéristiques magnoniques similaires à celles du réseau diamant : Là aussi, les modes dans l'état de référence sans défaut peuvent être classés en modes de surface, de sommet et de nanofil. Les différents types de configuration des sommets oscillent à des fréquences différentes, c'est pourquoi l'introduction d'un nouveau type de sommet dans le réseau entraîne l'apparition d'un nouveau pic caractéristique. En général, les sommets appartenant au même type de charge oscillent à la même fréquence, et la fréquence d'oscillation diminue lorsque la magnitude de la charge augmente. Cette observation est vraie pour tous les types de charge, sauf pour les sommets à charge nulle. Les sommets à charge nulle dans le treillis cubique oscillent à deux fréquences différentes, selon leur type ou leur variant.

Pour résumer, nous avons réalisé des simulations micromagnétiques par éléments finis sur la dynamique de l'aimantation à haute fréquence dans les nanocylindres de cobalt et dans plusieurs types de nanoarchitectures tridimensionnelles : Tétraèdres de type Sierpinski, structures Buckyballs, structures à réseau de type diamant et structures à réseau cubique. La flexibilité de la méthode d'éléments finis s'est avérée cruciale pour modéliser avec précision la géométrie compliquée et les surfaces courbes de ces structures. Nous avons identifié les différentes configurations d'équilibre dans les tétraèdres de Sierpinski, y compris un état de vortex déconnecté, et étudié comment ces états affectent les propriétés magnoniques. Nous avons observé une réponse en fréquence à large bande dans les étages supérieurs de la fractale. Dans les structures Buckyball, nous avons pu identifier quatre configurations de sommets différentes, dont deux états de charge simple obéissant à la règle de la glace et deux configurations de charge triple de type défaut, en fonction de la configuration magnétique des nanofils. La formation de ces charges de défaut peut être contrôlée en utilisant des champs magnétiques externes. La présence de ces défauts a un fort impact sur les propriétés haute fréquence de ces structures. Cette capacité à contrôler le spectre magnonique ouvre de nouvelles possibilités d'applications potentielles de ces structures en magnonique. De plus, cet effet peut être exploité comme un outil indirect pour détecter expérimentalement ces défauts par des mesures de fréquence. L'étude a été étendue aux Buckyballs creuses, qui présentent globalement un comportement similaire à celui de leurs homologues solides. Deux types de cristaux magnoniques artificiels ont également été étudiés. Le réseau en forme de diamant est constitué d'un réseau périodique de nanofils avec quatre nanofils se rencontrant à chaque som-met. Nous avons identifié trois configurations de sommets stables dans la structure du diamant, dont les charges nulles obéissant à la règle de la glace et deux types de charges doubles. Différents microétats peuvent être générés dans un tel treillis en saturant le treillis dans différentes directions. Le spectre de fréquence du réseau contient trois types de modes : les modes de surface, les modes de sommet et les modes de nanofils. Dans les modes de sommet, les sommets défectueux oscillent à une fréquence beaucoup plus basse que celle des sommets à charge nulle. Ces modes peuvent être contrôlés plus précisément à l'aide de champs externes. L'introduction de défauts géométriques dans le treillis a entraîné l'apparition de modes de fréquence supplémentaires. Pour vérifier la validité générale de ces résultats, un autre cristal magnonique artificiel a également été étudié, à savoir le treillis cubique. Ses sommets sont formés par l'intersection orthogonale de six nanofils. En raison de l'augmentation du nombre de coordination, davantage de types de configurations de sommets sont présents dans le treillis cubique. De plus, en raison des effets géométriques, différents variants de configurations de sommets avec la même charge ont été observées. Le spectre de fréquence du réseau cubique présente en grande partie le même comportement que celui du réseau de type diamant, en ce sens qu'ils contiennent tous deux des modes qui peuvent être clairement attribués aux oscillations de la surface, des sommets et des nanofils, et que les sommets avec différents types de configuration oscillent à des fréquences différentes.

# CONTENTS

# Chapter 1

Introduction

In the course of history, research in magnetism has contributed significantly to the development of human civilization. Humanity's fascination for magnetism started in ancient Greece with the discovery of the Lodestone and has evolved ever since and revolutionized the life of humans [1]. The invention of the magnetic compass enabled the great voyages and thereby the age of discovery. The invention of the dynamo eventually led to the large-scale production of electricity and all the benefits electricity brought with it. High-density magnetic data storage development paved the way for today's society with its abundance of information. Other applications like magnetocaloric refrigeration or high-power electromotors have also evolved from research in magnetism. Today, magnetism has penetrated practically all branches of science and technology.

Nanomagnetism [2] is the sub-branch of magnetism that concerns the study of the properties of magnetic materials at ever increasingly smaller dimensions. Research in nanomagnetism aims to manipulate, control, and exploit magnetic phenomena on the nanometer scale. Questions of interest in this scientific domain are how the magnetization adapts to finite-size effects arising in low-dimensional systems, such as, e.g. thin films, nanowires, or nanodots. Over the past decades, extensive studies have investigated ways to use external stimuli to control the magnetization in such systems. Methods to manipulate the magnetization on the nanoscale include the application of magnetic fields, spin-polarized currents, or approaches exploiting other coupling mechanisms like the magnetoelectric effect.

Along with scientific curiosity, nanomagnetism research has traditionally been motivated by the rapid progress in information technology and its demand for low-power, high-density [3], and non-volatile storage media. This demand skyrocketed with the discovery of the Giant magneto-resistance effect [4, 5] (GMR), for which P. Grünberg and A. Fert were awarded the Nobel Prize in Physics in 2007. This effect, along with the Tunnel magneto-resistance effect [6, 7] describes a significant change of the electric resistance in a magnetic multilayer system depending on the relative orientation of the magnetization direction of the layers. These discoveries started the field of spintronics [8, 9, 10], in which the interaction of the electron spin and the magnetization is of central importance. These studies led to a trend in nanomagnetism where much of the research was focused on thin films and interface effects.

In addition to its relevance for spintronic applications, the scientific interest in nanomagnetism extends far beyond the desire to improve device miniaturization and other technological aspects. On the sub-micron length scale, magnetic materials often develop unique and characteristic features, such as the formation of magnetic vortices [11], domain walls, skyrmions [12, 13, 14], or other topological structures. These fundamental magnetic structures can display particle-like properties and high stability. Their typical size, which is governed by material-specific length scales (the exchange lengths), is in the range of a few nanometers for usual ferromagnetic materials. These properties render such structures exciting candidates for shift-register type applications. Consequently, much effort has been made to understand how these magnetic structures are affected by finite-size effects. A prominent example is the magnetic racetrack memory[15, 16] device concept, whose basic principle is to use magnetic domain walls in nanowires as units of information that can be displaced within the device in a controlled way through electrical currents [17]. In this specific case, the nanopatterning of a magnetic material serves to guide particle-like magnetic structures (a more recent version of the racetrack memory device uses skyrmions [18, 19, 20] instead of domain walls) along a predefined path. But there are many other cases where finite-size effects on the submicron scale can significantly impact a nanomagnet's properties. Over more than two decades, the influence of the size and shape of magnetic nanoparticles has been a central topic in magnetism research [2]. Experimental and simulation studies have evidenced the delicate impact that details of a magnetic particle's geometry can have on its static, dynamic, and hysteretic properties. The patterning of nanomagnets is also important in magnonics, where suitably shaped magnets can guide high-frequency magnetic excitations (spin waves) to exploit interference effects for logical operations and data processing [21].

Most of these studies on the shape- and size dependence in nanomagnetism concerned patterned thin-film elements of varying thickness [2, 22]. However, recent advancements in direct write techniques such as Focused-electron-beam-induced deposition (FEBID) [1] [23, 24, 25, 26] or lithographic techniques such as two-photon-lithography (TPL) [2] [27, 28, 29], in combination with various deposition techniques

---

[1] Focused electron beam induced deposition: this is a direct write technique in which a precursor material injected as a gas is deposited on the surface of a substrate on the focus of an electron beam. Thus controlling the position of the beam arbitrary three dimensional patterns could be written.

[2] TTwo photon lithography: In this technique a femtosecond laser on the infrared range is focused on a photosensitive resist. the high intensity of the focused beam allows the absorption of two photons which results in a photosensitive polymerisation of the resist. This technique can thus be employed to fabricate three dimensional nanostructures

[30, 29, 31] have made it possible to fabricate an entirely new category of magnetic nanostructures by extending the patterning into three dimensions instead of two. The fabrication techniques offer possibilities to generate magnetic materials with nanometric feature size and of essentially arbitrary shape. These techniques can be used, for instance, to fabricate magnetic nano-architectures consisting of interconnected magnetic nanowires. In parallel to this development in nano-patterning techniques, there has also been spectacular progress in measurement and imaging techniques such as neutron tomography [32, 33], electron tomography [34], X-ray nano-tomography [35, 36] and laminography [37]. These techniques make it possible to probe and image complicated three-dimensional magnetization structures and their dynamical properties, even if they are buried deep inside the volume of a three-dimensional magnetic sample [38, 36]. In the past, such information was accessible only indirectly through comparisons with micromagnetic simulations [39], which were often combined with surface-sensitive experimental imaging techniques [40], or by reconstructing the contrast obtained from grazing-incidence XMCD transmission measurements. Due to this two-pronged development in fabrication and measurement techniques, three-dimensional (3D) nano-magnetism has recently evolved into an important research topic in magnetism [26, 41]. Similar to how the properties of the bulk material change under the influence of various size-, shape- and interface effects in two-dimensional magnetic structures, the patterning of these novel three-dimensional nano-architectures could bring out new magnetic phenomena and physical effects absent in two-dimensional systems. Such new features could result in new functionalities for various applications, such as high-density data storage, magnonics, non-conventional computing [26]. These developments could also result in a new type of magnetic metamaterials [42, 43]: ferromagnets with artificial properties generated by the peculiar three-dimensional structure and that do not occur naturally in the magnetic material. Such magnetic metamaterials would be a logical extension of existing optical and acoustic metamaterials made of suitably shaped periodic 3D nanoarchitectures.

The extension of nanostructures into the third dimension often gives rise to new types of magnetization structures such as, e.g., 3D vortex domain walls [37, 44, 38], Bloch point structures [45], 3D skyrmions [46, 47, 48], or Hopfions [49, 50]. These 3D spin structures with non-trivial topologies often have higher stability and improved susceptibility to magnonic excitations. Moreover, they may be manipulated through spin-polarized currents and therefore have the potential to be exploited for applications in three-dimensional spintronics. Hopfions, for example, could be ideal candidates for racetrack-type devices because owing to their vanishing gyrovector, they should not experience deflections by the Hall effect, which affects the motion of current-driven skyrmions [51]. Even conventional spin structures such as domain walls can have different dynamical properties in 3D nanostructures when compared to 2D. For example, a domain wall propagating in a 2D strip under the influence of an external field would become unstable above a particular speed due to Walker breakdown [52]. In contrast to this, previous studies with micromagnetic simulations [53, 54] have shown that winding the nanostrips in the form of a tube can suppress the Walker breakdown [55] and thereby permit domain-wall velocities as fast at $1000 \, \mathrm{m \, s^{-1}}$. These ultra-fast domain-walls can even reach the phase velocity of spin waves and thus create a Spin-Cherenkov effect [53, 56]. This phenomenons is similar to the existence of a "sonic-boom" in the case of supersonic projectiles. In addition, when compared to planar thin films, the curved surfaces of nanotubes lack space inversion symmetry, which can lead to various curvature-induced effects [57, 58, 59], such as curvature-induced Dzyaloshinskii–Moriya interaction (DMI) [60] and curvature induced effective anisotropy. It has been suggested that these curvature-induced phenomena, whose strength could be controlled by the particle geometry, could provide a new method for manipulating the magnetic properties of the nanostructures [61]. Such effects also have the potential to be exploited in three-dimensional high-density data storage (3D-HDD) media, in which information can be stored in the form of magnetic entities such as domain-walls, skyrmions or vortices. This field of research, named cuvilinear magnetism [57, 62, 63], is still in the developing stage and one can assume that several interesting physical phenomena related to these effects and the associated applications are yet to be discovered.

An interesting sub-category of three-dimensional magnetic nanostructures are interconnected nanowire networks. The aforementioned sophisticated nanopatterning techniques allow for the assembly of nanowire networks in the form of complex nanoarchitectures, e.g., magnetic buckyballs [30] or arrays of "nano-trees" [64, 65, 66]. In such nanoscale architectural forms, a particular geometric situation develops at the vertices, where three or more nanowires meet. This arrangement of nanowires leads to magnetic frustration, as in the case of artificial spin ices (ASI) 2.4. Hence, these structures can be regarded as three-dimensional artificial spin ice (3D-ASI) systems. Along with novel advanced fabrication and measurement techniques, micromagnetic simulations are necessary to understand the physics of these three-dimensional nanoarchitectures.

In this thesis, we investigate the three-dimensional magnetic structures formed in different types of such nanoarchitectures and study their dynamic properties, like the response to external fields and their high-frequency dynamics. We investigate these effects through finite-element micromagnetic simulations. Principally, we investigated artificial Buckyball nanoarchitectures, diamond-lattice-like architectures, and cubic-lattice-like architectures. The thesis is divided into six chapters. The first chapter is the introduction, which provides the study's context, motivation, and relevance. The second chapter introduces the theory of micromagnetism. In particular, it discusses the micromagnetic energy terms and the equation describing the magnetization dynamics. In the third chapter, the numerical implementation of the micromagnetic theory is briefly introduced, along with the numerical details of a post-processing Fourier analysis tool developed in this thesis. The fourth chapter consists of the micromagnetic investigation of magnetic structures and high-frequency modes formed in Cobalt nanodots. The fifth chapter summarizes our studies of the Buckyball nanoarchitectures where their magnetic configurations, hysteretic properties, and high-frequency response are discussed in detail. The sixth chapter reports on the studies on two types of artificial magnonic crystals (AMC): the diamond-type crystal and the cubic-lattice-type crystal.

# CHAPTER 2

Fundamental aspects of micromagnetism

Micromagnetism is a continuum theory for ferromagnetic materials. It describes the magnetization structure on a mesoscopic length scale and captures the typical magnetic features developing in the sub-micron range. It thereby provides the appropriate theoretical basis to study the physical properties of typical ferromagnetic nanostructures, including the more recent three-dimensional types. Micromagnetic theory bridges the gap between atomistic and quantum-mechanical models on the one hand and a macroscopic description through Maxwell's theory of electromagnetism on the other hand. In the latter, material-specific electromagnetic phenomena are described through volume-averaged quantities, such as permeabilities and susceptibilities. Instead of this macroscopic description, the "microscopic" features of the magnetization (hence the name) are central to micromagnetic theory, even though their size is typically rather in the nanometer than in the micrometer range. In addition to providing a framework for calculating the spatial structure of the magnetization, the theory of micromagnetism also addresses aspects of the magnetization dynamics unfolding typically on the pico- and nanosecond time scale. The theory of micromagnetism was pioneered by W. F. Brown Jr in 1940-41 when he published two papers [67, 68] on the theory of micromagnetism of ferromagnetic materials building. It builds upon the previous works of Akulov [69], Bloch [70] and, perhaps most importantly, by Landau and Lifshitz in 1935 [71]. Later contributions by Kittel [72], Stoner-Wohlfarth [73], Néel [74], Aharoni [75, 76], and Shtrikman have further contributed to the foundations of the theoretical framework. More recently, the rapid development in computation technology has established the micromagnetic theory as an important branch of modern magnetism. Nowadays, computational micromagnetism is an indispensable tool for the fundamental investigation of nanomagnetic phenomena such as magnetization structures, energetics and dynamic properties of nano- and mesoscale systems.

## 2.1 Theory of micromagnetism

The central quantity of interest in micromagnetic theory is a ferromagnet's spontaneous magnetization field. The magnetization is defined as the density of magnetic moments per unit volume, and it is represented as a continuous vector field $\boldsymbol{M}(\boldsymbol{r}, t)$. A central assumption of micromagentic theory is that the vector field of the magnetization is directional, i.e., the magnetization vector has constant length $|\boldsymbol{M}| = M_s = \text{const.}$ within the magnetic material, where $M_s$ is the saturation magnetization of the material. Only its direction is allowed to vary in time and space, not its magnitude. Micromagnetism, in general, deals with the determination of $\boldsymbol{M}(\boldsymbol{x}, t)$ or the reduced magnetization $\boldsymbol{m}(x) = \boldsymbol{M}(\boldsymbol{x})/M_s$ at each point $\boldsymbol{x}$. For the static state, the basis for this involves the minimization of the free energy of the system. The free energy (or the total energy) of a ferromagnet depends on the structure $\boldsymbol{M}(\boldsymbol{r})$ of the magnetization. The most important contributions to a magnet's total energy arise from various components, the most important of which are usually the exchange energy, the magneto-crystalline anisotropy energy, magnetostatic-dipolar energy, and the Zeeman energy. The theory of micromagnetism uses continuum expressions of these intrinsic energy terms, which all depend on the magnetic structure $\boldsymbol{M}(\boldsymbol{r})$. Each of these expressions can be formally derived from more microscopic pictures involving discrete spins $S_z$ by taking the continuum limit. The transition to the continuum picture is based on the definition of the magnetization $\boldsymbol{M}(\boldsymbol{r})$ as the density of magnetic moments. In the following section we briefly introduce the theoretical expressions used to calculate each of the energies.

### 2.1.1 Exchange energy

The exchange interaction describes the fundamental tendency of ferromagnetic materials to align neighboring magnetic moments parallel to each other. The microscopic origin of the ferromagnetic exchange interaction is purely quantum-mechanical and arises from Pauli's exclusion principle. It is a short-range interaction that, by favoring a parallel arrangement of neighboring spins, gives rise to the phenomenon of ferromagnetism.

There are different ways to derive the micromagnetic form of the ferromagnetic exchange energy density. One possibility is to take the classical atomistic Heisebenberg interaction as a microscopic starting point and to perform a transition to the continuum limit. In the Heisenberg model, the exchange interaction between two magnetic moments $\boldsymbol{S}_i(\boldsymbol{r_i})$ and $\boldsymbol{S}_j(\boldsymbol{r_j})$ is given by

$$H_{\text{ex}} = -2 \sum_{i \neq j} J_{ij}\left(\boldsymbol{r}_{ij}\right) \boldsymbol{S}_i\left(\boldsymbol{r}_i\right) \cdot \boldsymbol{S}_j\left(\boldsymbol{r_j}\right) \tag{2.1}$$

where $J_{i,j}$ is the quantum-mechanical exchange integral [77]. Assuming that the angle $\phi_{ij}$ between neighboring spins $\boldsymbol{S}_i(\boldsymbol{r_i})$ and $\boldsymbol{S}_j(\boldsymbol{r_j})$ is small, one can apply a Taylor expansion of eq. (2.1) which, after a few manipulations [78], leads to the form

$$e_{\text{exc}} = A \left[ (\boldsymbol{\nabla} m_x)^2 + (\boldsymbol{\nabla} m_y)^2 + (\boldsymbol{\nabla} m_z)^2 \right] \tag{2.2}$$

Where $A$ is the exchange stiffness constant of the material expressed in $\text{J m}^{-1}$. In this derivation, $A$ is related to the quantum-mechanical exchange integral $J_{i,j}$ [77] and depends, moreover, on the atomic lattice constant and the crystalline structure [78]. In the theory of micromagnetism, $A$ is assumed to be a position-independent constant of the material.

A different approach consists in applying a phenomenological principle based on the fact that the exchange interaction tends to preserve a homogeneous structure of the magnetization field $\boldsymbol{M}(\boldsymbol{r})$. Therefore, any inhomogeneity in the magnetic structure should lead to an increase in exchange energy. In this perspective, one can assume that the expression for the exchange energy density should be a function of the spatial derivatives of the reduced magnetization $\partial m_i / \partial x_j$ in order to describe the degree of local inhomogeneity in the vector field of the magnetization. Several categories of combinations of gradients can be discarded for symmetry reasons. For instance, one can assume that the exchange energy is invariant under time-inversion symmetry, i.e., an operation of the type $\boldsymbol{M} \to -\boldsymbol{M}$. Based on such considerations, one can derive the term

$$e_{\text{xc}} = \sum_{i,k,l} A_{kl} \frac{\partial m_i}{\partial x_k} \frac{\partial m_i}{\partial x_l} \tag{2.3}$$

as a generalized expression of such a phenomenological "inhomogeneity energy density" [79, 80]. In practice, the symmetric tensor $A_{kl}$ can be represented as a constant $A$, and the expression for the ferromagnetic exchange simplifies to eq. (2.2), which describes the exchange energy per unit volume in the continuum representation. According to eq. (2.2), any deviation from a perfectly homogeneous magnetization state leads to an increase of the exchange energy density and is thus penalized, while the energy term vanishes for a perfect homogeneous state.

### 2.1.2 Crystalline anisotropy energy

The magnetocrystalline energy is derived from the coupling between spin and orbital moments (L-S coupling) and the interaction of the ions and the crystal field [81]. In a crystalline ferromagnet, this relativistic effect results in the formation of one or more easy axes along which the magnetization preferably aligns, and hard axes which represent unfavorable orientations of the magnetization. This anisotropic effect is accounted for in an energy term in which any angular deviation of $\boldsymbol{m}$ from the easy axis in energetically penalized. The simplest case of magneto-crystalline anisotropy is that of uniaxial anisotropy, in which there exists one preferred direction of magnetization (the easy axis). The direction orthogonal to this axis is called the hard-axis. The uniaxial anisotropic energy density $e_{ku}$ can be expressed as follows as the function of the relative orientation the magnetization and the direction of the easy axis $\boldsymbol{k}$

$$e_{\mathrm{ku}} = -K_{\mathrm{u}1}(\boldsymbol{m} \cdot \boldsymbol{k})^2 + K_{\mathrm{u}2}(\boldsymbol{m} \cdot \boldsymbol{k})^4 \tag{2.4}$$

The constants $K_{u1}$ and $K_{u2}$ are the anisotropy constants of first and second nonvanishing order, respectively, which in SI units are expressed as $\mathrm{J\,m^{-3}}$. A preferential magnetization direction can also arise from the effect of the so-called *shape anisotropy*, which favors magnetization along the long axis of elongated particles. The shape anisotropy, however, is a purely magnetostatic effect that is unrelated to the crystalline structure of the material.

### 2.1.3   Zeeman energy

The magnetostatic energy of a ferromagnet in an magnetic field—the Zeeman energy—is the energy connected with the interaction of an externally applied field $\boldsymbol{H}_{\mathrm{ext}}$ with the magnetization $\boldsymbol{M}(\boldsymbol{r})$. The Zeeman energy density $e_{zee}$ can be expressed as follows.

$$e_{\mathrm{Zee}} = -\mu_0 \boldsymbol{H}_{\mathrm{Zee}} \cdot \boldsymbol{M} \tag{2.5}$$

According to eq. (2.5), the value of Zeeman energy is minimal (most negative) when the magnetization is aligned parallel to the direction of the external field, while the orientation anti-parallel to the applied field is energetically penalized.

### 2.1.4   Demagnetizing energy

The demagnetizing energy is the magnetostatic energy originating from the interaction of the magnetization with its own dipolar field. The latter is also called the stray field or the demagnetizing field. This interaction is often the most difficult term to calculate in micromagnetism. The magnetostatic dipolar field results from the sum of all magnetic dipole moments that constitute the magnetic structure. Within the discrete or atomistic description, each magnetic moment represents a magnetic point dipole that contributes to the sample's dipolar field. The individual magnetic moments can be expressed as

$$\boldsymbol{\mu_i}(\boldsymbol{r_i}) = g_L \mu_B \boldsymbol{S}(\boldsymbol{r_i}) \tag{2.6}$$

where $g_L$ is the Lande g-factor

Then the dipolar field can be expressed as the sum of individual dipoles,

$$\boldsymbol{H}_{dip.}(\boldsymbol{r}) = \frac{1}{4\pi} \sum_i \left( \frac{\boldsymbol{\mu_i}(\boldsymbol{r_i})}{R^3} - \frac{3 \left( \boldsymbol{\mu_i}(\boldsymbol{r_i}) \cdot \boldsymbol{R} \right) \cdot \boldsymbol{R}}{R^5} \right) \tag{2.7}$$

where $\boldsymbol{R} = \boldsymbol{r} - \boldsymbol{r_i}$ and $R = |\boldsymbol{R}|$.

The micromagnetic continuum theory does not use such dipole sums. Instead, the continuum expression of the dipolar field involves integrals over the so-called magnetostatic charge distribution [82, 83, 79]. A suitable starting point for deriving this expression is the constitutive equation [84, 85], which relates the magnetic field $\boldsymbol{H}$, the magnetic induction $\boldsymbol{B}$, and the magnetization $\boldsymbol{M}$:

$$\boldsymbol{B}(\boldsymbol{r}) = \mu_0 \boldsymbol{H}(\boldsymbol{r}) + \mu_0 \boldsymbol{M}(\boldsymbol{r}) \tag{2.8}$$

In the absence of electrical charge and displacement currents, $\boldsymbol{j} = 0$ and $\partial \boldsymbol{D}/\partial t = 0$, Maxwell's equations yield that the magnetostatic field $\boldsymbol{H}$ is irrotational [84]:

$$\boldsymbol{\nabla} \times \boldsymbol{H} = 0 \tag{2.9}$$

Hence, $\boldsymbol{H}$ can be expressed as the gradient of a scalar potential $U(\boldsymbol{r})$

$$\boldsymbol{H}(\boldsymbol{r}) = -\boldsymbol{\nabla} U(\boldsymbol{r}) \tag{2.10}$$

where $U(\boldsymbol{r})$ is the magnetostatic scalar potential. Equating for $\boldsymbol{H}$ in equation (2.8) and taking the divergence yields

$$\boldsymbol{\nabla} \cdot \boldsymbol{B} = \mu_0(-\nabla^2 U + \boldsymbol{\nabla} \cdot \boldsymbol{M}) \tag{2.11}$$

The left-hand side vanishes according to Maxwell's equation about the absence of magnetic monopoles $\boldsymbol{\nabla} \cdot \boldsymbol{B} = 0$. Inserting this in 2.11 we obtain the Poisson equation

$$\nabla^2 U = -\rho(\boldsymbol{r}) \tag{2.12}$$

Where $\rho(\boldsymbol{r}) = -\boldsymbol{\nabla} \cdot \boldsymbol{M}$ is the magnetostatic volume charge density. From eq. (2.12), it becomes clear that the dipolar field vanishes in an infinitely extended sample with homogeneous magnetization. The dipolar field arises as a result of spatial inhomogeneities of $\boldsymbol{M}_s$, which can be related either to divergences of the magnetic orientation or to a change of the modulus of $|\boldsymbol{M}_s|$, as it occurs, e.g., across the surface of a magnetic sample. The general solution of the Poisson equation (2.13) is given by

$$U(r) = \frac{1}{4\pi} \int_{\Omega} \frac{\rho\left(\boldsymbol{r}'\right)}{|\boldsymbol{r} - \boldsymbol{r}'|} \mathrm{d}^3 \boldsymbol{r}' + \frac{1}{4\pi} \int_S \frac{\sigma\left(\boldsymbol{r}'\right) \cdot \mathrm{d}f'}{|\boldsymbol{r} - \boldsymbol{r}'|} \tag{2.13}$$

Where $\Omega$ is the volume the magnetic body is occupying and $S$ is the surface of this volume. Equation (2.13) shows that the magnetostatic potential $U(\boldsymbol{r})$ can be decomposed into the sum of two integrals; one representing the contributions of the volume charges and the other those from the surface charges:

$$\rho = -\boldsymbol{\nabla} \cdot \boldsymbol{M} \text{ volume charge density}$$
$$\sigma = \boldsymbol{M} \cdot \boldsymbol{n} \text{ surface charge density.}$$

These charge density distributions are the sources of the magnetostatic field $\boldsymbol{H}_{\mathrm{dip}} = -\nabla U$. The magnetostatic energy contribution arising from the dipolar field is similar to the Zeeman term, where the external field $\boldsymbol{H}_{\mathrm{Zee}}$ is replaced by the internal field $\boldsymbol{H}_{\mathrm{dip}}$, but divided by two:

$$e_{\mathrm{dip}} = -\frac{1}{2}\mu_0 \boldsymbol{H}_{\mathrm{dip}}(\boldsymbol{r}) \cdot \boldsymbol{M}(\boldsymbol{r}) \tag{2.14}$$

The factor $\frac{1}{2}$ is typical for self-energy terms. In this case, it reflects the fact that the magnetic moments of the sample are not only the sources of the dipolar field, but that they are at the same time influenced by it. In the context of dipole sums, the origin of this factor $1/2$ can thus be understood from the necessity to exclude a double summation[1].

---

[1] This can also be explained using the magnetostatic reciprocity theorem [82, 76]: The energy of a magnetic moment $\mu_A$ in the dipolar field of another moment $\mu_B$ is identical to the energy of the other moment $\mu_B$ in the dipolar field of the magnetic moment $\mu_A$. Obviously, this energy should only be counted once for a pair of magnetic moments.

The total dipolar energy $E_{dip}$ can be computed by integrating $e_{\text{dip}}$ over the sample volume $\Omega$:

$$E_{\text{dip}} = -\frac{\mu_0}{2} \int_\Omega \boldsymbol{M}(\boldsymbol{r}) \cdot \boldsymbol{H}_{\text{dip}}(\boldsymbol{r}) dV \tag{2.15}$$

The energy density of the magnetostatic field is

$$e_{\text{dip}} = \frac{\mu_0}{2} \boldsymbol{H}_{\text{dip}}^2 \tag{2.16}$$

It can be shown [86, 76] that the total energy stored in the dipolar field is equal to the sample's magnetostatic self-energy:

$$-\frac{\mu_0}{2} \int_\Omega \boldsymbol{M}(\boldsymbol{r}) \cdot \boldsymbol{H}_{\text{dip}}(\boldsymbol{r}) dV = \int_{\text{(entire space)}} \frac{\mu_0}{2} \boldsymbol{H}_{\text{dip}}^2 dV \tag{2.17}$$

From equation 2.15 it becomes obvious that the total dipolar energy of a ferromagnetic body is always positive. Considering the negative sign on the left-hand side, this also means that the dipolar field $\boldsymbol{H}_{\text{dip}}$ is oriented mainly antiparallel to the magnetization $\boldsymbol{M}$ throughout the sample. For this reason, the dipolar field is also called the *demagnetizing* field.

Since the dipolar energy is non-negative, it is minimized by eliminating its sources, i.e., by arranging the magnetization field in such a way to minimize surface charges and volume charges. This is known as Brown's pole avoidance principle. It is the fundamental principle which leads to the formation of flux-closure structures like magnetic vortices or other divergence-free domain patterns in thin-film elements [87], and it provides the reason for the effect known as shape anisotropy. The latter describes the tendency of a homogeneously magnetized elongated particle to orient the magnetization along the longest axis, thereby minimizing the amount of magnetic surface charges. As we further observe in this thesis, the pole-avoidance principle also appears in the tendency to form ice-rule obeying vertices in artificial spin ice structures.

### 2.1.5 Total energy

From all these energy contributions, the total energy of the ferromagnetic body can be calculated by integrating the sum of the individual energy densities over the sample volume.

$$E_{\text{tot}} = \int_\Omega \left( e_{\text{exc}} + e_{\text{ani}} + e_{\text{zee}} + e_{\text{dip}} \right) dV \tag{2.18}$$

This term $E_{tot} = E[\boldsymbol{m}(\boldsymbol{r})]$ takes into account the main micromagnetic contributions, and it depends on the spatial distribution of the magnetization field $\boldsymbol{M}(\boldsymbol{r})$. An equilibrium state of the magnetization $\boldsymbol{M}_0$ is given when the energy functional $E_{tot}$ is a minimum, i.e., $\delta E(\boldsymbol{M}) = 0$, such that any variation $\delta\boldsymbol{m}$ of the magnetization $\boldsymbol{M} = \boldsymbol{M}_0 + \delta\boldsymbol{m}$ results in an increase in $E_{tot}$. Taking into consideration only variations $\delta\boldsymbol{m}$ consistent with the constraint of constant norm of the magnetization, Brown derived the equilibrium condition of vanishing torque [86, 88],

$$\boldsymbol{m} \times \boldsymbol{H}_{\text{eff}} = 0 \tag{2.19}$$

in combination with a surface condition, which in its simplest form is

$$\boldsymbol{n} \cdot \nabla \boldsymbol{m} = \frac{\partial \boldsymbol{m}}{\partial \boldsymbol{n}} = 0 \tag{2.20}$$

where $\boldsymbol{n}$ is the surface normal vector. In a more general case, Brown's static boundary condition (2.20) should also include torques arising from surface anisotropy contributions, which we neglect here [89]. This condition has been further extended for the stationary high-frequency dynamic case by Guslienko and Slavin [90]. The effective field $\boldsymbol{H}_{\text{eff}}$ is defined through the variational derivative of the energy functional with respect to the magnetization:

$$\boldsymbol{H}_{\text{eff}} = -\frac{1}{\mu_0 M_s} \frac{\delta E_{tot}}{\delta \boldsymbol{m}} \tag{2.21}$$

In an explicit form, the effective field $\boldsymbol{H}_{\text{eff}}$ can be written as

$$\boldsymbol{H}_{\text{eff}} = \frac{2A}{\mu_0 M_s} \nabla^2 \boldsymbol{m} - \nabla U_{\text{dip}} - \frac{1}{\mu_0 M_s} \frac{\partial e_{\text{an}}}{\partial \boldsymbol{m}} + \boldsymbol{H}_{\text{ext}} \tag{2.22}$$

where the specific expression for the effective field of the magnetocrystalline anisotropy depends on the

Thus the system is in an equilibrium when the magnetization $\boldsymbol{m}(\boldsymbol{r})$ is aligned with the effective field, resulting in the torque to vanish. Solving micromagnetic problems–in general–involves the calculation of the effective field $\boldsymbol{H}_{\text{eff}}$ in order to determine the the distribution of the magnetic vector field, i.e., the magnetization structure. This is typically done by solving the Landau-Lifshitz-Gilbert equation discussed in the next section. Explicit solutions based on energy minimization have been obtained for a few fundamental examples such as domain-walls [83, 91, 92], nucleation problems [93, 94], domain patterns [87, 95], and the law of approach to ferromagnetic saturation [96, 97].

Generally, more than one state exists which satisfies 2.19 (meaning that the minimum can be a local minimum) and unique solutions exist only for certain simple examples.

The different energy contributions describe competing interactions, and the minimization of the total energy usually represents a compromise between the different tendencies arising from the individual energy terms. For instance, the tendency to form magnetic flux-closure structures, favored by the magnetostatic interaction, is in competition with the ferromagnetic exchange interaction, which tries to prevent any inhomogeneity. Since one interaction is of short and the other of long range, the balance between these effects can be described by characteristic lengths, the so-called exchange lengths [83], which will be discussed later.

## 2.2 Magnetization dynamics: The Landau-Lifshitz-Gilbert equation

In micromagnetism, the time-dependent dynamics of the magnetization vector is described by the Landau-Lifshitz-Gilbert (LLG) equation. The derivation of the LLG equation starts from the classical equation describing the rotational motion of a rigid body of angular momentum $\boldsymbol{P}$ under an applied torque $\tau$,

$$\frac{dP}{dt} = \tau \tag{2.23}$$

The magnetic analogue of this equation is given by,

$$\tau = \frac{dL}{dt} = \boldsymbol{M} \times \boldsymbol{H}_{\text{eff}} \tag{2.24}$$

The gyromagnetic ratio $\gamma_0$, given by

$$\gamma_0 = \frac{g_L |e| \mu_0}{2m} = 2.212\,76 \times 10^5\,\mathrm{m\,A^{-1}\,s^{-1}} \tag{2.25}$$

connects the angular momentum $\boldsymbol{L} = \frac{-\boldsymbol{M}}{\gamma_0}$ to the magnetization $\boldsymbol{M}$. Here, $e$ and $m$ are the charge and the rest mass of electron, respectively, and $g_L$ is the Landé $g$-factor [98]. Substituting for $\boldsymbol{L}$, we obtain the following equation which describes the undamped, time-dependent precession of $\boldsymbol{M}$ in the case of a torque exerted by $\boldsymbol{H}_{\text{eff}}$

$$\frac{d\boldsymbol{M}}{dt} = -\gamma_0 \tau = -\gamma [\boldsymbol{M} \times \boldsymbol{H}_{\text{eff}}] \tag{2.26}$$

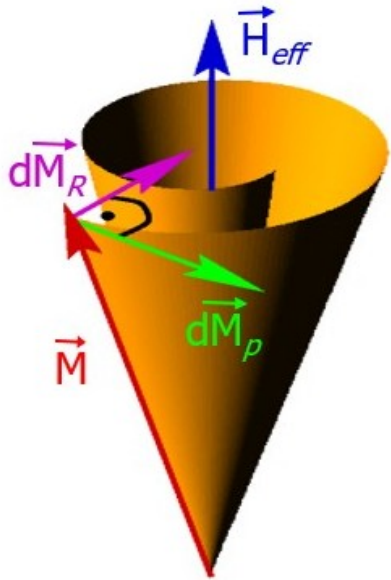

Figure 2.1: Schematic description of the magnetization dynamics described by the LLG equation. The sketch represents the damped precessional motion of the magnetization vector $\boldsymbol{M}(r,t)$ around the effective field $\boldsymbol{H}$. The precessional term is shown in red and the damping term in blue.

According to this equation, $\boldsymbol{M}$ precesses around the effective field with a frequency $\omega = -\gamma \boldsymbol{H}_{\text{eff}}$ without any dissipation. The negative sign in the frequency implies that the sense of rotation is anti-clockwise with respect to a positive $\boldsymbol{H}_{\text{eff}}$. In a realistic scenario, this motion is subjected to various dissipative interactions such as orbital couplings, spin-diffusion, or interactions with impurities. To account for such dissipative effects, L.D. Landau and E.M. Lifshitz [71] introduced a phenomenological damping constant $\alpha_L$ and added a corresponding damping term to the precession term to obtain the Landau-Lifshitz (LL) equation

$$\frac{d\boldsymbol{M}}{dt} = \gamma_L \left[ \boldsymbol{M} \times \boldsymbol{H}_{\text{eff}} \right] - \frac{\alpha_L}{M_s} \left[ \boldsymbol{M} \times \left[ \boldsymbol{M} \times \boldsymbol{H}_{\text{eff}} \right] \right] \tag{2.27}$$

The second term in equation (2.27) represents the relaxation motion of the magnetization towards $\boldsymbol{H}_{\text{eff}}$, so that after a finite time the motion comes to a stop as the magnetization aligns with the effective field and the torque vanishes. Note that both the relaxation and the precession term describe changes $d\boldsymbol{M}/dt$ of the magnetization that is perpendicular to $\boldsymbol{M}$, which ensures the constraint of constant modulus $|\boldsymbol{M}|$.

In 1955, Gilbert [99, 100] suggested an alternative form of the Landau-Lifshitz

equation (2.27), motivated by the reasoning that not only the relaxation motion, but also the precessional part of the dynamics should be affected by dissipation. In the case of strong damping, it appears particularly important to consider damping effects also in the precessional part of the dynamics. By introducing a velocity-dependent dissipation term and the Gilbert damping constant $\alpha_G$, he presented an alternative, implicit equation which accounts for the general effect of damping on the magnetization dynamics. This equation allows to accommodate also the case of strong damping

$$\frac{d\boldsymbol{M}}{dt} = \gamma_G \left[ \boldsymbol{M} \times \boldsymbol{H}_{\text{eff}} \right] + \frac{\alpha_G}{M_s} \left[ \boldsymbol{M} \times \frac{d\boldsymbol{M}}{dt} \right] \tag{2.28}$$

The seemingly different equations for the magnetization given by Landau-Lifshitz (2.27) on the one hand and by Gilbert (2.28) on the other hand can be reconciled and interpreted as merely two different representations of the same equation. If the Gilbert form is reformulated in an explicit form, taking in to account that $\boldsymbol{M} \cdot d\boldsymbol{M}/dt = 0$, one obtains the so-called Landau-Lifshitz-Gilbert equation, given by

$$\frac{d\boldsymbol{M}}{dt} = \frac{\gamma}{(1 + \alpha_G^2)} \left[ \boldsymbol{M} \times \boldsymbol{H}_{\text{eff}} \right] + \frac{\alpha_G \gamma}{(1 + \alpha_G^2) M_s} \left[ \boldsymbol{M} \times \left[ \boldsymbol{M} \times \boldsymbol{H}_{\text{eff}} \right] \right] \tag{2.29}$$

In spite of the formal equivalence of the equations, the Gilbert formulation is often considered to be more realistic, and in the modern literature only the dimensionless value $\alpha_G$, not $\alpha_L$, is used to describe a magnetic material's damping properties. One can therefore drop the subscript $G$ when referring to the the damping constant $\alpha$. In dynamic processes, one can distinguish low-damping cases, which are characterized by values of $\alpha \ll 1$, while values of $\alpha \gtrsim 0.5$ describe cases of high damping. Assuming such unrealistically large damping values can be useful in numerical simulations if the goal is to quickly calculate an equilibrium structure.

The general approach to finding a solution to a micromagnetic problem involves the numerical integration of the LLG equation in time, given an initial configuration and a number of input parameters describing the material properties, the geometry of the ferromagnetic body, and additional simulation parameters such as external fields. Details of the numerical methods are discussed in the next chapter.

## 2.3 Basic micromagnetic structures and effects

In the case of an equilibrium structure, as briefly mentioned before, the magnetization distribution of a ferromagnetic body develops a state where a compromise is established between the different competing energy terms so as to minimize the total energy as much as possible. This behavior results in the formation of a few types of fundamental micromagnetic structures. In the following section, a few fundamental structures which will be later used in the thesis are briefly presented.

### 2.3.1 Magnetic domain walls

A magnetic domain wall is the interface region between two oppositely magnetized neighboring domains. The rotation of the magnetization direction occurring across the domain wall ensures a smooth transition of the magnetization within a specific distance known as the domain wall width. Based on the type of transition, several different kinds of domain walls can be distinguished. The first kind is the Bloch-domain wall which describes the smooth transition between two oppositely magnetized domains so that within the domain wall the magnetization is gradually rotated in a plane separating the two domains. The width $b$ of a Bloch wall is given by

$$b = \pi l_{\mathrm{k}} \tag{2.30}$$

Where $l_k$ is the *exchange length* of the anisotropy, an intrinsic length scale of the material. Further below we will discuss another exchange length, $l_s$, which is due to the competition between ferromagnetic exchange and magnetostatics. In an anisotropic ferromagnetic material, the value of $l_k$ depends on the competition between the exchange interaction and the magnetocrystaline anisotropy energy. Accordingly, $l_k$ can be expressed as [92]

$$l_{\mathrm{k}} = \sqrt{\frac{A}{K_u}} \tag{2.31}$$

The energy density of a Bloch-wall, that is, the energy stored in a unit area of the wall, can be expressed as

$$\epsilon_b = 4\sqrt{AK} \tag{2.32}$$

The magnetic structure of a Bloch wall evades the formation of volume charges ($\rho = -\nabla \cdot \boldsymbol{m}$) in expense of the formation of surface charges ($\sigma = \boldsymbol{n} \cdot \boldsymbol{m}$). Bloch walls are commonly observed in anisotropic magnetic particles with lateral dimensions considerably exceeding the domain wall width.

In soft-magnetic thin films, the magnetostatic energy contribution by the surface charge development plays a decisive role. Hence a smooth 180 degree rotation of the magnetization between two oppositely magnetized in-plane domains develops differently than the previously described Bloch wall. In such cases, the tendency to suppress magnetic surface charges results in the formation of Néel, walls [74], in which the magnetization rotates in the film plane. The central part of the domain is thereby magnetized perpendicular in the film plane, perpendicular to the domain wall. In the case of Néel walls, the energy density and the wall width are determined by the competition between demagnetizing energy and exchange energy. The domain wall width $n$ of Néel wall is also characterized by an exchange length:

$$n = \pi l_s \tag{2.33}$$

$$l_s = \sqrt{\frac{2A}{\mu_0 M_s^2}} \tag{2.34}$$

and the energy density of the Néel wall is given by

$$\epsilon_n = \sqrt{\frac{2A}{M_s}} \tag{2.35}$$

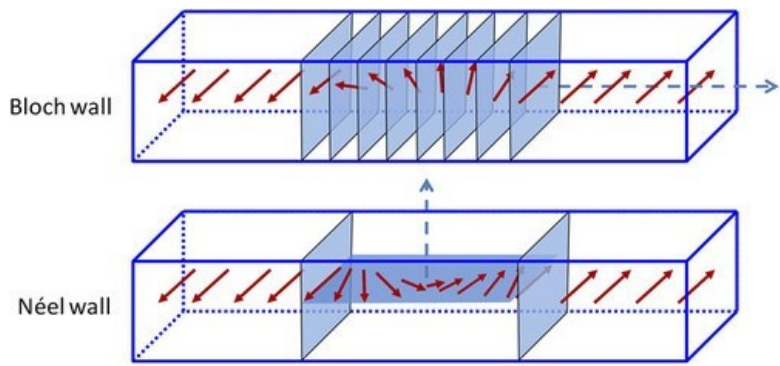

Figure 2.2: Comparison between a Bloch and a Néel wall. In Bloch wall the magnetization is rotated 180 degree around the normal of the domain wall while in the case of a Néel wall the magnetization is rotated in the plane of the film.[101]

In the case of thin elongated nanostructures, such as soft-magnetic thin cylindrical nanowires, a different type of domain wall is observed. In these systems, the magnetic structure is characterized by domains in which the magnetization aligns parallel to the wire axis because of the dominant effect of shape anisotropy. The transition regions from in such wires are called head-to-head (tail-to-tail) domain walls [102, 103]. If parts of the nanowire are magnetized in opposite axial directions, a head-to-head or tail-to-tail domain wall structure is formed at the interface.

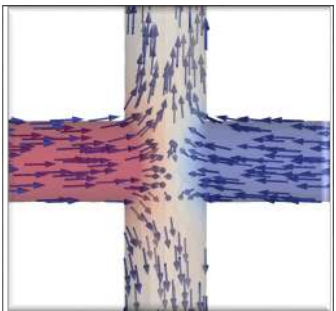

Figure 2.3: Micromagnetic simulation of a junction of four Permalloy nanowires of thickness 14 nm. The opposite wires are magnetized in anti-parallel direction and a combination of a head-to-head and tail-to-tail domain wall is formed at the interface junction

In such elongated structures, different types of head-to-head walls can form. If the thickness or diameter remains below above a critical limit, the head-to-head domain wall is of transverse type, whereas in the case of larger widths or thicknesses, vortex-type head-to-head domain walls are formed. In the case of cylindrical nanowires, there is also the case of head-to-head walls in the form of Bloch-point domain walls [104, 105, 106, 107], which typically form in the case of wire diameters exceeding about 60 nm.

The exchange length of a material is not only useful to describe the typical extension of micromagnetic structures, but is also an important parameter in the context of micromagnetic numerical simulations. To perform accurate simulations of magnetization structures that can develop within a ferromagnetic system, it is necessary to use a sufficiently small spatial discretization. This is necessary in order to avoid numerical discretization errors that can lead to completely erroneous results like, e.g. the "domain wall collapse" described by Donahue [108]. As a rule of thumb, the maximum distance between two neighboring descretization points (the cell size) should always remain below the smallest of the set of exchange lengths of the modeled material. For the materials used in the simulations of this thesis, the exchange lengths are in the order of 4 nm to 5 nm, and consequently the finite element models which we used are ensured to have maximum cell sizes below these limits.

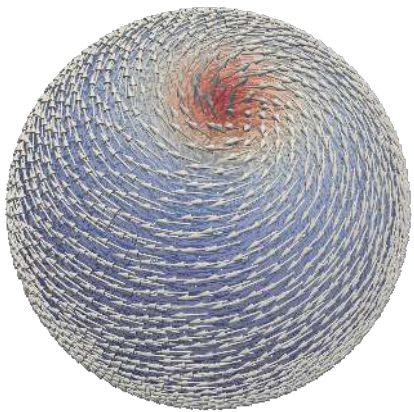

Figure 2.4: Micromagnetic simulation of a three-dimensional vortex formed in a Permalloy nanosphere of radius 100 nm

### 2.3.2 Magnetic vortices

Magnetic vortices are swirling structures of the magnetization that efficiently close the magnetic flux. Accordingly, they usually form as a result of the magnetic system's tendency to minimize the demagnetizing energy by forming divergence-free structures. In two-dimensional films and nanodots, the magnetic structure of a vortex is characterized by the in-plane circulation of the magnetization around the core of the vortex. At the center of the vortex, the magnetization points normal to the plane of the vortex. For topological reasons, magnetic vortices are highly stable structures once they are formed. The size of the vortex core is defined by the interplay between the demagnetizing and exchange energy terms and is thus typically depends on the exchange length $l_s$, although the precise value can change with the thickness of the structure. The core radius of a vortex in a film of thickness $h$ can be approximated as [109],

$$r_{core} = 0.68 l_s \left( \frac{h}{l_s} \right)^{1/3} \tag{2.36}$$

### 2.3.3 Shape anisotropy: axial magnetization of elongated structures

Similar to how the magneto-crystalline anisotropy of a ferromagnetic body can result in a preferred direction of magnetization, the geometric shape of the body can also give rise to a preferred orientation of magnetization. This effect in which elongated objects exhibit a tendency to be magnetized along the long direction is called shape anisotropy.

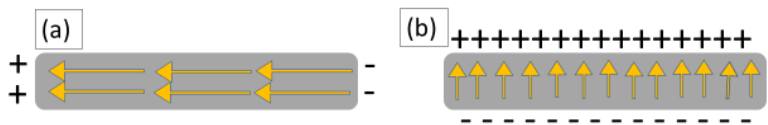

Figure 2.5: Shape anisotropy of an elongated structure (a) magnetization along the easy direction resulting in a comparatively lower surface charge formation (b) the energetically unfavorable configuration, resulting in a higher demagnetising energy due to the formation of higher amount of surface charges.

Even though the effects are similar, shape anisotropy has a completely different origin from that of magneto-crystalline anisotropy. While the latter is attributable to the spin-orbit coupling, the shape anisotropy is caused by the tendency of the body to minimize the demagnetizing energy by reducing the formation of magnetic surface charges ($\sigma = \boldsymbol{n} \cdot \boldsymbol{m}$). Consider an elongated ferromagnetic body, as visualized in figure 2.5. The preferred orientation of the magnetization is along the long axis of the body as depicted in (a) because this arrangement results in the formation of comparatively less surface charges than the structure shown in (b), where the magnetization is oriented along the short axis. Therefore, the magnetic configuration (a) has a lower demagnetization energy than the second case. In soft-magnetic materials, this effect results in the formation of a shape dependent direction of preferred magnetization along the long axis. Because of this effect, elongated elements such as cylindrical nanowires, nanotubes, or rectangular strips and wires with high aspect ratio are preferably magnetized along the long axis in the absence of a strong external fields. This behavior is important for the studies discussed in this thesis concerning interconnected nanowire networks. The competing interaction of axially magnetized nanowires at the meeting points at the vertices gives rise to emergent effects and artificial spine ice (ASI) behavior.

The shape anisotropy effect also occurs in extended films, where it favors an in-plane magnetization. This tendency can be in competition with, e.g., a unixaial magneto-crystalline anisotropy, if the crystalline easy axis is perpendicular to the surface normal, thus favoring an out-of-plane magnetic orientation. This type of competition can be used to quantify the relative strength of the magnetocrystalline anisotropy, and it is at the basis of the definition of the dimensionless quality factor $Q = 2K_u/(\mu_0 M_s)$. A value of $Q \simeq 1$ denotes a strongly anisotropic material, while low-anisotropy, i.e., soft-magnetic materials have $Q \ll 1$.

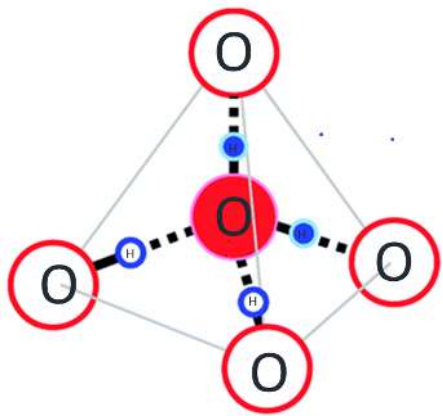

Figure 2.6: Crystalline structure of water-ice. Each Oxygen atom is surrounded by four Hydrogen atom: two are bound by covalent bond and the remaining two by hydrogen bonds with the neighboring $H_{2O}$ molecule

## 2.4 Artificial spin ices

Water ice exists in about 18 different crystalline structures [110]. In all these structures, neighboring water molecules are linked together by strong hydrogen bonds. That is, each oxygen atom in a water-ice is surrounded by four hydrogen atoms, out of which two are bonded by covalent bonds and the other two realize a hydrogen bond with the neighboring water molecules 2.6. This state, in which two of the four hydrogen atoms are closer to and two are farther away from the oxygen atom result in a "two-in/two-out" type of arrangement, which is known as the Bernal-Fouler rule or simply as the ice-rule [111]. Each oxygen atom in the crystal can exist in six different configurations obeying the two-in/two-out rule, resulting in a sixfold degeneracy per oxygen atom. The collective degeneracy of the system grows exponentially with increasing number of oxygen atoms. In 1935, [112, 113] Linus Pauling made use of this ice-rule to explain the zero-point entropy of water [114, 115]. The ice-rule gained broader significance since the appearance of spin ice systems [116]. Magnetic spin ices are materials consisting of a tetrahedral arrangement of magnetic ions, whose low energy state is a "two-in/two-out" state analogous to the water ice. This ground state arrangement in magnetic spin ices gives rise to the inability to satisfy the energy minimization conditions for all the competing interactions from the neighbors, which results in frustration. Artificial spin ices are [117, 118, 119, 120, 121] patterned magnetic structures, which attempt to mimic these frustrated interactions. Artificial spin ices can be regarded as metamaterials made of elongated single domain units, arranged in a particular way so that their magnetostatic interaction at the vertex points can lead to frustration. Two well-known examples are the square ASI [122], in which four single-domain units are arranged in a "+" shape to form a square grid, and the Kagome lattice [123, 124] in which three such units are arranged at each vertex to form a hexagonal grid. A geometric description of the square and Kagome lattices is given in figure 2.7. Since all the individual units in such an ASI are of single-domain type, they can be assigned an Ising-like dipole moment. We can identify various different vertex configurations based on the total number of moments oriented towards or away from the vertex. The square ASI can have 16 different vertex configurations, out of which the first six obey the ice rule. The vertex configurations which does not follow the ice rule are known as magnetic defects. Such magnetic defects are also sometimes referred to as emergent monopoles [125, 126, 127, 128, 122, 129, 130]. Artificial spin ices are one of the most advanced sub-branches of nanomagnetism, and owing to their re-configurability [131, 132] they have potential applications in a large number of fields such as magnonic logic computation [133, 134, 135, 136, 137], data storage [138], reservoir computation [139, 140], stochastic computation [141], re-configurable magnonic devices [131] and platforms for re-programmable magnonic resonators and crystals [119]. Until recent years, research on artificial spin ice systems was limited to two dimensional structures[117, 118, 120, 119, 142, 131]. Accordingly, much attention was focused on systems where patterned nano-elements are arranged in a single plane or in multiple two-dimensional planes [143]. With the recent advancements in three-dimensional fabrication techniques, such as FEBID [144, 24, 25, 26, 145] and TPL [27, 29, 28], a new generation of nanostructures can be formed - three dimensional networks of interconnected nanowires such the Buckyball network fabricated by Donnelly *et al.* [30] or the three dimensional diamond lattice fabricated by Keller *et al.* [64] and Andrew May *et al.* [65]. These structures consist of a three dimensional network of interconnected nanowires and they can be fabricated in a way so that these individual nanowires exist in a single domain-state, magnetized along the wire axes. Thus, they can have the basic properties of artificial spin ices and these structures can be classified as the first generation of three-dimensional artificial spin ice structures [66].

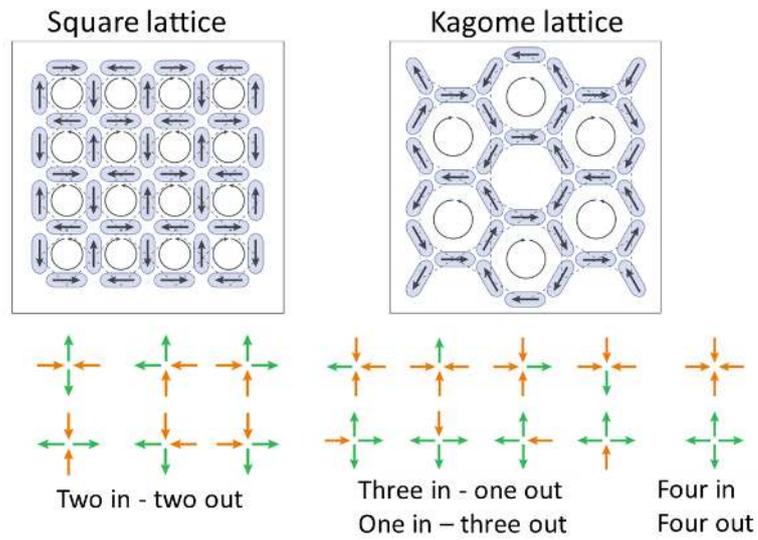

Figure 2.7: The two familiar artificial spine ice lattices square ASI and Kagome ASI along with the 16 theoretically possible vertex configurations in a square ASI [119]

# CHAPTER 3

Finite element modeling, simulation and
post-processing

## 3.1 Introduction

One of the first works in computational micromagnetic simulation was published in 1969 by A.E LaBonte [146], in which he numerically calculated the domain-wall formation in a ferromagnetic film. Five decades after this start of computational studies on micromagnetic structures, following an enormous leap in computation technology as well as significant advances in the field of numerical techniques and applied mathematics, micromagnetic simulations have evolved into an indispensable and branch of modern day magnetism. Today, computational micromagentism is a powerful and highly reliable tool for studying magnetization structures and their dynamical properties in nanoscale systems. Micromagnetic simulations have the advantage that they can be used to obtain information which is often inaccessible with standard experimental techniques, such as the internal magnetization structure in complicated three-dimensional systems, comparisons of the energy of different magnetization states, or the study of ultra-fast dynamic process with high spatial and temporal resolution.

Today, more than a dozen of free and closed micromagnetic simulation tools exist [147]. They can be divided into two major categories based on the numerical technique implemented as finite difference method (FDM) and finite element method (FEM) [148]. In finite-difference methods, the geometry is subdivided into equidistant cubic cells. This regular arrangement of the unit cells allows for the calculation of the demagnetizing fields by fast Fourier transform techniques, which can be computed very efficiently and accurately. This type of numerical technique is well suited for the simulation of extended films, rectangular strips and other regular geometries, as the equidistant finite-difference grid can accurately model these kinds of shapes. Although these methods are usually fast and accurate, they are not very efficient when complicated geometries such as curved surfaces are to be treated. Due to the effect of magnetic surface charges, an accurate representation of the sample shape is essential in micromagnetic simulations. In the case of irregular geometries, the "staircase approximation" needed in the FDM can introduce errors in micromagnetic simulations that persist even at small cell sizes [149].

Another difficulty of FDM methods is that they require the simulation of the entire prism of the volume in which a discretized structure is embedded. This is inefficient, e.g., in the case of interconnected wire networks with a lot of empty space, which occupy only a very small fraction of the volume over which they extend. In such cases, the finite-element method based simulation tools becomes relevant and may in fact be the only way with which certain systems can be reasonably modeled. In FEM, the geometry is subdivided into irregular cells, which in the three-dimensional case are usually tetrahedral elements of varying sizes. In contrast to the regular grid of a FDM, the position and density of the discretization points in FEM are freely adjustable, which results in a much more accurate description of arbitrary geometries. In addition to the method's intrinsic geometric flexibility, which allows for the precise modeling of complex geometries, there is no need in the FEM to include the non-magnetic volume between magnetic particles in the simulations. This superiority of FEM makes them the ideal method for the simulation of three-dimensional nanowire networks.

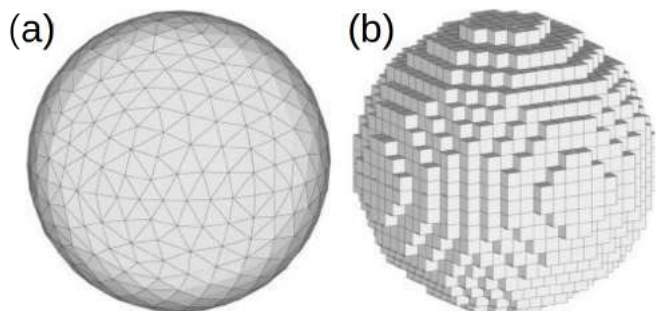

Figure 3.1: Comparison of a finite-element mesh of a sphere (a) with a uniform grid approximation of the same shape. The geometric flexibility of the finite element method method allows for a more accurate description of the curved surface while using a lower number of degrees of freedom (i.e., number of discretization cells) [150].

All simulations in this thesis are performed using the finite element micromagnetic simulation tool `tetmag` [151] developed by Riccardo Hertel at the IPCMS Strasbourg. In this chapter, we briefly illustrate the basic working principle behind this software package. The `tetmag` code requires various input parameters for each simulation to define the details of the problem. The input files define, e.g., the micromagentic material parameters including the saturation magnetization $M_s$, exchange stiffness constant $A$, the value of crystalline anisotropic constants and, if applicable, external parameters such as external field, pulse parameters *etc*. Most fundamentally, the simulation requires a finite-element mesh describing the geometry of the system. Open-source packages like GMSH [152] or Netgen [153] are available for the generation of FE meshes according to the geometry of choice. These codes implement several advanced algorithms for the generation of unstructured grid meshes from the geometric description, such as the

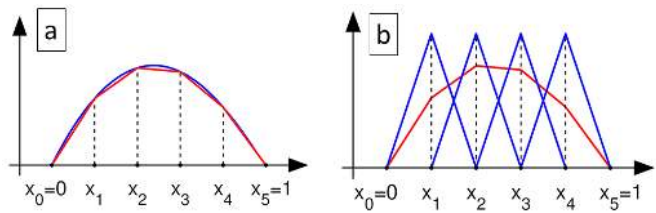

Figure 3.2: Example of a shape function in 1D. (a) A continuous function (blue) is approximated by a piece-wise linear function (red) (b) The function is represented by means of a basis function (blue) and their linear combination (red)

Delaunay and the Advancing-front method.

## 3.2 Finite element formulation

A fundamental aspect of the finite element method consists in interpolating an unknown function inside each finite element based on the computed values at the discretization nodes [154, 155]. Consider a tetrahedral element $e$ with four nodes $i = 1, 2, 3, 4$. The value of a function $u(r)$ inside the element can be approximated as,

$$u(x) \simeq \hat{u}(x) = \sum_{i=1}^{4} N_i^e(x) \tilde{u}_i^e \tag{3.1}$$

where $\tilde{u}_i^e$ is the value of the function at the nodes and $N_i^e$ are the element's shape functions. The shape functions serve as basis functions in the finite-element formulation. For each node $i$ of an element $e$, there is a shape function with the property of having a value equal to one at the node, and zero at all other nodes of the element:

$$N_i^e(\boldsymbol{x}_j) = \delta_{ij} \tag{3.2}$$

where $\delta_{ij}$ is the Kronecker symbol and $\boldsymbol{x}_j$ is the position of the node $j$ of element $e$. Outside the element $e$, the element's shape functions are equal to zero. The mist common and simple choice are linear shape functions, in which the value varies linearly with the position within each element $e$. They can be expressed as

$$N_i^e(r) = \frac{1}{6V_e}(a_i + b_i x + c_i y + d_i z) \tag{3.3}$$

Here $V_e$ is the volume of the element $e$. These coefficients depends only on the shape of the elements. For a tetrahedral element, the coefficients can be expressed as follows [156], where $i = 1, 2, 3$ and $4$.

$$a_i = \frac{1}{6V_e} \begin{vmatrix} x_{i+1} & y_{i+1} & z_{i+1} \\ x_{i+2} & y_{i+2} & z_{i+2} \\ x_{i+3} & y_{i+3} & z_{i+3} \end{vmatrix} \qquad b_i = -\frac{1}{6V_e} \begin{vmatrix} 1 & y_{i+1} & z_{i+1} \\ 1 & y_{i+2} & z_{i+2} \\ 1 & y_{i+3} & z_{i+3} \end{vmatrix}$$

$$c_i = -\frac{1}{6V_e} \begin{vmatrix} x_{i+1} & 1 & z_{i+1} \\ x_{i+2} & 1 & z_{i+2} \\ x_{i+3} & 1 & z_{i+3} \end{vmatrix} \qquad d_i = -\frac{1}{6V_e} \begin{vmatrix} x_{i+1} & y_{i+1} & 1 \\ x_{i+2} & y_{i+2} & 1 \\ x_{i+3} & y_{i+3} & 1 \end{vmatrix}$$

The volume $V_e$ of the element $e$ can be calculated as

$$V_e = \frac{1}{6} \begin{Vmatrix} 1 & x_1 & y_1 & z_1 \\ 1 & x_2 & y_2 & z_2 \\ 1 & x_3 & y_3 & z_3 \\ 1 & x_4 & y_4 & z_4 \end{Vmatrix} \tag{3.4}$$

These operations can be interpreted as a coordinate transformation from the Cartesian to the local, barycentric system. One of the advantages of the finite element method is the freedom of choice of the location of these discretization points, which allows for the accurate description of any type of geometries. The functions are represented by means of basis shape functions and mathematical operations such as differentiation and integration can be executed on those basis functions.

### 3.2.1 Differentiation

Given a function $u(\boldsymbol{r})$ such that,

$$u(\boldsymbol{r}) = \sum_{i=1}^{4} N_i^e(\boldsymbol{r}) \tilde{u}_i^e \tag{3.5}$$

The derivative of $u$ with respect to x, y and z can be written as

$$\frac{\partial u}{\partial x} = \sum_{i=1}^{4} u_i \frac{\partial N_i(x,y,z)}{\partial x} \tag{3.6}$$

$$\frac{\partial u}{\partial y} = \sum_{i=1}^{4} u_i \frac{\partial N_i(x,y,z)}{\partial y} \tag{3.7}$$

$$\frac{\partial u}{\partial z} = \sum_{i=1}^{4} u_i \frac{\partial N_i(x,y,z)}{\partial z} \tag{3.8}$$

$$\tag{3.9}$$

Hence for the element $e$

$$\nabla u^e(x,y,z) = \sum_{i=1}^{4} u_i^e \nabla N_i^e(x,y,z) \tag{3.10}$$

the derivatives of $u(\boldsymbol{r})$ are constant within an element as the basis function $N_i$ are linear functions in space. From eq. (3.3), the derivatives of the shape function can be expressed as,

$$\frac{\partial N_i}{\partial x} = b_i, \qquad \frac{\partial N_i}{\partial y} = c_i, \qquad \frac{\partial N_i}{\partial z} = d_i \tag{3.11}$$

Where $a_i$, $b_i$ and $c_i$ are matrices given by 3.2. Substituting these in 3.10, we get the following expression for $\nabla u(\boldsymbol{r})$,

$$\nabla \tilde{u}^e(x,y,z) = \begin{pmatrix} b_1^e & c_1^e & d_1^e \\ b_2^e & c_2^e & d_2^e \\ b_3^e & c_3^e & d_3^e \\ b_4^e & c_4^e & d_4^e \end{pmatrix} \begin{pmatrix} u_x^e \\ u_y^e \\ u_z^e \end{pmatrix}$$

This can be expressed as

$$\nabla \tilde{u}^e(x,y,z) = \underline{\underline{\mathbf{B}}}^e \tilde{u}^e(x,y,z) \tag{3.12}$$

The matrix $\mathbf{B}^e$ depends only on the spatial discretization of the element $e$ and is independent of the function $u(\boldsymbol{r})$. In the implementation of a finite-element solver, $\mathbf{B}$ can be set-up and stored in the preprocessing stage, before starting the simulation, and used whenever such a mathematical operation is required.

### 3.2.2   Integration

As in the case of differentiation, the integral of a function can be expressed by means of integrals over the basis functions. Using the representation of a function $u(\boldsymbol{r})$ according to eq. (3.1), the integral of $u(\boldsymbol{r})$ over a tetrahedral element $e$ of volume $V_e$ can be written as

$$\int_{V_e} \tilde{u}^e(x) dV = \int_{V_e} \sum_{i=1}^{4} \tilde{u}^e(x) N_i^e(x) dV = \sum_{i=1}^{4} \tilde{u}^e(x) V_e \int_{V_e} \frac{N_i^e(x)}{V_e} dV \qquad (3.13)$$

The second part of the integral is simple: $\int N_i^e dV = V_e/4$. This denotes the fraction of the volume of the tetrahedron shared by each node $i$. Thus, eq. (3.13) becomes

$$\int_{V_e} \tilde{u}(x) dV = \frac{1}{4} V_e \sum_{i=1}^{4} u_i(x) \qquad (3.14)$$

## 3.3   Computation of the effective field

We can use these principles to numerically treat micromagnetic problems with the FEM. The methods described here are detailed in Ref. [157]. They have been originally implemented by R. Hertel in the code `TetraMag` [158], which preceded the `tetmag` software used in this thesis.

The effective field, which is defined in eq. (2.22), is calculated by taking the variational derivative of the energy functional with respect to the reduced magnetization,

$$\boldsymbol{H}_{\text{eff}} = \frac{2A}{\mu_0 M_s} \nabla^2 \boldsymbol{m} - \frac{1}{\mu_0 M_s} \frac{\partial e_{\text{an}}}{\partial \boldsymbol{m}} + \boldsymbol{H}_{\text{ext}} - \nabla U \qquad (3.15)$$

The effective field can be expressed as

$$\boldsymbol{H}_{\text{eff}} = \boldsymbol{H}_{\text{ani}} + \boldsymbol{H}_{\text{exc}} + \boldsymbol{H}_{\text{ext}} + \boldsymbol{H}_{\text{dem}} \qquad (3.16)$$

where the exchange field $\boldsymbol{H}_{exc}$ is given as

$$\boldsymbol{H}_{exc} = \frac{2A}{\mu_0 M_s} \nabla^2 \boldsymbol{m} \qquad (3.17)$$

The anisotropic field $\boldsymbol{H}_{ani}$ is given as

$$\boldsymbol{H}_{\text{ani}} = -\frac{1}{\mu_0 M_s} \frac{\partial e_{an}}{\partial \boldsymbol{m}} \qquad (3.18)$$

Each component of the effective field is computed individually at the $N$ discretization nodes.

### 3.3.1 Exchange field: weak formulation

The exchange field is given by eq. (3.17). In addition, specific conditions at the surface boundary $\partial S$ need to be considered which, as derived by Brown [67] and Rado-Weertman [89] are i

$$\frac{\partial \boldsymbol{m}}{\partial \boldsymbol{n}}\bigg|_{\partial S} = 0 \qquad (3.19)$$

n the case of vanishing surface torques (zero surface anisotropy). In the FEM representation the exchange field is calculated by using the Galerkin method. As a first step, both sides of eq. (3.17) are multiplied by test functions $\psi_i(r)$ and integrated over the problem domain. Furthermore, the test functions are represented in the basis of the element shape functions. This procedure yields, after some manipulation, a set of $3N$ equations for the exchange field $\boldsymbol{H}_{\text{exc},i,r}$. Here the index $i = 0, 1, 2 \ldots N$ represents the node number and $r = x, y, z$ are the Cartesian components.

$$\int_V \psi_i H_{\text{exc},r} dV = \int_V \frac{2A}{\mu_0 M_s} \psi_i \nabla^2 m_r \, dV \qquad (3.20)$$

Integrating by parts gives

$$\int_V \psi_i H_{\text{exc},r} dV = -\frac{2A}{\mu_0 M_s} \int_V \nabla \psi_i \cdot \nabla m_r \, dV + \frac{2A}{\mu_0 M_s} \int_{\partial S} \psi_i \nabla m_r \cdot n_r dS \qquad (3.21)$$

The partial integration removes the second derivatives involved in the original equation and yields an equation, called the weak form, which contains only first derivatives in space. The surface integral in the second part of the equation vanishes according to the condition (3.19) as the normal derivative of the magnetization vanishes at the surface boundary. Thus, the remaining part can be written as the sum of $x$, $y$, and $z$ components and the volume integration can be written as the sum over the total number of elements $E$. For simplicity, we consider only the $x$ component

$$\sum_{e=1}^{E} \int_{V_e} \psi_x(x) H_{\text{exc},x}(x) dV = -\frac{2A}{\mu_0 M_s} \sum_{e=1}^{E} \int_{V_e} \nabla \psi_x(x) \nabla m_x(x) dV \qquad (3.22)$$

where the integration is now carried over the volume $V_e$ of the $e^{th}$ element. As showed in the previous section, the test functin $\psi$, reduced magnetization $\boldsymbol{m}(x)$ and $\boldsymbol{H}_{exc}$ can be expressed in terms of the shape functions.

$$\psi_x^e(x) \simeq \sum_{i=1}^{4} \psi_x^i N_i^e(x)$$

$$m_x^e(x) \simeq \sum_{j=1}^{4} m_x^j N_j^e(x) \tag{3.23}$$

$$H_{exc,x}^e \simeq \sum_{l=1}^{4} H_{exc,l}(x) N_l^e(x)$$

Then equation 3.22 can be written as

$$\sum_{e=1}^{E} \sum_{i=1}^{4} \sum_{l=1}^{4} \psi_x^i H_{x,l} \int_{V_e} N_{e,i} N_{e,l} dV = -\frac{2A}{\mu_0 M_s} \sum_{e=1}^{E} \sum_{i=1}^{4} \sum_{j=1}^{4} \psi_x^i m_{x,j} \int_{V_e} \nabla N_{e,i} \nabla N_{e,j} dV \tag{3.24}$$

The expression can be further simplified by the use of two characteristic integrals. The element matrix

$$M_{i,j}^{(e)} = \int_{V_e} \psi_i \psi_j \, dV \tag{3.25}$$

Is called the *mass matrix*. The matrix element of the mass matrix are given by the equation

$$\int_{V_e} N_i N_j dV = (1 + \delta_{i,j}) \frac{V_e}{20} \tag{3.26}$$

We now approximate the $M_{i,j}^e$ with a diagonal matrix by means of the "mass-lumping" approximation

$$\tilde{M}_{i,j}^e = \delta_{i,j} \sum_{j=1}^{4} M_{i,j}^e \tag{3.27}$$

The diagonal elements of $\tilde{M}_{i,j}^e$ are obtained by summing the elements of the corresponding row. Thus,

$$\tilde{M}_{i,j}^e = \frac{V_e}{20} \sum_{j=1}^{4} (1 + \delta_{i,j}) = \frac{V_e}{20} (2 + 1 + 1 + 1) = \frac{V_e}{4} \tag{3.28}$$

It is worth noting that the integral on the right-hand side

$$\int_{V_e} \nabla N_{e,i} \nabla N_{e,j} dV = K_{i,j}^e = \sum_{\alpha}^{x,y,z} c_{i\alpha}^e c_{j\alpha}^e V_e \tag{3.29}$$

contains the stiffness matrix, $K$, which is of central importance in FEM formulations. Substituting for these integrals and re-arranging, the $x$ component of the exchange field $H_{exc,x}$ can be written as

$$H_{x,l} = -\frac{2A}{\mu_0 M_s} \frac{4}{\sum_{e=1}^{k} V_e} \sum_{e=1}^{k} \left( \sum_{\alpha}^{x,y,z} c_{i\alpha}^n c_{j\alpha}^n \right) m_{x,j} V_e \tag{3.30}$$

In matrix form, this can be expressed as

$$H_\alpha^i = \underline{\underline{\mathbf{A}}}_{ij} \cdot m\alpha^j \tag{3.31}$$

where $\alpha = x, y, z$ for each of the three Cartesian components and $\underline{\underline{\mathbf{A}}}_{ij}$ is given by

$$\underline{\underline{\mathbf{A}}}_{ij} = -\frac{2A}{\mu_0 M_s} \frac{4}{\sum_{e=1}^{k} V_e} \sum_{e=1}^{k} \sum_{\alpha}^{x,y,z} c_{i\alpha}^e c_{j\alpha}^e V_e \tag{3.32}$$

### 3.3.2 Anisotropy field

The anisotropic field $H_{ani}$ due to a second order uni-axial anisotropy can be expressed as follows,

$$\boldsymbol{H}_{ani,i} = -\frac{-2K_u}{\mu_0 M_s} \cdot k_{u,i} \cdot (\boldsymbol{m}_i \cdot k_{u,i}) \tag{3.33}$$

Where $i = 0, 1, 2...N$ is the node index and $k_{u,i}$ is the uniaxial anisotropy axis direction at the $i^{th}$ node.

### 3.3.3 Demagnetizing field: FEM/BEM method

The dipolar field or demagnetizing field, which accounts for the long-range magneto-static interactions, is usually the most complicated component of the effective field to compute in terms of both computation time and mathematical complexity. The demagnetizing field $\boldsymbol{H}_{\text{dem}}$ can be expressed as $\boldsymbol{H}_{\text{dem}} = -\nabla U(r)$, where $U(r)$ is the magnetostatic scalar potential at the point $\boldsymbol{r}$. In principle, the potential $U(r)$ can be calculated from the explicit formula [84, 82]

$$U(r) = \int_{\Omega} \boldsymbol{M}(r) \cdot \nabla_{r'} G(r, r') dV \tag{3.34}$$

where $\Omega$ is the volume containing the ferromagnetic material and $G(r, r')$ is the Green's function

$$G(r, r') = \frac{1}{r - r'} \tag{3.35}$$

The direct integration of equation 3.34 requires $O(N^2)$ matrix elements, where $N$ is the number of discretization points. For a very small problem with 10,000 nodes this approach would already require nearly one GB of memory, and the quadratic dependence on $N$ makes it impossible to use such an approach in modern simulations like those of this thesis, which may involve millions of finite elements. The calculation and storage of such large matrices is prohibitively expensive and slow. For a typical time-dependent simulation, the demagnetizing field has to be calculated several tens of thousand times, and thus the direct integration approach would be much too slow and is never used in simulations.

Instead of a direct integration, $U(\boldsymbol{r})$ can be determined by solving the Poisson equation. In order to account for the boundary conditions of the magnetostatic potential, the FEM can be combined with the boundary element method in the form of a finite-element/boundary-element (FEM-BEM) scheme as described by Fredkin and Koehler [159, 160]. Solving the Poisson equation is numerically less expensive than a direct integration. Instead of having to store the $O(N^2)$ discretized Green's function, the formalism requires the solution of a linear set of equations described by the $O(N)$ stiffness matrix introduced in section 3.3.1 is stored.

A practical difficulty of solving the magnetostatic problem in this way is connected with the boundary conditions of $U(r)$, which are defined at infinity [161] according to $U(r) \rightarrow 0$ for $r \rightarrow \infty$. This open-boundary condition suggests a requirement to calculate the potential also in elements far outside the volume containing the magnetic material $(\Omega)$, hoping that a suitable cut-off radius can be defined at which the potential can be set to zero. Such an approach would be inaccurate an undesirable, as it would introduce the need to include a large number of uninteresting external elements. This number of unwanted elements would grow rapidly with size in the three-dimensional case.

Such a situation can be circumvented by FEM-BEM schemes like the Fredkin-Koehler method [159]. Essentially, the boundary element method maps the boundary conditions in the outer region onto a surface integral, which needs to be coupled to the FEM solution in the inner region. The method is accurate and removes the need to extend the computational region beyond the ferromagnetic volume. The disadvantage of this method is that the surface integral involves a densely populated matrix, which in spite of being much smaller than in the case of a direct integration can still limit the applicability of the method in the case of large problems. The size of the BEM matrix scales approximately with the order of $O(N^{4/3})$ in the case of a spherical particle, or more generally with $O(N_b^2)$ if $N_b$ is the number of nodes at the surface – the boundary nodes. In our finite-element simulation tool, the obstacle related with the size of the BEM matrix is removed by the use of $\mathcal{H}^2$-type hierarchical matrices [162, 163, 164]. By using $\mathscr{H}^2$ matrices one can convert the quadratic complexity of the surface integral to a linear one. As demonstrated recently, we can thereby achieve a reduction of the matrix size by about $99\,\%$ in problems with over $10^6$ surface nodes [151]. This significant reduction of the numerical costs, obtained while preserving a high accuracy, enables us to calculate problems with an extensive size on machines with modest specifications.

The Poisson equation for the magnetic scalar potential $U(r)$ is given by

$$\Delta U(\boldsymbol{r}) = -\rho \tag{3.36}$$

Where $\rho = -\nabla \cdot \boldsymbol{M}$ is the magnetostatic charge density. The normal vomponent of the magnetostatic field is discontinuous at the boundary surface $\partial(\Omega)$ of the volume containing the magnetic material $(\Omega)$, which yields the condition

$$\frac{\partial U^{in}(r)}{\partial n} - \frac{\partial U^{out}(r)}{\partial n} = \boldsymbol{M} \cdot \boldsymbol{n} \tag{3.37}$$

for the normal derivatives of the potential at the urface, where the superscripts "in" and "out" represent the inner and outer limits of the derivative at the boundary surface $\partial\Omega$, respectively, and $\boldsymbol{n}$ is the normal vector oriented towards the outside. To solve for (3.36) inside the magnetic volume $\Omega$, we follow the ansatz by Fredkin and Koehler to split $U(r)$ into two parts $U(r) = U_1(r) + U_2(r)$, each with specific properties. The first part $U_1$ is the solution of the Poisson equation inside the magnetic volume $\Omega$

$$\nabla^2 U_1(r) = -\nabla \cdot \boldsymbol{M}, \quad r \in \Omega \tag{3.38}$$

with Neumann-type boundary condition at the surface $\partial\Omega$

$$\frac{\partial U_1}{\partial n}\bigg|_{\partial\Omega} = \boldsymbol{M} \cdot \boldsymbol{n} \tag{3.39}$$

The second part $U_2$ solves the Laplace equation

$$\nabla^2 U_2 = 0 \tag{3.40}$$

.The normal derivative of $U_2$ at the boundary $\partial\Omega$ is continuous:

$$\frac{\partial U_2^{in}(r)}{\partial n} - \frac{\partial U_2^{out}(r)}{\partial n} = 0 \tag{3.41}$$

This technique of splitting $U$ into two parts allows us to calculate $U$ by only having to solve inside the region $\Omega$, which is otherwise not possible as for the original problem the boundary condition at the surface $\partial\Omega$ is unknown.

The first part $U_1$ is calculated by solving the Poisson-Neumann problem (3.38) by converting it to the weak form, as discussed before in the case of the exchange field

$$\int_V \psi \Delta U_1 dV = \int_V \psi \nabla \boldsymbol{M} dV \tag{3.42}$$

Integrating by parts,

$$\int_V \nabla \cdot \psi \cdot \nabla U_1 dV - \int_{dS} \psi (\nabla U_1 - \boldsymbol{M}) \cdot \boldsymbol{n} dS = \int_V \nabla \psi \cdot \boldsymbol{M} dV \tag{3.43}$$

Using the Neumann boundary condition (3.39) for the gradient of $U_1$ at the boundary $\partial\Omega$ one obtains

$$\int_v \nabla \psi \cdot \nabla U_1 dV = \int_V \nabla \psi \cdot \boldsymbol{M} dV \tag{3.44}$$

From this, $U_1$ can be computed by expanding the terms in the basis defined by the shape functions. Next, we consider the Laplace equation of the second part $U_2$:

$$\nabla^2 U_2 = 0 \tag{3.45}$$

A unique solution of the Laplace equation is obtained ifhe Dirichlet boundary conditions are provided, i.e., the value of $U_2(r)$ at $r \in \partial\Omega$. This information can be obtained from the following relation

$$U_2(r) = \frac{1}{4\pi} \int_{\delta\Omega} U_1(r') \frac{\partial G(r, r')}{\partial n} ds \tag{3.46}$$

Even though equation (3.46) is valid everywhere inside the body, its computation at all discretization point is computationally expensive. Hence, it is integrated only at the boundary surface to obtain the necessary Dirichlet boundary condition to solve 3.45. This results in the following integral connecting the values of $U_1$ with those of $U_2$ at the surface [160]:

$$U_2(r) = \frac{1}{4\pi} \int_{\partial\Omega} U_1(r') \frac{\partial G(r, r')}{\partial n} ds + \left( \frac{\Theta(r)}{4\pi} - 1 \right) \qquad (3.47)$$

Here $\Theta(\boldsymbol{r})$ is the solid angle subtended at the surface point $\boldsymbol{r}$. The integration according to eq. 3.47 is implemented numerically by a collocation approach, as described in Ref. [151].

With this collocation scheme, the discretized representation of the integral (3.47) takes the form

$$\underline{U_2} = \underline{\underline{P}} \cdot \underline{U_1} \qquad (3.48)$$

where $\boldsymbol{P}$ is a densely populated matrix of size $N_b \times N_b$ ($N_b$ is the number of nodes at the surface). The $O(N_b^2)$ scaling of this dense matrix is the aforementioned bottleneck of the Fredkin & Koehler approach. For large problems, the memory requirements for the storage of this matrix can become prohibitively high. To circumvent this, a hierarchical matrix compression method is employed.

### $\mathscr{H}^2$ matrix compression

Sub-matrices of rank $t \times s$ can be approximated by low rank matrices. In factorized form,

$$\underline{\underline{K}}\Big|_{t \times s} \approx AB^T, \quad A \in R^{t \times k}, \quad B \in R^{s \times k} \qquad (3.49)$$

There exist several different types of algorithms for matrix compression. Analytical methods such as by Taylor expansion, interpolation *etc.* are highly reliable but they result in matrices with ranks higher than necessary[165, 166]. On the other hand algebraic methods offer high efficiency [167] but may lead to unreliable results. Hybrid matrix compression [168] combines an initial analytic approximation and then an algebraic compression in order to get the benefits of both techniques. Our micromagnetic code uses $\mathscr{H}^2$-matrix compression [151] provided by the open-source matrix compression algorithm *H2Lib* [169]. Using this method, a remarkable reduction of the memory requirements by up to about 99 % can be achieved. This reduction moreover increases the computation time of the matrix-vector product. It offers a nearly linear $O(N_b)$ scaling of memory with the number of surface nodes $N_b$. We can thereby rapidly calculate magnetostatic fields even for large problems that, without such a compression, would theoretically require RAM sizes in the TB range. Using the $\mathscr{H}^2$ compression scheme, the size of the matrices required to treat such problems can be reduced to a few GB.

## 3.4   Time integration of the LLG equation

In section 2.1.5 we discussed the different energy contributions in micromagnetism. A stable equilibrium magnetization state can be obtained by minimizing the total energy of the system. If we are instead interested in obtaining the time dependent evolution of magnetization $\boldsymbol{M}(r,t)$, we must numerically integrate ("solve") the discretized LLG equation. The magnetic state $M(r, t + \delta t)$ at time $t + \delta t$ is obtained by the time integration of the LLG equation based on the magnetic state $\boldsymbol{M}(r,t)$ at the time $t$. For this, the LLG equation is discretized to $3N$ set of partial differential equations

$$\frac{\partial \boldsymbol{m}_{i,l}}{\partial t'} = -\frac{1}{1+\alpha^2}\boldsymbol{m}_{i,l}\times\boldsymbol{h}_{i,l} - \frac{\alpha}{1+\alpha^2}\boldsymbol{m}_{i,l}\times(\boldsymbol{m}_{i,l}\times\boldsymbol{h}_{i,l}) \qquad (3.50)$$

where the subscripts $i = 1, 2, 3...N$ are the node index numbers, $l = x, y, z$ are the three Cartesian components, $\boldsymbol{m} = \boldsymbol{M}/M_s$ is the reduced magnetization, $\boldsymbol{h} = \mu_0\boldsymbol{H}_{\text{eff}}$ is the reduced effective field, and $t' = \frac{|\gamma|}{\mu_0}t$ is the reduced time. Once this system of ordinary differential equations (ODE) is set up, numerical ODE solvers can be used to integrate for $\boldsymbol{m}(r,t)$ in time. Various ODE solvers, which we will discuss later, are available as open-source in the form of libraries.

The simplest approach for finding the magnetization at time $t + \delta t$ would be an explicit time integration of the LLG equation based on the value of $\boldsymbol{m}$ and $\boldsymbol{h}_{\text{eff}}$ at the time $t$. In this approach, which does not account for the change in the effective field caused by the change in magnetization within the time step $\delta t$, the calculation can become unstable. The time integration therefore employs schemes in which the effective field is continuously updated with any change in $\boldsymbol{m}(r,t)$ at each integration step. This is especially important in the case of the exchange field, where even a small variation in the magnetization distribution can manifest as a considerable change in the effective field.

To be solved with standard library solvers, equation (3.50) can be expressed as an initial value problem (IVP) in the following form,

$$\frac{dy_i}{dt} = f_i(y_1, y_2, y_3...y_{3N}, t), \quad i = 1, 2, 3\ldots L, \quad y_i(t = 0) = W \qquad (3.51)$$

Where $y_i$ are the three Cartesian components of the magnetization $\boldsymbol{m}(r,t)$ and $f_i$ is the right-hand side of equation 3.50 which should be supplied as a routine to the solver. The solver takes the initial $W$ value, that is the discretized magnetization state at $t = 0$, as an input. We should also provide the value of the time step $\delta t$, which is the time-step after which the solver returns the computed magnetization to the user. In the case of adaptive time steps, the solver may internally sub-divide this user-defined time step $\delta t$ further into finer intervals. The value selected by the user for the time step is therefore in fact the *maximum* time step that the routine performing the integration is allowed to take before an update of the effective fields is enforced. To save computational resources, `tetmag` provides the option to "freeze" (i.e., not to update) the demagnetizing field within this time range $\delta t$. With time steps in the range of $100\,\text{fs}$ this approximation is usually very accurate. In any case, the remaining effective field terms, in particular the exchange field, are updated continuously. Once the solver returns the value of $\boldsymbol{m}(r)$ at the time $t + \delta t$ time, the demagnetizing field is calculated again and the effective field is updated. This process is repeated until convergence is obtained, or until a user-defined time range is simulated. During the simulation, it is important to chose a suitable value for the time-step $\delta t$. A too large time-step can lead to instabilities, while a time step too small unnecessarily increases the computation time required for the simulation.

The time integration of the LLG is a critical part of the code, and in spite of the internally performed adaptive adjustment of time steps, the task cannot be performed in a fully automated manner. The user must provide the maximum time step $\delta t$ in order to define the range of admissible time steps $\Delta t$ that are used for the LLG integration. The optimum value for this input parameter can be determined based on the value of damping term $\alpha$ used for the simulation. Typically, if we are interested in finding the minimum-energy relaxed magnetic state for a particular set of conditions and details of the precessional magnetization dynamics are unimportant, we may use a large value of alpha (0.5-0.6) and accordingly a large time step ($\delta t = 1\,\text{ps}$). On the other hand, if we are interested in the time-dependent dynamics of the system with a realistic low value for the damping term ($\alpha = 0.01$), an accordingly smaller time step ($\delta t = 0.1\,\text{ps}$) must be chosen.

Our micromagnetic simulation software `tetmag` has an option for choosing between two different solvers to treat this initial-value problem with $3N$ coupled degrees of freedom. The first one is a Dormand-Prince $5^{th}$ order Runge-Kutta method based solver provided by the ODEINT package [170] included in the C++ Boost libraries [171]. The second one is CVODE, an Adams-Moulton predictor-corrector method based solver provided by the Sundials library [172]. The first solver is mostly used for fast relaxation simulations where a large time-step can be used. The second (CVODE) solver is more stable but slightly slower. It is mostt suitable for the accurate simulation of time-dependent dynamics with a low value of the Gilbert damping. Both solvers can exploit acceleration through massive parallelization, both thread-based as well as GPU based.

## 3.5 Simulation method: working of tetmag

All the simulations discussed in this thesis are are carried out with our finite element simulation software **tetmag**, which is developed and maintained by Riccardo Hertel at the IPCMS Strasbourg. Although the development of the micromagnetic finite-element code was not apart of this thesis, in this section we briefly discuss about the working and the capabilities of `tetmag`. The main part of the `tetmag` code is written in C++ [173] in an object-oriented form. This modular structure of the code allows for a flexible development, making it possible to insert, improve or modify specific classes of the code according to new simulations demands without having to re-write or to modify existing parts of the code. This approach makes the code robust and easily maintainable. For several central and performance-sensitive tasks, `tetmag` makes use of high-level libraries like Eigen [174] or Boost [171]. The code is thread-parallelized with OpenMP [175], which makes `tetmag` suitable for large-scale computations on high-performance clusters. A good part of the results discussed in this thesis have been carried out on the high-performance computer center of the University of Strasbourg [176]. Moreover, `tetmag` can also achieve a significant speedup through GPU acceleration. This is achieved by implementing certain classes in the CUDA language [177] and making use of the Thrust framework [178] for GPU acceleration. The choice of whether to use GPU acceleration or thread-based CPU parallelization can be made by the user through a simple interface. For large-scale problems, where the number of elements exceeds about $5 \times 10^6$, the simulation rate can be increased considerably by enabling GPU acceleration. The `tetmag` software takes as input the following parameters for each simulation

- an ASCII file listing the material parameters

- a text file containing details of the simulation configuration

- optional: a file containing the initial magnetization structure of the problem

The finite element mesh can be generated by means of free and open-source codes like Gmsh [152], Netgen [153], FreeCAD [179] *etc.* In the text file describing the simulation configurations, the user can choose specify the input parameters such as the value of Gilbert damping (typically with values $0.01 \leq \alpha \leq 1$), the total simulation time, various types of external fields, the type of solver to use, whether to use CPU or GPU for the calculation, and the initial magnetization state. The initial magnetization state is the configuration from which the time evolution of the simulation state starts. This can be a trivial state like a homogeneously magnetized state or artificial configurations like a randomized initial state, where the magnetization at each discretization point is oriented in a random direction in space. The initial state can also be user-prepared, for example one can use the relaxed state obtained from a previous simulation. Once these input ingredients are prepared, a simulation can be initiated. The `tetmag` software will find the value of the magnetization vector $\boldsymbol{m}(r, t)$ at each discretization point at each time-step by integrating the initial value problem in time, as discussed in the previous section. The output is a time series of magnetization states for each time step. The standard output for `tetmag` is given in files of "vtu" format, which contain information about the finite element mesh and on the magnetization vector field at each point. These files can be viewed and analyzed with the powerful ParaView software [180], which is free and open-source. Details about the post processing of the simulated data are briefly discussed in the next section.

## 3.6 Post processing: Extraction of frequency modes with a windowed inverse Fourier transform

Oscillatory magnetization processes, at which the magnetization vibrates in the GHz range, play an important role in magnonics and in the analysis of the intrinsic modes of a nanomagnet. Such magnetization oscillations can be excited by applying a short perturbation to a relaxed magnetic structure. The signals obtained from the ensuing simulated dynamics may contain contributions from two or more distinct and superposed frequencies. To disentangle such oscillations, and in general, to get a deeper understanding about the dynamical properties of the system, it is necessary to filter out the individual modes and to analyze their spatial profile separately.

The first part of this thesis was involved with the development of post-processing tools to identify and isolate individual high-frequency modes from the simulated results of an oscillatory magnetization dynamics through Fourier analysis. In this subsection, the theory and implementation of this technique is briefed. The Fourier transform $g(\omega)$ of a function $f(t)$ is defined as

$$g(\omega) = F(f(t)) = \frac{1}{\sqrt{2\pi}} \int_{-\infty}^{\infty} f(t)e^{i\omega t}dt \qquad (3.52)$$

And the inverse function, inverse Fourier transform is defined as

$$f(t) = F^{-1}(g(\Omega)) = \frac{1}{\sqrt{2\pi}} \int_{-\infty}^{\infty} g(\omega)e^{-i\omega t}d\omega \qquad (3.53)$$

The function $F$ – the Fourier transform function described in equation (3.52) – describes a transformation from the time domain to the frequency domain. If $f(t)$ is a periodic function of time, the corresponding Fourier transform $g(\omega)$ contains information about $f(t)$ that can be decomposed into individual sinusoidal signals of different amplitude. This transformation is widely used in digital signal processing (DSP) to extract individual frequency components from a mixed signal. The discrete Fourier transform (DFT) is the discrete equivalent of the continuous Fourier transform, defined as

$$X(\omega_k) \triangleq \sum_{n=0}^{N} x(t_n) e^{i\omega_k t_n}, \quad k = 0, 1, 2, \ldots, N-1 \qquad (3.54)$$

Where $x(t_n)$ is the input signal, $X(\omega_k)$ is the complex-valued spectrum and $N$ is the number of samples in the input signal.

The original signal, or any particular part of the signal, can be re-created from $X(\omega_k)$ by the inverse function, i.e., the inverse discrete Fourier transform (IDFT), which is defined as

$$x(t_n) = \frac{1}{N} \sum_{k=0}^{N-1} X(\omega_k) e^{-j\omega_k t_n}, \quad n = 0, 1, 2, \ldots, N-1 \qquad (3.55)$$

To investigate the magnonic modes of any system, the relaxed magnetization configurations are excited with various techniques and the resulting magnetic ring-down; that is the oscillatory, small angle precessional relaxation from the excited to equilibrium state is recorded as a function of time $t$. The recorded time-dependent magnetization usually contains a superposition of several frequency modes, and it can be difficult to understand the dynamic processes associated with each frequency through a simple visual inspection. Hence, in the initial stages of the studies of this thesis, a Fourier analysis tool was developed which could identify and filter out the individual modes from these superposed oscillations. The tool was written in Python [181] in an object-oriented way. For the numerical calculation of the Fourier transform and the inverse Forier transform, the FFT library of the Numpy package [182] was employed. The tool takes as input the VTU files produced by the `tetmag` simulation software. The time evolution signal of the magnetization at each node point is obtained by reading the corresponding data from all the VTU files. The code will then extract the frequency modes contained in these signals as explained above. The local power spectra at each node $i$ can be defined as

$$p_i(\omega) = |g_i(\omega)| = |F(\boldsymbol{m}_i(t))|, \quad i = 1, 2, 3 \ldots N, \quad l = x, y, z, \quad (3.56)$$

where $F(f)$ is the discretized Fourier transform. It contains the information about the different frequency components and their relative intensity at each node $i = 1, 2, 3..N$. To obtain a global view of the entire structure, the power spectra at all the discretization nodes are summed, yielding an average spectrum

$$P(\omega) = \sum_{i=1}^{N} p_i(\omega) \quad (3.57)$$

It is to be noted that the average of the discretized power-spectra can be completely different from the power-spectrum of the averaged magnetization. The spatially averaged power spectrum $P(\omega)$ contains information about the frequency of various modes and its average intensity in the whole structure. This can be displayed as a frequency vs. intensity plot, and the various frequency components and their relative intensity can be visually identified. The dynamical profile of a desired individual frequency can then be filtered by means of discretized windowed-inverse Fourier transform at each node. This is achieved by suppressing the remaining frequency contributions in the Fourier transform $g_i(\omega)$ and then carrying out an inverse Fourier transform (iDFT) back to the time domain. This results in the extraction of the oscillation of $m(r, t)$ at a particular frequency at each node. This process is repeated for $N$ nodes to get the overall dynamics at the desired frequency.

The Forier analysis tool was modified and improved continuously throughout the thesis period such as to carry out various problem-specific and advanced analyses. For example, to look into the modes developing at artificial nanowire networks, the tool was modified to include an option which could extract some specific geometrical locations of a lattice. This made it possible to selectively analyze the behavior of nanowires, spheres and other units isolated from the rest of the lattice. The flexibility of the python language made these modifications effortless and efficient. The working principle of the Fourier extraction tool is illustrated in figure 3.3.

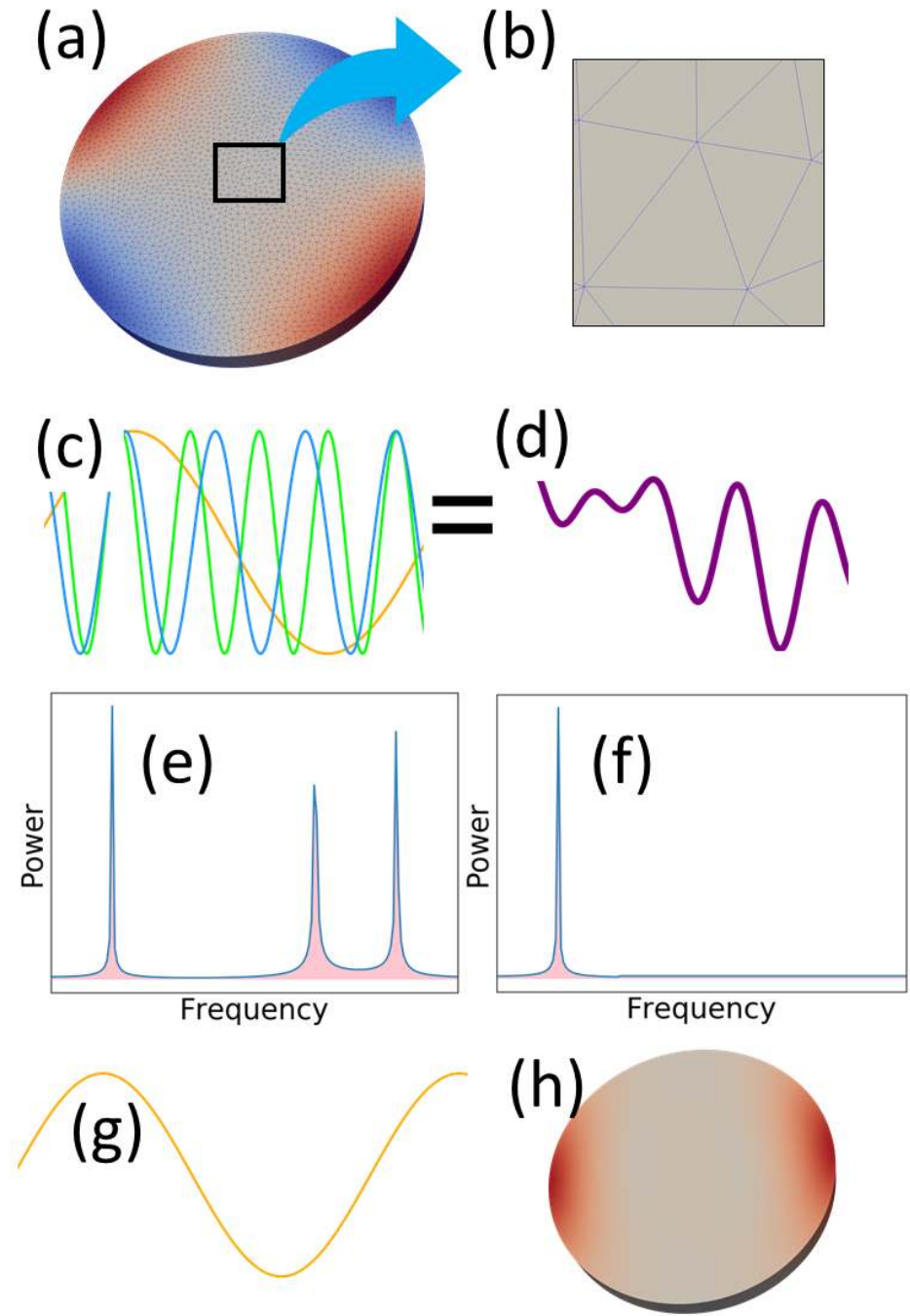

Figure 3.3: (a) Snapshot of the oscillation of $\boldsymbol{m}(r,t)$, defined at each node of the finite-element mesh (b). The time-series of $\boldsymbol{m}_i(r,t)$ of the magnetization at a node $i$ is a superposition of several oscillations (c,d). The oscillation at the $i^{th}$ node can be Fourier- transformed to a power spectrum, from which three modes can be identified. (f) A windowed power-spectrum is generated by filtering out the first mode and suppressing the higher modes. From this windowed Fourier spectrum, an inverse Fourier transform (g) yields the isolated dynamic profile of the desired frequency at this node. Repeating this process for all the $N$ nodes will give the overall spatial profile of the desired frequency mode (h).

# CHAPTER 4

Micromagnetic simulations of high-frequency
modes in Cobalt nanodots

Before studying the three-dimensional network of nanowires, we investigated the magnetization structures and the high-frequency response of a more traditional geometry: Cobalt nanodots. This study was motivated by a collaboration with the experimental group of David Schmool of the University of Versailles. The experimentalist colleagues fabricated Cobalt-Silver nanodots arrays with a nanodot radius 100 nm and a thickness of approximately 50 nm for the Cobalt layer and 30 nm for the capping silver layer. A vector network analyzer-ferromagnetic resonance (VNA-FMR) analysis of the nanodot array is shown in Fig. 4.1. The FMR measurement revealed the presence of two modes, whose frequency is seen to be increasing with the applied field. By means of simulations, we investigated the magnetic structure formed in these nanodots and we could identify the mode profiles leading to the two peaks found in the FMR.

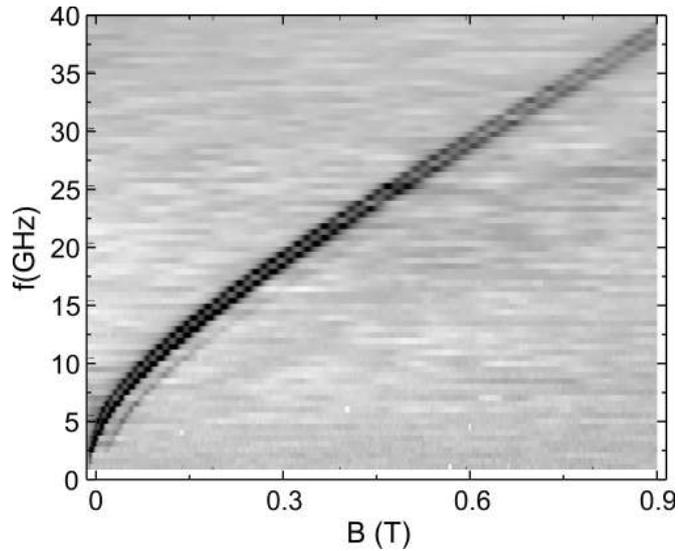

Figure 4.1: VNA-FMR data of Co/Ag nanodot arrays, with field applied along the plane of the nanodots [183]

Due to limitations in the experimental techniques, the exact values of the material parameters such as saturation magnetization $M_s$ and the anisotropy were unknown, also due to other reasons such as a possible oxidation of the Co dots and possibly interdiffusion processes of the Co-Ag layers. To account for these uncertainties, we performed simulations with slightly varying thickness and radius varied. A series of simulations were conducted by systematically varying these parameters in an appropriate range. By comparing the results with those of the experiments we could deduce the material parameters of the nanodots.

## 4.1   Finite element modeling and simulation

The finite-element mesh of the nanodot was prepared using the open-source mesh generation software GMSH [152]. The radius of the disc was varied from 75 to 100 nm and the thickness was varied from 20 to 50 nm. The maximum mesh-size ($lc$) was maintained below 4 nm to make sure that the cell size was always below the exchange length of the material.

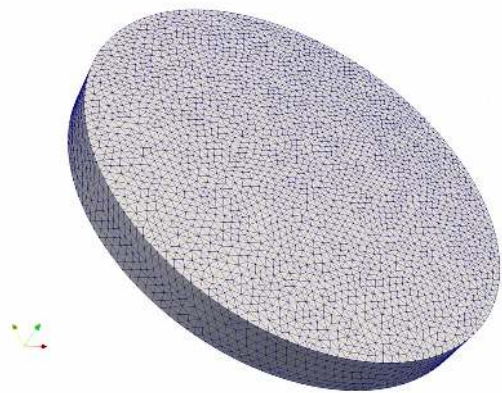

Figure 4.2: Finite element mesh of the nano-disc with thickness 20 nm and radius 100 nm. The unstructured grid contains 10,750 nodes and 59,780 tetrahedrons

The saturation magnetization $\mu_0 M_s$ of bulk Cobalt is usually 1.7 T [184]. However, fitting the FMR data to the analytical models [183] provided a value of 1.1 T for the saturation magnetization. Additionally, magnetometry measurements indicated a value of 0.34 T, which is four times less than that of the bulk value. The saturation magnetization was systematically varied from 0.343 T to 1.7 T. A value of $3.0 \times 10^{-11}$ J m$^{-1}$ was used for the exchange stiffness constant $A$. The value of uniaxial anisotropy was varied from 0 to 18 kJ m$^{-3}$.

The relaxed magnetization structures at various fields were calculated by simulating the magnetization structure until the value of maximum torque falls below a pre-defined value. These relaxed states were then excited with various methods:

1. With a Gaussian field pulse [185] or

2. By applying a small enough field at a small angle to slightly lift the magnetization from the equilibrium and then letting it relax to the original field [186].

The magnetic ring-down of these excited states are recorded and the individual frequency modes are isolated an analyzed using the techniques described in the previous chapter. The results of the FMR were simulated by varying the in-plane field from 100 mT to 900 mT.

## 4.2   Magnetization structure

A simulated hysteresis loop of the nanodisc with thickness 50 nm and radius 100 nm is shown in Fig. 4.3. Relaxing the field of 200 mT to 200 mT from a uniformly saturated state results in a "onion"-type [187] magnetic state, see Fig. 4.3(c), in which the center part of the disc is magnetized along the field direction, while the magnetization at the edges curves along the surface to form a structure resembling the cross-section of an onion. As the field is further reduced, the outermost layer that is curved along the surfaces grow in size. Below a field of 2 mT a vortex state develops, which stays stable until a negative field of about 130 mT is applied. Simulated magnetization structures of the disc at various fields are displayed in figure 4.4. As a next step, we conducted dynamic simulations in an attempt to reproduce the experimental FMR results. Using the perturbation methods described above, small-angle precession modes of the disc were stimulated at various fields.

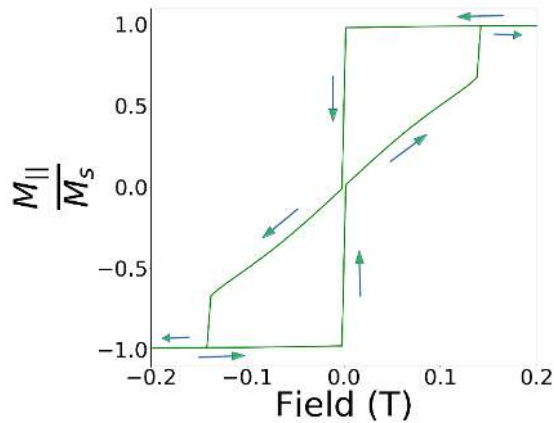

Figure 4.3: Simulated hysteresis loop of the Cobalt nanodot thickness 50 nm and radius 100 nm and a saturation magnetization of 1.1 T

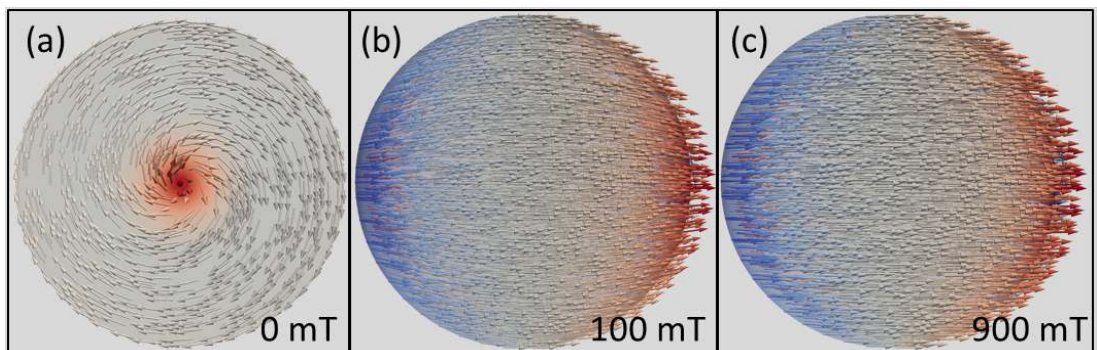

Figure 4.4: Relaxed magnetization structure of the nanodot with thickness 50 nm and radius 100 nm at different field values. The arrows represent the local direction of the magnetization, and the color code indicates the value of the $z$ component of magnetization. (a) Vortex state at 0 mT field, (b) "onion-state" at a field of 100 mT, (c) nearly saturated state at 900 mT.

## 4.3 Magnetization dynamics

The magnonic response of the disc at 400 mT is shown in Fig. 4.5. There exist two principal modes. The lower frequency mode at around 15.7 GHz is the edge mode, caused by symmetric oscillations of the head/tail part of the onion structure. The higher-frequency mode at around 20.3 GHz is caused by the oscillation involving both the central part and the edges of the disc. These individual frequency peaks are identified and isolated using the Fourier extraction tool as explained in the previous chapter. The extracted spatial profile of the two frequency modes are displayed in Fig. 4.6. These results qualitatively agree with the observation of two major frequency peaks in the FMR measurements. Various parameters such as the thickness, radius, the saturation magnetization, and the strength of the anisotropic field were varied to match the simulated frequency with the observed values.

Vibrating sample magnetometry measurements of the Co dot array yielded a saturation magnetization of 343 mT, a value that is much lower than the one of bulk Cobalt, which is 1700 mT. Such a reduction of the saturation magnetization $M_s$ may be caused by a combination of aging and inter-phase diffusion processes in the sample. To investigate the impact of the value of the saturation magnetization on the high-frequency oscillations, we performed simulations in which we varied $M_s$ from 343 mT to 1700 mT. According to information obtained from the colleagues, we also included a crystalline anisotropy corresponding to a $Q$-factor of 0.01. The resulting frequency response was compared with the experimental data 4.7. It was observed that the frequency of the first mode slightly decreases and that the frequency of the second mode increases linearly with an increase in $M_s$. Based on the frequency values and the relative separation between the two modes, we could assume that the most probable value of the saturation magnetization is about 1100 mT.

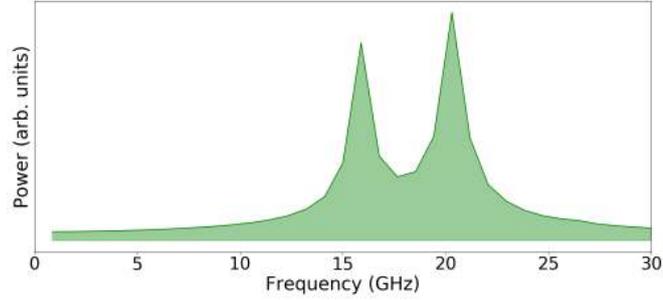

Figure 4.5: Power spectrum showing the typical frequency response of the disc at 400 mT at a damping of 0.1 and a simulation time period of 10 nanoseconds.

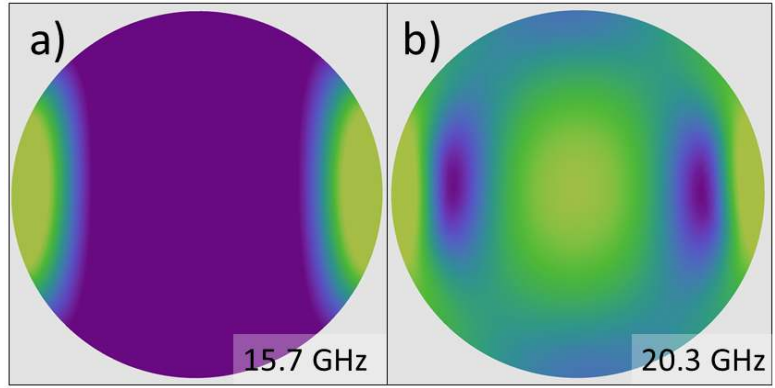

Figure 4.6: Extracted spatial profile corresponding to the oscillations at the two main frequencies. Yellow colored region are the anti-nodes, where the oscillation amplitude is maximal and the blue colored areas are the inactive regions (arbitrary units are used). (a) Edge mode oscillation of the lower-frequency mode (b) Oscillation of the higher-frequency mode.

Based on these results, the value of saturation magnetization was fixed and various other parameters such as anisotropy constant are varied. It was observed that a thickness of 20 nm, radius 100 nm, and a uniaxial anisotropy of 18 kJ m$^{-3}$ resulted in the closest agreement to the experimental results. Using these parameters, the field dependence of the frequency response shwon in Fig. 4.8 was simulated.

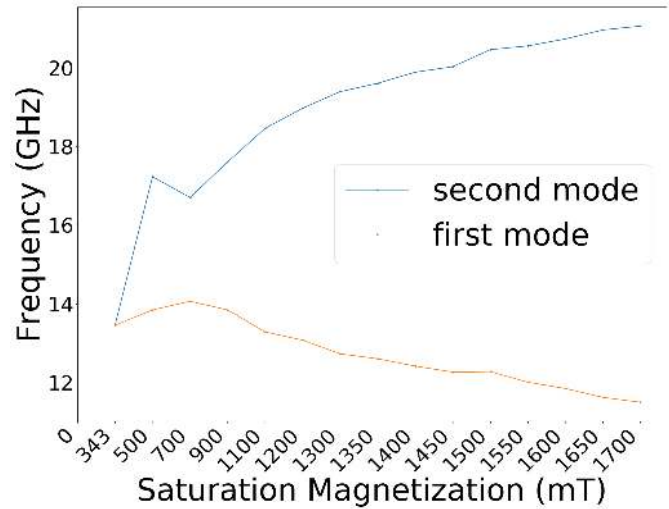

Figure 4.7: The variation of the frequency of the modes with change in saturation magnetization

## 4.4    Conclusion

To summarize, two dimensional arrays of Cobalt nanodots were fabricated by the collaborators and the field dependent frequency response was measured. Two prominent modes were identified in this and to investigate the dynamical profile of these modes finite element simulations were carried out. Owing to limitations in computational resources simulations were carried out on a single nanodot, instead of the whole array. We could identify the micromagnetic structure formed in these nanodots at various fields and also could identify the magnetic profile of the two modes. The field dependent frequency response was re-created through simulation which had qualitative agreement with the simulation results. Also, due to some constraints in experimental techniques, there were minor uncertainties in the material parameters. By comparing the frequency of the modes obtained in the simulation with that of the experimental values the material parameters could be indirectly obtained. As a future extension to this study, the whole array of nanodots could be simulated instead of a single one.

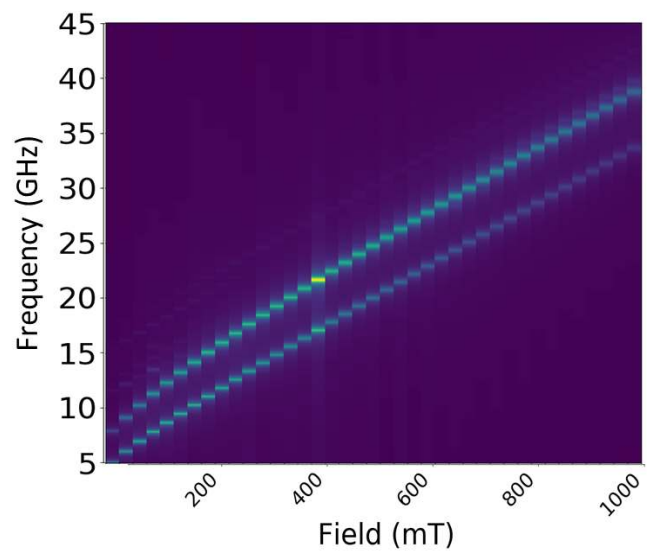

Figure 4.8: Variation of the frequency response of the disc with an external field

# Chapter 5

Three-dimensional Sierpinski fractal structures

## 5.1 Introduction

Geometric fractals are self-similar structures which are scale invariant [188]. From the crystalline structure of snowflakes to the distribution of veins on plant leaves, to the surface structure of mountains – fractals are abundant in nature. Usually, fractal structures are naturally formed as a result of a growth processes[189]. In the field of magnetism, it has been observed that, under specific conditions, magnetic domains and domain walls can form fractal structures [190, 191, 192, 193]. Numerous studies have been carried out in the past to investigate the effects of the formation of fractal domain walls on the magnetization reversal [194, 195, 196]. Additionally, the magnetic properties of 2D ferromagnetic nanostructures with self-similar fractal geometries have gained some attention [197, 198, 199, 200]. The interaction of coupled, self-similar oscillators in such structures can potentially give rise to interesting collective phenomena, such as a wide band absorption in fractal antennae [201]. In this section, we discuss magnetization structures formed in a three-dimensional Sierpinski gasket structure, their magnetization reversal, and their oscillatory high-frequency dynamics.

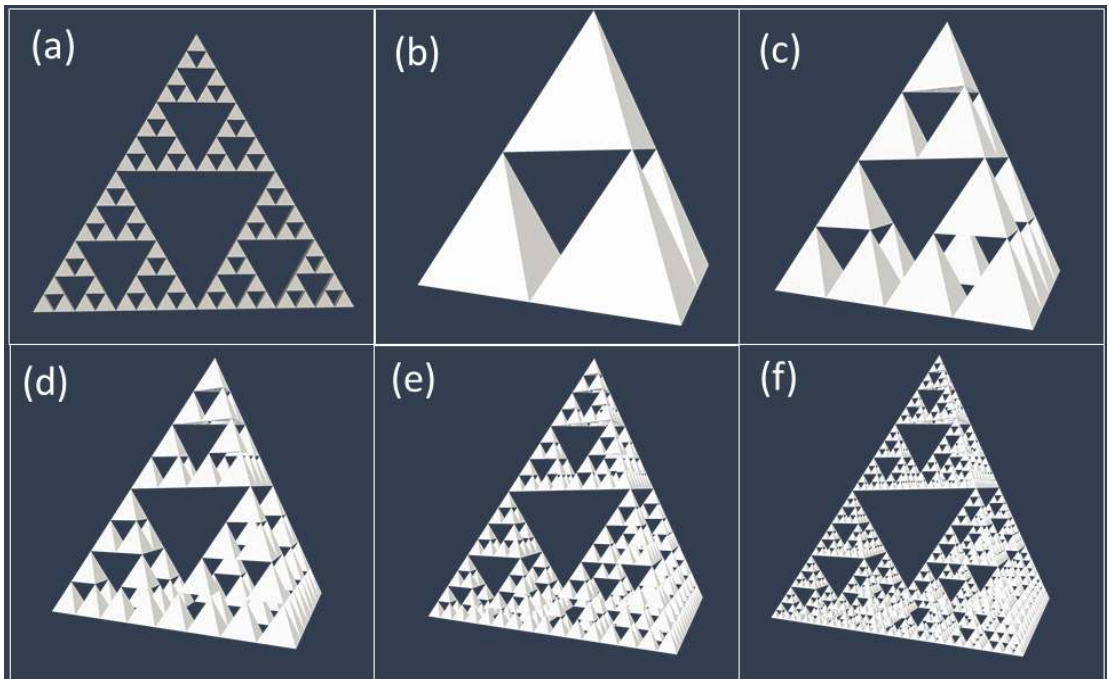

Figure 5.1: (a) A two-dimensional Sierpinski triangle (b) Stage 1 of the 3D fractal composed of four tetrahedrons of side length 256 nm (c,d,e,f) Stage 2,3,4 and 5 of the fractal structure respectively.

## 5.2 Methodology

A 3D Sierpinski structure is made of four tetrahedrons recursively arranged on the four corners of another tetrahedron. This can be considered as a three-dimensional variation of the more familiar two-dimensional Sierpinski triangles shown in Fig. 5.1(a). Our stage one fractal structure was a large tetrahedron of side length 512 nm which was, in turn, made of 4 smaller tetrahedrons of side length 256 nm. The successive stages of 2,3,4 and 5 are made by recursively dividing the individual tetrahedrons into further self-similar structures. Thus, the $n^{th}$ iteration of the fractal had $4^n$ tetrahedrons of side length $\frac{512}{2^n}$. Our final iteration, that is stage 5 of the fractal, contains 1024 tetrahedrons of side length 16 nm. The finite-element models of each iteration of the fractal are displayed in Figs. 5.1(b)-(f).

For the numerical simulation we, used material parameters corresponding to those of Permalloy, with a saturation magnetization of $M_s = 8 \times 10^5 \, \mathrm{A \, m^{-1}}$ and an exchange stiffness of $A = 1.3 \times 10^{-11} \mathrm{J \, m^{-1}}$ and zero crystalline anisotropy. These material parameters results in an exchange length of 5.6 nm and, correspondingly, all finite-element meshes are made with a mesh-size remaining below this value. The maximum cell size was set to 4.0 nm. The stage one structure consisted of nearly 100,000 nodes and nearly 500,000 finite element cells (which are also tetrahedrons), and the stage five structure had over 36,000 nodes and more than 70,000 finite element cells. Since the total side length of the whole structure was kept constant at 512 nm and the because of the particular way in which the tetrahedrons are distributed, the total surface area of the structure remains constant at $\sqrt{3} \cdot 512^2$ for all the stages. The total volume decreases logarithmically with each stage, resulting in a corresponding increase of the surface-to-volume ratio, see Fig. 5.2.

Stable magnetic states developing in these structures are obtained by first saturating the structure along the $z$ direction under a sufficiently strong magnetic field, which is then gradually relaxed to zero. The hysteretic properties of these structures were obtained by applying an external field of 500 mT along the $z$ direction and then reducing the field in appropriate small steps. The magnonic response of the various equilibrium states are obtained by exciting the relaxed states by various techniques and recording the magnonic ring-down. The frequency components are then investigated by means of an inverse windowed Fourier transform.

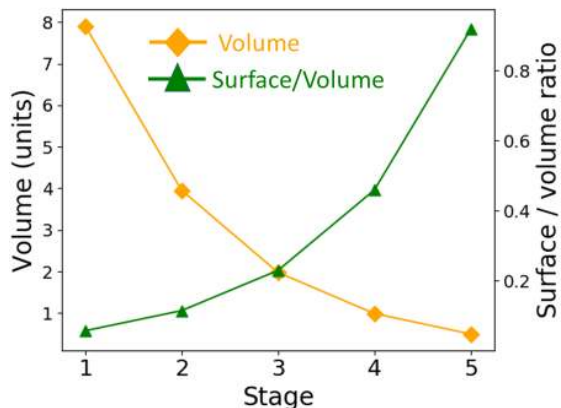

Figure 5.2: Change in the volume and surface/ volume ratio with change in stage

## 5.3   Magnetization structure at zero field

In the structure corresponding to the first stage of the fractal, which contains four tetrahedrons of side length 256 nm, the zero-field remanent magnetization structure, obtained after saturating the sample along the $z$ axis, contains four vortices whose vortex planes lie along the face of the tetrahedron parallel to the direction of saturation. As we move further inwards towards the opposite vertex from the vortex plane, the magnetization gradually progresses to an "A"-shaped structure. In the second stage of the fractal, the side length of the individual tetrahedrons are halved and most of the tetrahedrons individually exist in the vortex-A-type magnetization structure observed in the first stage except three of the tetrahedrons which individually exist in a Λ like structure and collectively form a disconnected vortex, in which a flux-closure structure is obtained by the magnetization of neighboring tetrahedrons curling to form a loop. From the third stage onwards, the size of the individual tetrahedrons are too small for the formation of magnetic vortices in a single tetrahedron and they all exist in a Λ like structure and neighboring tetrahedrons collectively forming disconnected vortices observed in the second stage. In this case, the local cluster of tetrahedrons with a size corresponding to the side length of the second stage acts as the container of these vortex units. In the fourth and fifth stages, the local cluster is further sub-divided into smaller units which individually exists in almost a single domain state and they collectively form the disconnected vortex structure.

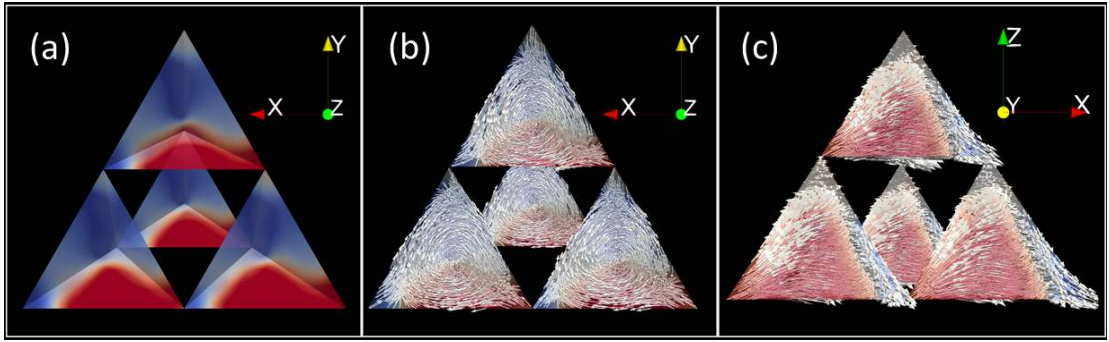

Figure 5.3: Relaxed magnetization structure at zero field of the stage one fractal (a) $M_x$ component (b) Arrows represents the direction of the reduced magnetization $\boldsymbol{m}$ at each discretization point, vortex structures devolved at the faces perpendicular to the direction of saturation can be identified. (c) Perspective view on the "A"-like structure.

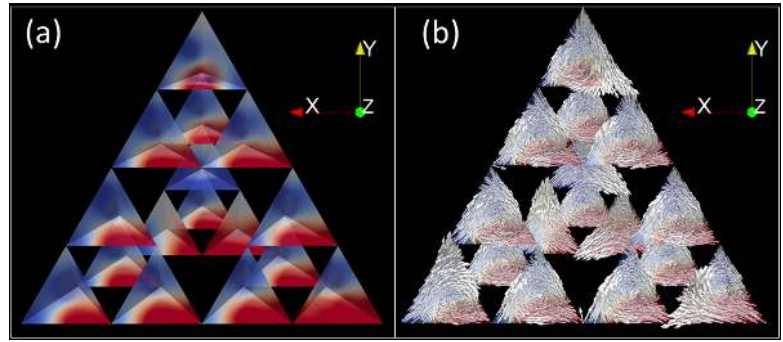

Figure 5.4: Magnetization structure of the second stage (a) $M_x$ component (b) arrow representation. Formation of a disconnected vortex structure can be observed on the central cluster.

## 5.4 Hysteretic properties

To study the quasistatic evolution of the magnetic fractal structures in a magnetic field of varying strength we simulated the hysteresis loops of these systems. An external field of $500\,\mathrm{mT}$ is applied along the $z$ axis, thereby saturating the structure along the $z$ direction. The field is then reduced in small steps till $-500\,\mathrm{mT}$ and the volume-averaged magnetization component of the magnetization vector along the applied field direction is recorded. The compiled results of the hysteretic simulations of all the stages are summarized in Fig. 5.8. One can observe that a very diverse shape of hysteresis loops can be obtained depending on changes in the stage of the fractal. For the first stage, the structure exists in a saturated state at a field of $500\,\mathrm{mT}$, as the external field is lowered we observe a gradual, smooth decrease of the reduced magnetization $\langle M \rangle / M_s$ with the external field. This behavior is caused by the large size of the tetrahedrons in the first stage which permits smooth switching of the magnetization from a saturated to a vortex state. This behavior remains the same for the second stage as well since the individual tetrahedrons in the first and second stage undergoes a similar magnetization reversal. From the third stage onwards we can clearly identify two different regions on the $M$-$H$ curve . In the initial part of the curve, the magnetization remains almost constant up to a particular field and beyond this point we observe an abrupt drop. This deviation is caused by the nucleation of the disconnected vortex structures at the respective fields. The field at which these vortices are nucleated can be called as the nucleation field. Stages four and five exhibit similar behavior. We observe a continuous decrease in the nucleation field as we increase the stage of the fractal. The variations in the remanence and the coercive field with each stage could be attributed to the difference in the resulting equilibrium micromagnetic states formed in the individual tetrahedral units at each stage. In terms of the area defined by the *M-H* loop, we can see a small increase from stage one to two and a steady decrease on the following stages.

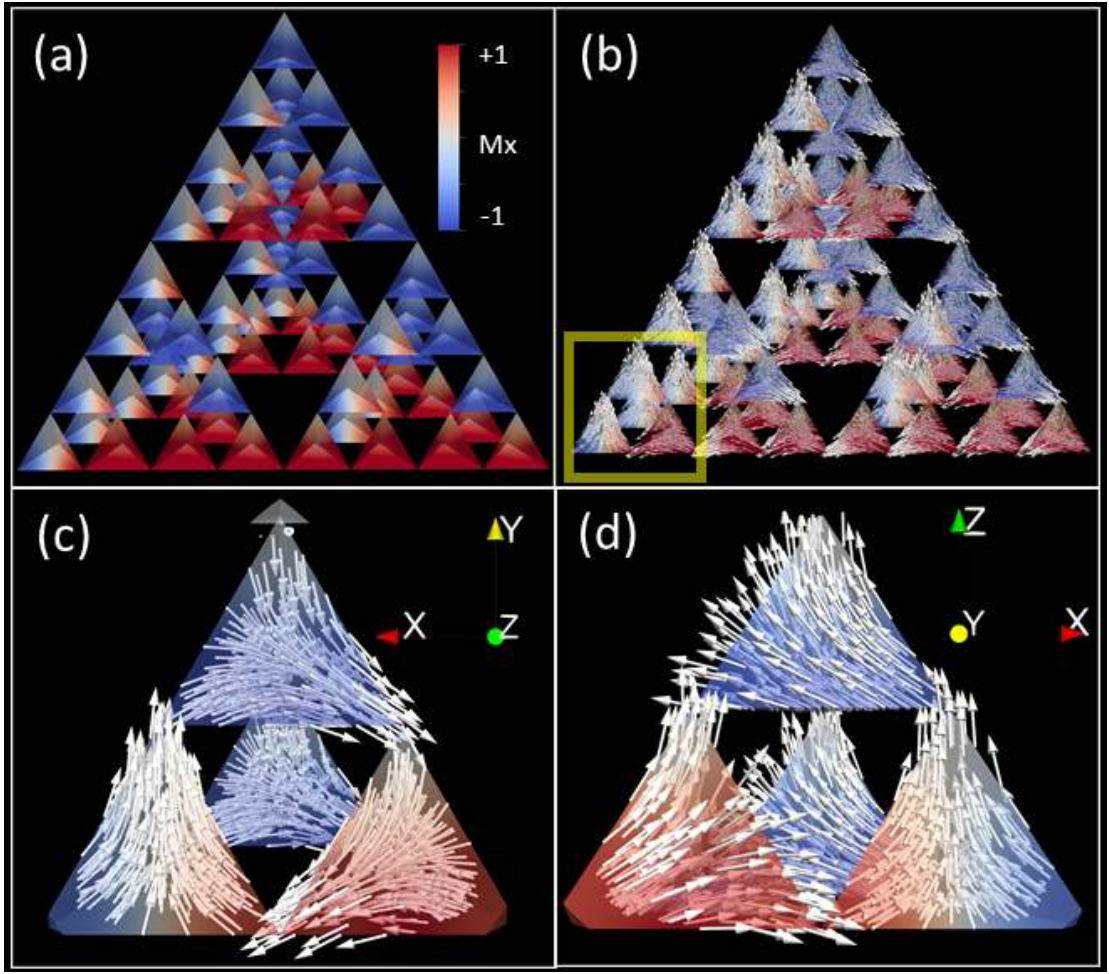

Figure 5.5: Magnetization structure at the third stage: All the individual tetrahedrons exist in an 'Λ' like structure. (c) Zoomed-in view of the cluster forming a disconnected vortex structure. (d) A different perspective showing the "A"'-shaped structure of the cluster.

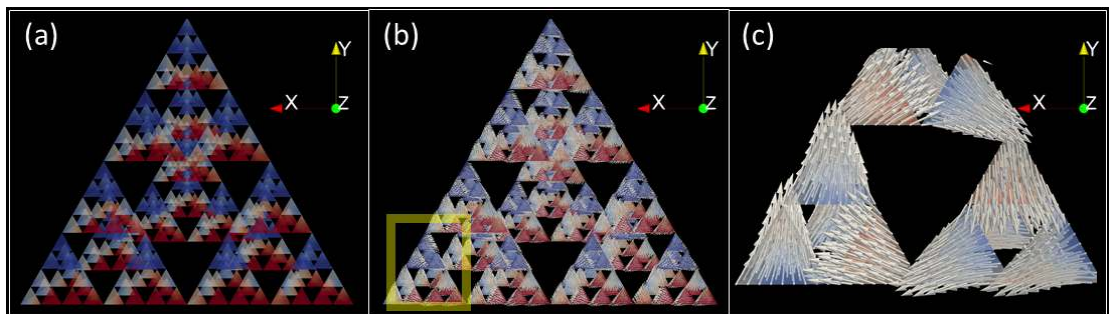

Figure 5.6: Magnetic structure of the fractal at the fourth stage: the local clusters forming the disconnected vortices are sub-divided into smaller units, however the structure retains its shape.

## 5.5 Magnonic excitations

To investigate the high-frequency properties of these fractal structures, we simulated the small-angle precession modes of these structures at various fields by exciting the relaxed magnetization states by appropriate methods and then recording the magnetic ring-down. A comparison of the frequency response stages one to five is given in figure 5.9. In the first stage we can identify two major peaks: one low frequency mode at 700 MHz and a higher mode at 2.4 GHz. These modes are caused by the synchronous gyration of the vortex cores. A time evolution of the variation in the $M_z$ component at a frequency of 700 MHz which is extracted using the inverse windowed Fourier transform is given in figure 5.10.

In the second stage, we can identify two regions of oscillation. Several low frequency modes from 1.1 GHz to 1.9 GHz and a comparatively weaker modes around 4.7 GHz. Similar to the first case, the lower modes are caused by the activity of the vortex structures and the higher frequency modes are caused by the activity of the three tetrahedrons which are in an "A"-shaped state. Out of all the five stages, the third stage has the least effective flux closure at zero field as not all the tetrahedral clusters form the disconnected vortex structures. As a result, we can observe a shift towards higher frequency of the modes in the third stage compared to the other stages. We can also observe that the width of the absorption/emission band is broadened considerably as we move towards the third and fourth stages. In the fourth and fifth stage, one can observe a shift of the modes towards lower frequencies. Due to the formation of flux-closure ring structures in the higher stages there exists a strong dipolar coupling between tetrahedrons of the same cluster. Most of the magnonic activity of the last two stages are concentrated on the corner tetrahedrons of local clusters which are effectively outside of the flux closure loops structures.

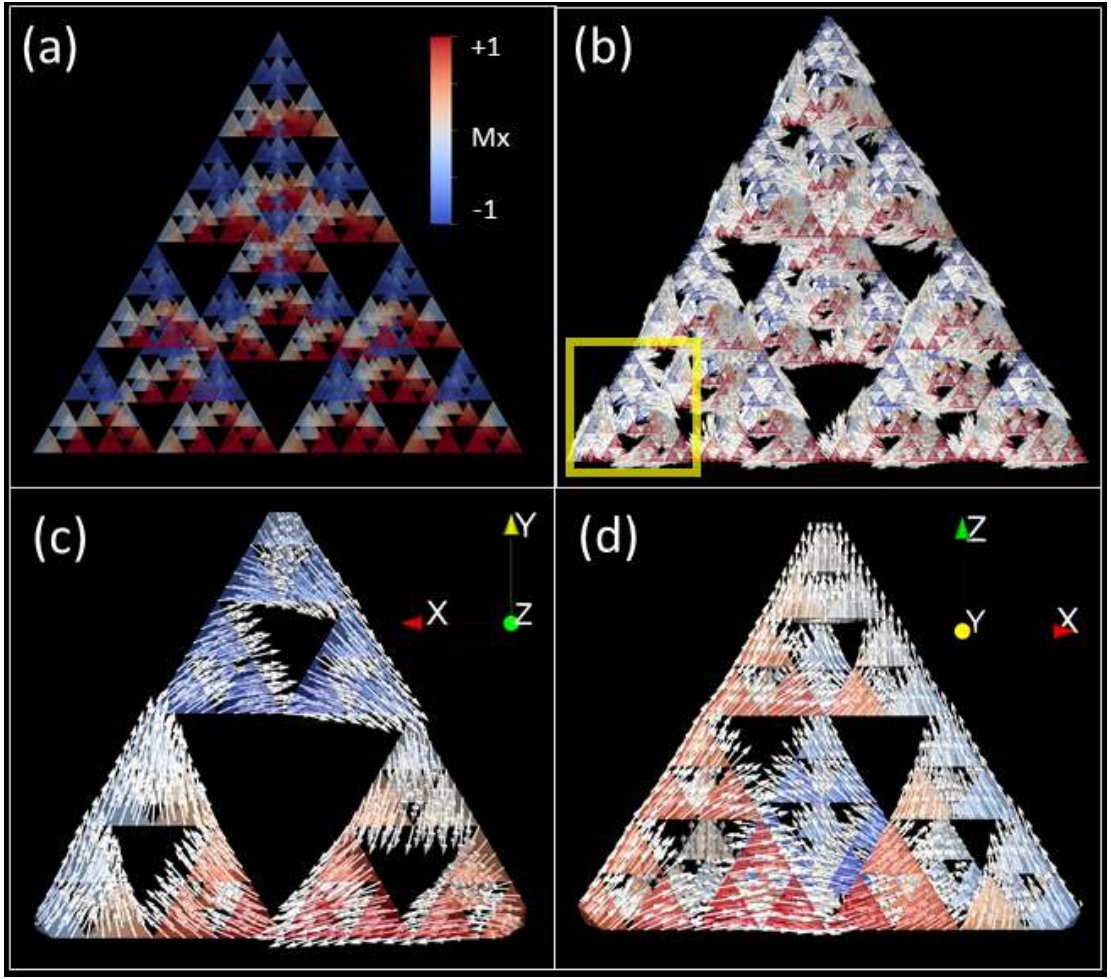

Figure 5.7: Stage 5, the individual tetrahedrons are small enough to form a single domain structure (c) Zoomed-in view showing the local-clusters retaining the vortex structure (d) and the lateral "A"-shaped structure.

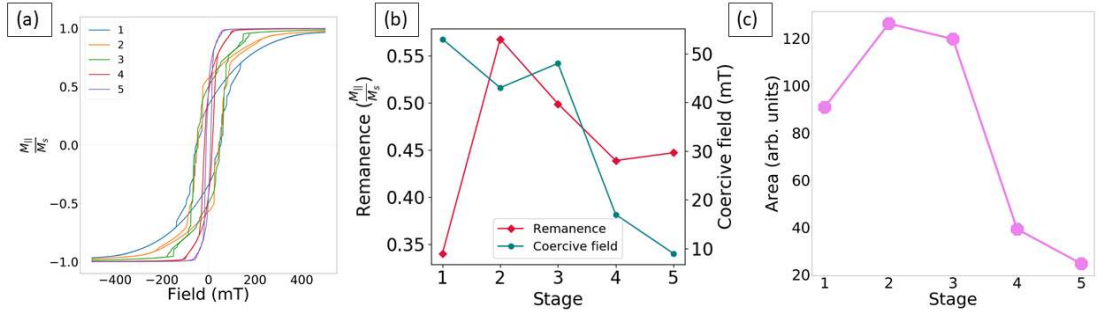

Figure 5.8: (a) Hysteresis curves of stage one to five (b) Variation of coercive field and the remanence with each stage. (c) Variation of the area of the $M$-$H$ loop in arbitrary units.

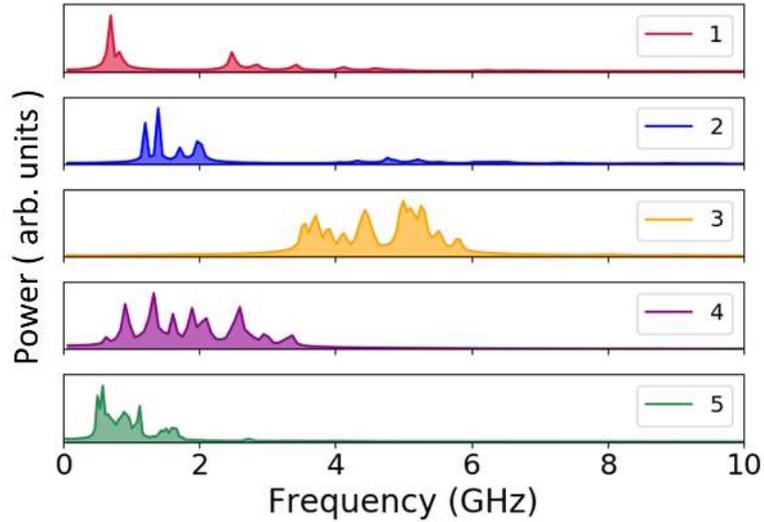

Figure 5.9: Fourier spectra of the different stages at zero field.

To investigate the field dependence of frequency response, an external field was applied along one of the corners of the tetrahedron and the magnonic response investigated. The procedure is repeated for various field strengths from $100\,\mathrm{mT}$ to $900\,\mathrm{mT}$ and the results are compared, see Fig. 5.11. In the presence of a sufficiently strong external field, the tetrahedrons are magnetized along the field direction and they no longer are in a vortex like state we discussed before. This shift in magnetic structure is reflected in the frequency response.

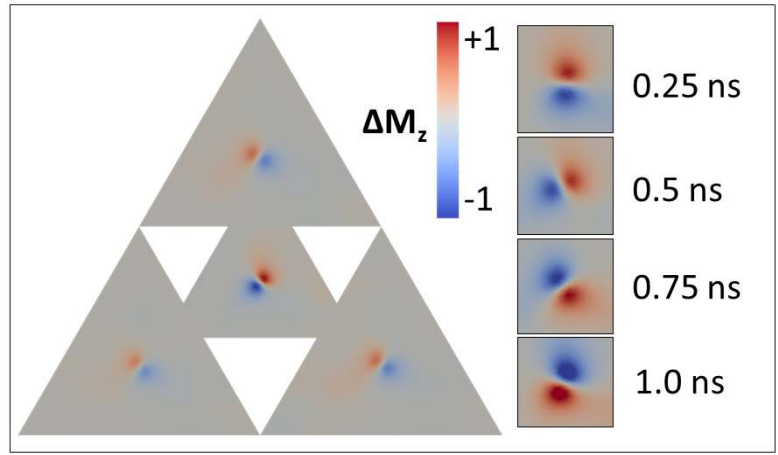

Figure 5.10: Vortex core gyration of the stage one fractal at 700 MHz. The color code represents the change in the $z$ component of magnetization in time.

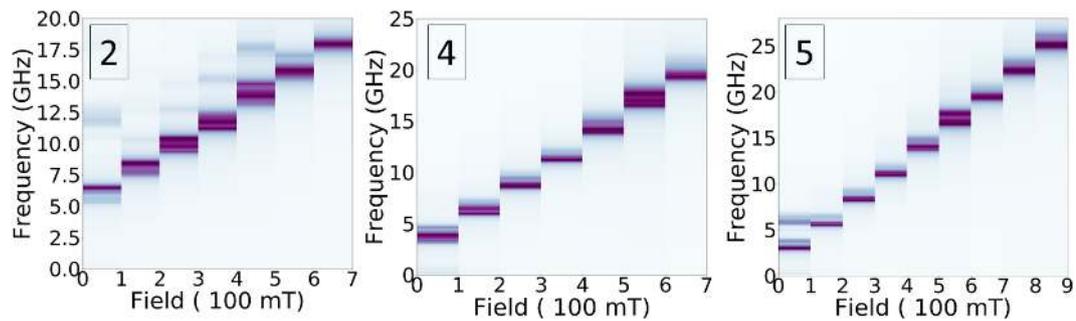

Figure 5.11: Simulated heatmaps showing the field dependence of stage 2,4 and 5.

## 5.6 Chapter summary

A three-dimensional Sierpinski fractal structure is made of four tetrahedrons arranged at the four corners of a larger tetrahedron, higher stages of the fractal are generated by dividing each tetrahedron into four sub-tetrahedrons. We simulated five stages (1-5) of fractal with a side length of 512 nm. The individual tetrahedron units in the first two stages of the fractal were large enough so that their equilibrium magnetization structure at zero field consisted of the formation of a three -dimensional vortex structures. Each higher iteration results in the side length of the tetrahedrons getting halved. From third stage onward, the individual tetrahedrons are too small to individually contain a magnetic vortex and all the units exist in a single domain state. In third stage onward, flux closure is achieved by the formation of a vortex-like structure where the magnetization of the neighboring tetrahedral units circles to form a closed structure. The field response of these structures were studied by simulating the hysteresis loops. The stage of the fractal had a considerable impact on the magnetization reversal and the frequency response of these structures. The magnonic response of the relaxed states were studied by exciting these states by a Gaussian pulse and analyzing the resulting magnetic ring-down for a time period of 10 ns or more. The self-similar geometry and the existence of a wide range of length scales in these fractals had a considerable effect on their magnonic signature. The first stage of the fractal exhibited a typical frequency spectrum that included a pronounced sub-GHz peak corresponding to the gyration of the vortex core. In addition, it contained a few weaker high-frequency modes corresponding to the oscillation of the vertices of the tetrahedrons. From the third stage onward, we could see the emergence of a wide band of intertwined peaks corresponding to the activity of magnetization structures of varying length scales replacing the individual, well defined peaks found at lower stages. On the fourth stage, this wide-band nature of the response was retained and the median frequency of the band was shifted towards a lower frequency region. In the fifth stage, due to the small size of individual units, we can see a departure from the fractal behavior seen in third and fourth stages as the width of the band is considerably decreased.

In short, we could see that the fractal nature of these structures had a considerable impact on their magnetization structures, and most importantly their wide-band magnonic response. Even though experimental techniques at this stage has not evolved to a point where these kinds of three dimensional structures can be fabricated, we believe the results of this short study will be helpful for future investigations into these types of geometries.

## Acknowledgement

The C++ code which was used for the generation of the finite element models were based on a previous code written by Swapneel Amit Patak for the generation of two dimensional Sierpinski triangles. I acknowledge and express my gratitude for Swapneel for this contribution.

# Chapter 6

Artificial Buckyball nano-architectures

## Introduction

As already discussed in the introduction, three-dimensional magnetism has recently emerged as a major topic of interest in magnetism [26, 41], producing novel phenomena which were absent in two-dimension. This includes curvature-induced effects [202, 203] and three-dimensional magnetic structures such as skyrmion tubes [204], Bloch points [104], or Hopfions [205]. Recent advancements in three-dimensional patterning techniques such as focused-ion-beam-induced deposition (FEBID) [25, 144, 24, 64] or two-photon lithography (TPL) [27, 28], in combination with various other deposition methods, has made it possible to fabricate complex three-dimensional nanoarchitectures.

An interesting category of three-dimensional nanostructures are artificial ferromagnetic Buckyball architectures [30, 206]. Their peculiar geometric arrangement leads to competing magnetic interaction at the vertices, resulting in frustration as already known in the case of artificial spin ice structures [120, 117, 118, 119]. The artificial buckyball geometries can be considered as a three-dimensional and interconnected variation of the familiar Kagome lattice [123], which has vertex arrangements quite similar to those of the buckyballs in the sense that in both cases the vertices represent the intersection of three wires meeting at equal angles. In contrast to the buckyball, the hexagonal lattice has a hexagonal arrangement whereas the Buckyball contains both hexagons and pentagons. Because of this similarity with the hexagonal lattice, the buckyball geometry can be considered as a model system which represents a transition from two-dimensional artificial spin ices (2D-ASI), which were extensively studied to a third dimension. Three-dimensional artificial spin ice structures (3D-ASI) hold the potential for many applications such as data-storage, neuromorphic computing [207, 208], spintronic devices *etc.* Therefore, the understanding about the physics of such structures is of great importance. In this section, we present results of extensive micromagnetic simulations, especially regarding the magnetization structure, the frustrated states at vertices, the hysteretic properties, and their high-frequency magnetization dynamics.

## 6.1  Geometry and finite element modelling

The structure is made of 90 cylindrical nanowires interconnected at 60 vertices. At the vertices, spheres are added to form a smooth and interconnected three-dimensional spherical network of hexagons and pentagons. Three cylindrical nanowires meet at each vertex to form a three-dimensional Y-type junction with planar angles 108° and 120°.

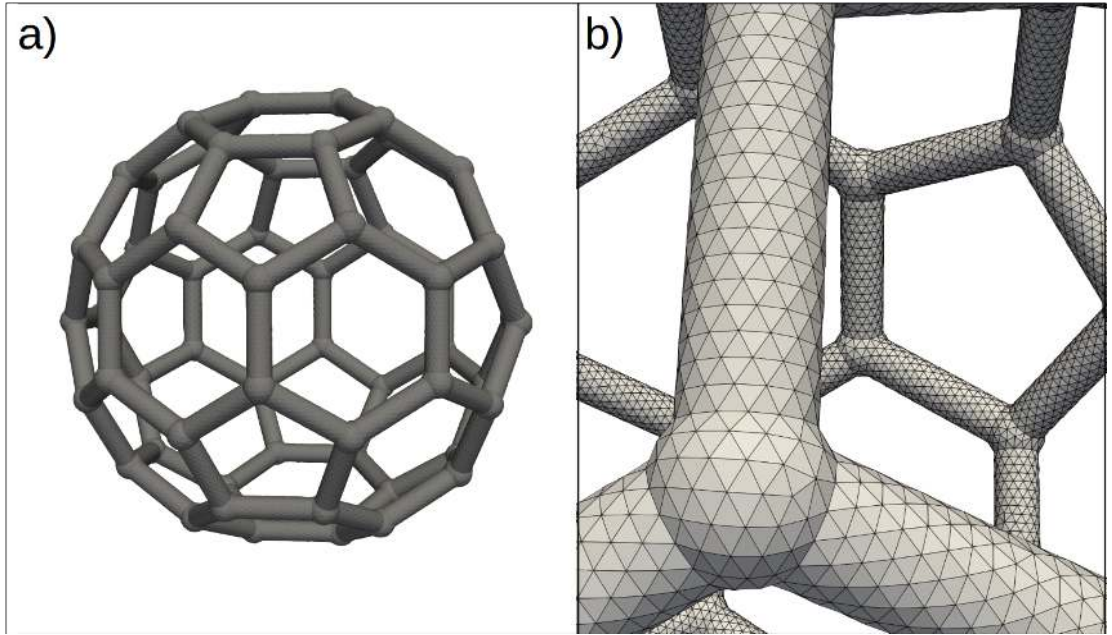

Figure 6.1: (a) a finite element model (FEM) of the Buckyball (b) Zoomed-in view of the mesh demonstrating the accurate modelling of the geometry and its curved surfaces.

An example of the finite element mesh used for the simulations is displayed in Fig. 6.1. The mesh is generated by the python API of the open-source FEM tool Netgen/Ngsolve[153] in MSH-3.0[152] format and is visualized using the open-source visualization tool Paraview/VTK [209]. Owing to the geometric flexibility of the finite element method, the curvature and three-dimensional structure of can be accurately modelled. To investigate the variation of the properties of the structure with change in dimension, the diameter of the buckyball was varied from 130 nm to 1.3 μm by keeping the ratio between the length of the nanowires ($L$) the radius of the nanowires ($R$) and the radius of the spheres at the vertices ($S$) fixed as $L : R : S =$ 25:3:4. Thus, the smallest Buckyball (130 nm) had a nanowire length of $L = 50$ nm, wire thickness 12 nm and sphere diameter 16 nm, while the largest structure had nanowires of length 250 nm and thickness 60 nm, and a sphere size 80 nm. To ensure a smooth geometric approximation of the curved surfaces and to accurately model the magnetization structure, the largest cell size in the finite element mesh ($lc$) was always fixed below $R/2$, and moreover remained below the exchange length of the material used. Thus a maximum cell size of 3 nm was used for the smaller structures and a cell size of 5 nm was used for the larger models. As a result of this, the smallest structure had over 120,000 finite elements and the largest had nearly 2 million elements.

## 6.2 Material parameters

Since structures of this type are are usually patterned by means of FEBID, to get a realistic result we used material parameters corresponding to FEBID-deposited Cobalt (FEBID-Co)[210] with an exchange stiffness constant $A = 1.5 \times 10^{-11} \, \mathrm{J \, m^{-1}}$ and saturation magnetization $\mu_0 M_s = 1.2 \, \mathrm{T}$ where $\mu_0$ is the vacuum permeability, we assumed zero crystalline anisotopy for the material. These material parameters result in an exchange length $l_{\mathrm{exc}} = \sqrt{2A/\mu_0 20 \bar{M}_s^2} = 5.1 \, \mathrm{nm}$. As mentioned in the previous section, the mesh size was always made sure to be smaller than this length. To save simulation time, a value of 0.5 was used for the Gilbert's damping parameter $\alpha$ for the simulation of the static structures, where dynamic effects are not of interest. In contrast, a value of $\alpha = 0.01$ was used for simulations on the high-frequency magnetization dynamics in theses systems.

## 6.3 Simulation method and post-processing

We have investigated principally the static equilibrium magnetization structure at zero field, their dynamic high-frequency modes and their hysteretic behaviour. The equilibrium magnetization structure is obtained by integrating the LLG equation in time, as explained in the previous chapter, until a predefined convergence criterion is reached. This convergence criterion can either be reached when the total energy ceases to change as the simulation proceeds, a total simulation time exceeding a particular limit (we often limit the calculations of the magnetic ring-down simulations to a total simulation time of 10 ns to 20 ns), or a decline of the magnitude of the effective torque in the structure falling below a user-defined threshold value at every discretization point. For our simulations, we took the third approach to simulate static structures, such that the simulation was allowed to run until the maximum value of the effective torque[1] was below $1.0 \times 10^{-4}$. From our experience with `tetmag` we know that this parameter, which is proportional to the magnitude of the maximum local torque $|\boldsymbol{m} \times \boldsymbol{h}_{\text{eff}}|$, is sufficiently low to guarantee that the static structure obtained represents an energetically minimum state.

---

[1] This is a numerical value which is proportional to the local torque experienced by the magnetization under the effective field. The magnetization vector and the effective fields are normalized for numerical convenience and hence the absolute value of this torque does not have any real physical significance and this is only used as an internal numerical control parameter.

In the next step, high-frequency magnetic oscillations are excited with the methods already described in section 4.1: The first option is to apply a suitable low-amplitude Gaussian field pulse in the ps range. Exposing the sample to such a pulse represents a small perturbation that temporarily shifts the static structure of the magnetization out of equilibrium. The magnetization then relaxes back to the converged state within a few nanoseconds while generating a high-frequency small-angle precession dynamics [185]. The second option used in this work is based on a small misalignment of the effective field. In this method, the equilibrium state at a particular field is statically perturbed by applying a small field at a small angle with respect to the original field. Once the equilibrium of the perturbed state is obtained, the additional small field is removed, so that the structure dynamically relaxes back to the converged state in the original field. Like in the case of the Gaussian field pulse, the relaxation dynamics of the magnetic ring-down process towards the equilibrium contains the relevant information on the high-frequency modes [186].

In all cases, the magnitude of the perturbation was carefully chosen so that it does not cause any irreversible effect such as magnetization switching or any other major change in the static magnetic structure. The dynamics of the magnetic ring-down process of the excited state back to the energetic equilibrium is numerically recorded. This data is then processed by using our Fourier analysis tool, such as to extract the profiles corresponding to different oscillation modes. The hysteresis loops are simulated like in the first case: the external field is varied in small steps, making sure that a new equilibrium state is reached for each step.

## 6.4 Static magnetic configurations

The cylindrical nanowires were thin enough, so that at zero field they all tend to be axially magnetized. Although head-to-head and tail-to-tail domain walls are observed on larger dimensions, they usually only form as a transient, non-equilibrium state during hysteresis calculations. A micromagnetically interesting situation arises at the vertices, where three such nanowires meet and the magnetization of each wire has to adapt so as to accommodate the contributions from the neighbouring branches and, also to the change in geometry, see Fig. 6.2. This competing interaction at the vertices gives rise to locally frustrated magnetic configurations. As seen on further discussion, these frustrated states affect the overall properties of the array.

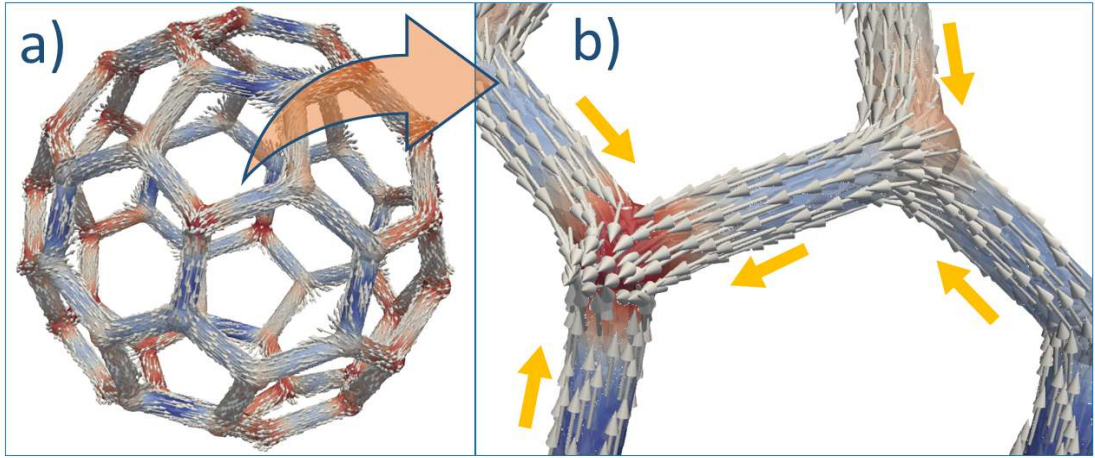

Figure 6.2: (a) Magnetization structure at zero field of a Buckyball of wire length 100 nm and thickness 24 nm (b) Zoomed in view showing the axial magnetization of the wires and the magnetization at the vertices, yellow arrows represent the total magnetization direction of each nanowire

To explain the frustration of the vertices, it is convenient to represent the magnetization direction of each wire with an Ising-like dipole moment [125]. Since odd number of wires are meeting at each vertex, it is impossible to arrange the direction of each dipoles to satisfy all competing interactions. This situation inherently results in vertices with varying degrees of frustration. In general, we can classify the magnetic configurations at the vertices into four types based on the relative orientation of these dipoles (Fig. 6.3) with respect to the vertex.

(a) one-in / two-out state (-1)

(b) two-in / one-out state (+1)

(c) three-in state (+3)

(d) three-out state (-3)

Because of time-inversion symmetry, (a) and (b) are equivalent, and so are (c) and (d). Based on the signed sum of the moments coming in or leaving out, we can assign a charge value for each type. For example, in case (a), since one moment is coming in (+1) and two are leaving out (−2), we can assign it a value of −1 [211]. Similarly, we can assign charge values of +1, +3, and −3 to the (b),(c) and (d) types, respectively. Out of these four types, the first two with −1/+1 charges obey the ice-rule and we call them **single charges**. Accordingly, the ice-rule violating −3/+3 vertices are denoted as **triple charges**.

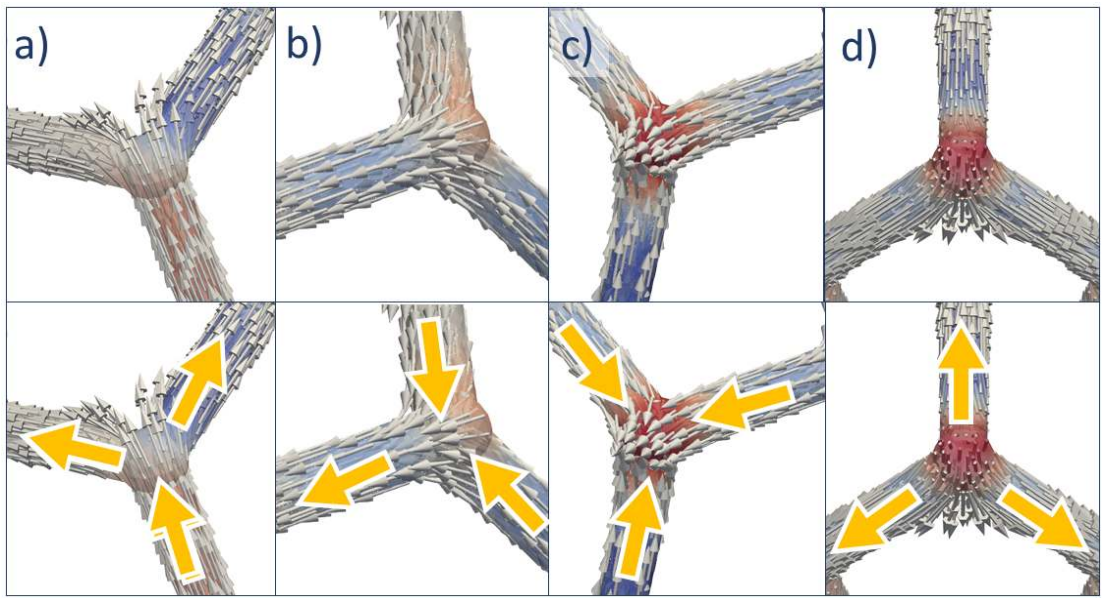

Figure 6.3: Four types of magnetic configurations at the vertices of a buckyball structure: (a) one-in / two-out (b) two-in / one-out (c) three-in (d) three-out.

The four types of vertices differ by the magnitude and the sign of the magnetostatic volume charge they carry. That is, different vertices with the same type of configuration have almost identical magnetostatic volume charge density $\rho$. The magnitude of $\rho$ indicates whether the vertex is of single- or triple-charge type, and the sign indicates whether total flux is carried towards or away from the vertex. Due to the peculiar magnetization structure, the triple charges have a higher density of magnetization volume charges compared to the single charges. Hence, by plotting the magnetostatic volume charge density $\rho = -\boldsymbol{\nabla} \cdot \boldsymbol{M}$ we obtain a more convenient way to display the different types of vertex configurations than by applying the conventional visualization method of plotting the Cartesian components of $\boldsymbol{M}$, or by plotting local values of $\boldsymbol{M}$ with arrows (Fig. 6.4). We can thereby visually identify quite easily the group to which each vertex belongs [1].

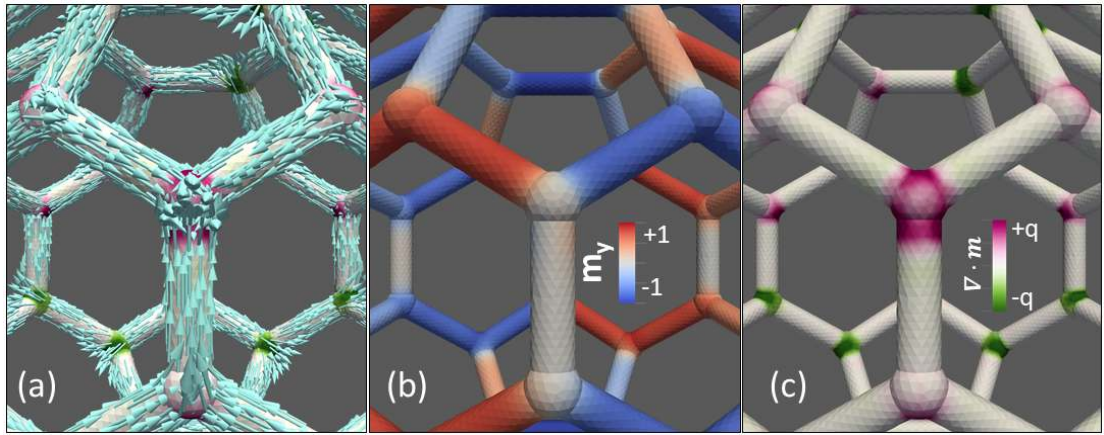

Figure 6.4: Comparison of different visualization methods: (a) The magnetization vector at each discretization point is represented by arrows, (b) the $y$ component of the magnetization is plotted with a color code, (c) The value of $\nabla \cdot \boldsymbol{m}$ at each point is plotted. By plotting the divergence of the magnetization, the different types of vertex configurations can be easily identified by means of the color the magnitude of the charge is given by the intensity of the color. Bright purple and green spots indicate $+3, -3$ charges respectively and mild spots indicate $+1/-1$ charges.

Moreover, if we assign an index number to each of the 60 vertices we can prepare a small code that calculates the average value of $\rho$ at each vertex. By plotting this computed value, we can easily identify and quantify the different kinds of charges are present in the system in an automated way, without visual inspection of the magnetic structure. An example of such a representation is shown in Fig. 6.5. One can clearly recognize the formation of different pleateaus, each corresponding to a specific vertex configuration type.

### 6.4.1 Properties of triple-charges

In the context of artificial spin ice lattices, the triple charges can be considered as defect structures with monopole-like properties [125, 126, 127, 128, 122, 129, 130].

---

[1]We point out that the visualization represents the divergence of the magnetization, while the magnetostatic volume charges are defined with an opposite sign, $\rho_m = -\nabla \boldsymbol{m}$. For the sake of simplicity, we don't always adhere to this sign convention in the discussion of the charges at the vertices. To highlight the properties of the vertex configurations, the presence of regions with divergence of different sign and magnitude is more important than the sign convention.

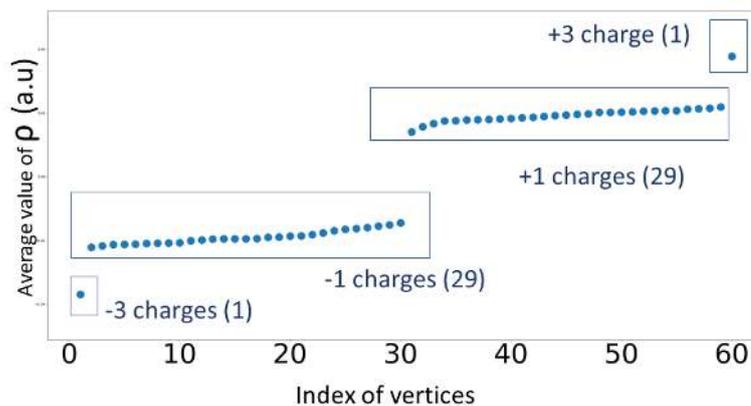

Figure 6.5: Plot of the average value of $\rho$ with vertex indices, we can see that there exists four types of vertices and the number of vertices in each type.

In single-charged vertices, since two of the wires have opposite moment orientation with respect to the vertex, the micromagnetic energy is minimized more efficiently [117, 118, 119, 123]. The energetic considerations concern primarily magnetostatics, because in a single-charge state two out of the three charges are compensated, but also the exchange interaction; an aspect that is not of importance in disconnected ASIs. Here, since the nanowire structure is interconnected, the comparatively low degree of magnetic inhomogeneity developing at the single-charge vertices contributes to their structure being energetically favorable. In contrast, none of these energy minimization criteria are satisfied in a triple-charged vertex configuration, which results in a comparatively higher energy density. Consequently, the triple charged vertices display a stronger local magnetic frustration. As a direct result of these two properties, the triple charged vertices are statistically less favourable. Simulations which were started from a randomized initial state usually yields relaxed zero-field states with two or three triple-charges out of the 60 vertices. A randomized initial state is an artificial configuration in which the magnetization vectors at each descretization point are aligned in a random direction. Starting a simulation from such an initial state is useful to obtain results which are independent of memory effects of the material.

Despite their comparatively higher energy density, the triple-charged vertices are stable at zero field. In all the geometries we studied, it requires a field of at least 30 mT to annihilate a triple charge structure. The micromagnetic structure of the triple charged vertices shows a strong size dependence, in contrast to the single charged states, whose structure remains almost the same in all size ranges. The magnetization structure of triple-charge vertices shows a gradual progression from a triple head-to-head domain wall structure in the smallest buckyballs to a three-dimensional vortex configuration in the largest structure, see Fig. 6.6.

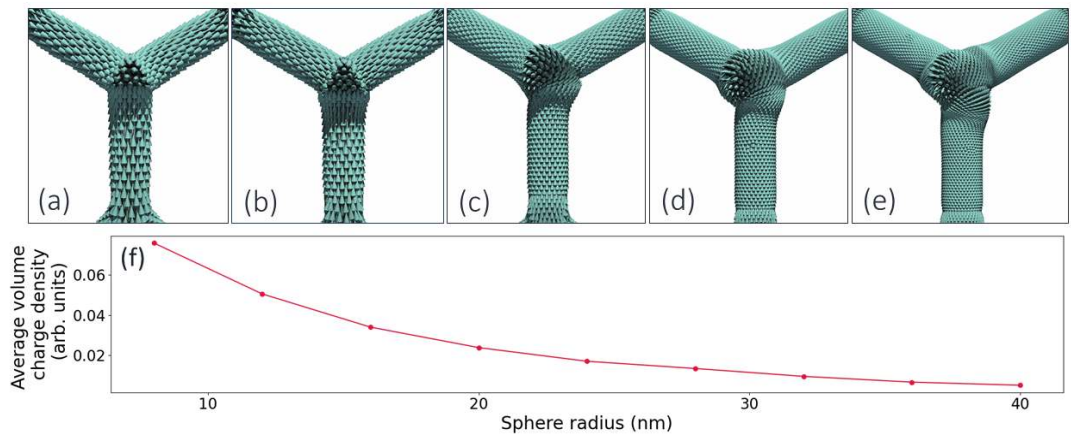

Figure 6.6: Progression of the triple-charge with size of the sphere (a) 16 nm (b) 32 nm (c) 48 nm (d) 64 nm (e) 80 nm (f) Comparison of the evolution of the average value of the magnetostatic volume charge density at the spheres for the triple charge vertices with an increase in sphere radius

### 6.4.2 Controlled generation and removal of triple-charges

The existence of defect charges is one among many driving forces for the research in ASI. More often than not, the formation and various other behaviours of these defect charges are stochastic in nature, and hence difficult to control experimentally. In the case of buckyballs, the different three-dimensional orientations of the nanowires provides an easy method for the control of the formation and dissolution of these triple charges, namely by applying external magnetic fields of suitable strength at an appropriate direction. An analogous behaviour for a controlled generation and removal of defects through a homogeneous external field is usually not possible in 2D-ASI systems. This possibility to manipulate magnetic defects is therefore an example of how the three-dimensional nature of the nanostructure leads to a qualitatively different and potentially significant behavior that is not found in the two-dimensional counterparts.

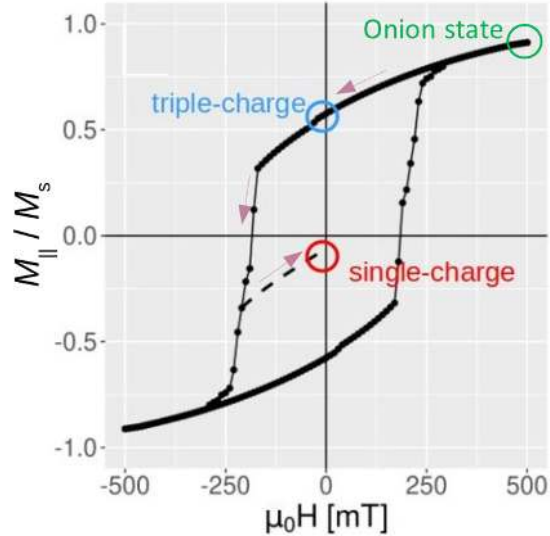

Figure 6.7: Simulated hysteresis loop, including a minor loop, of a buckyball with $L : R : S = 100\,\text{nm} : 12\,\text{nm} : 16\,\text{nm}$.

The insertion and removal of triple-charge vertices can be demonstrated by simulating hysteresis curves (Fig. 6.7) of these structures. The field is swept from $500\,\text{mT}$ to $-500\,\text{mT}$ in $10\,\text{mT}$ steps, and the evolution of the magnetic structure is analyzed. The field is applied along an axis which connects two diametrically opposite vertices. The choice of these pair of vertices is arbitrary. Due to the spherical symmetry of the buckyball, any pair of vertices on opposite sides behaves identically.

At $500\,\text{mT}$, the Buckyball is in an onion-like (Fig. 6.8) state, where the two vertices along the axis at which the field is applied form the tip and the root of the onion. This state is characterized by the nanowires being magnetized at an angle with respect to the axial direction as the Zeeman energy contribution overpowers the shape anisotropy of the wires. When the applied field is reduced, the magnetization in the individual nanowires gradual aligns towards the local axis of the nanowires in a reversible way, thereby adapting to the geometry provided by the nanowire network. As the field is further reduced to zero, the magnetization structure progresses towards an ASI behaviour into a state characterized by an axial magnetization of the individual nanowires. But the remanent structure is also influenced by the initial onion-like state, which it partially preserves. Along with the alignment of the wires towards their axis, the tip and the root of the onion structure evolve into a pair of +3 and −3 triple charges. This pair is gradually formed upon reduction of the external field remain stable even when a field in the opposite direction is applied. The zero-field remanent state hence contains a pair of triple charges at these two vertices. For convenience, we call this the **triple-charge state**.

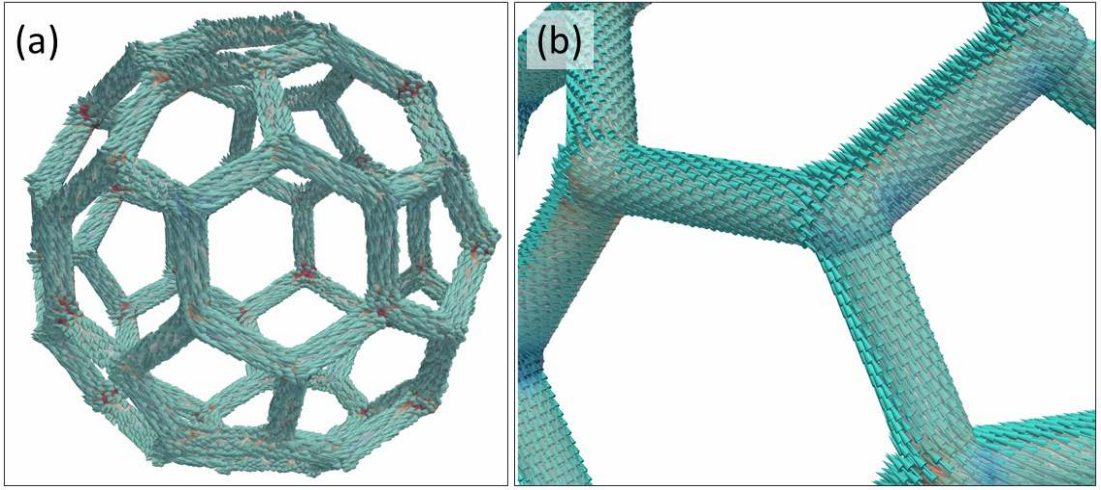

Figure 6.8: (a) Onion state at 500 mT (b) Zoomed-in view showing a canted magnetization in the wires.

As the hysteresis loop is continued by applying a gradually increasing field in the opposite direction, the triple charges remains stable for until a particular value. On further increase of the field until the coercive field, the triple charges are dissolved by the formation of a domain wall emitted from the triple charge vertex, which results in the flipping of the magnetization in one of the bars. This flipping is carried forward through the connected vertices setting up a chain-reaction type switching leading to the reorientation of the magnetization in a number of neighboring nanowires and a corresponding rearrangement of the charge distribution in the associated vertices. As the applied (negative) field strength is increased, the field-induced dissolution of the triple charge vertices is followed by the reversal of individual nanowires at various switching fields, depending on the orientation of the wires with respect to the applied field. These switching events are irreversible magnetization processes, and they can be observed as Barkhausen-type like steps in the hysteresis loop. The reversal of different nanowires occur at different field strengths, depending on the orientation of the wires with respect to the field direction (Fig. 6.9). The resulting configuration that forms as a result of this chain of events is free of any triple charges. It contains only ice-rule obeying single-charge vertices (+1, −1). The dissolution of triple charges by this process is also an irreversible process. Even if the field is reduced back to zero from the point of annihilation, the dissolved triple charges are not regenerated. Hence, by going to a zero-field state through a suitable minor loop, we can obtain a stable zero-field configuration which does not contain any triple charges. This state can be denoted as the **single charge state**. By simply applying a suitable sequence of external fields, we can thus obtain two qualitatively different states at zero field: Either the triple charge state which contains a pair of +3, −3 defect charges, or the single-charged state which is free of such defect structures.

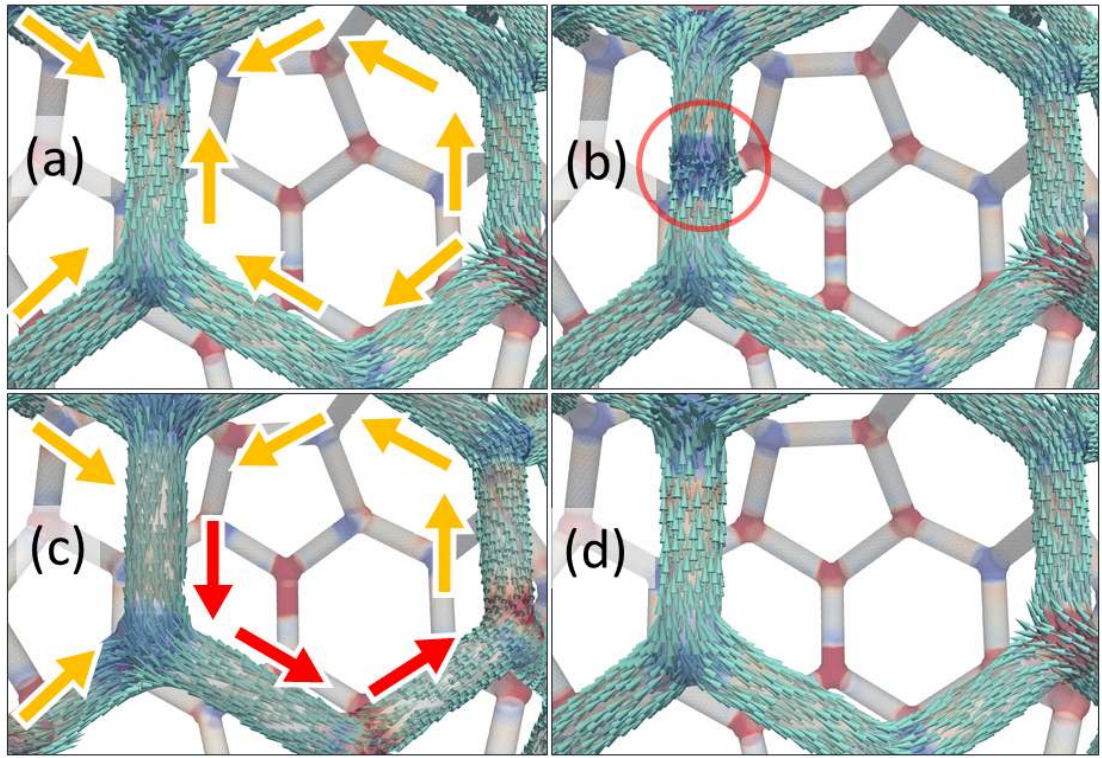

Figure 6.9: Different steps in the annihilation of the triple charges. Initially, the triple charge vertex at 0 mT is locate on the upper left (a). The yellow arrows represent the magnetization directions of each wire. (b) Transformation to a single-charge vertex *via* the emission of a domain wall at −50 mT. (c) Final state free of triple-charges, with rearranged charge distribution. Branches that are flipped are highlighted with red arrows. (d) Single charged state at 0 mT obtained through a minor loop.

The hysteresis curves were simulated for a total of ten geometries with wire lengths ranging from $L = 25\,\text{nm}$ to $250\,\text{nm}$. The results are analyzed and compared (see Fig. 6.10). The step-wise nature of the hysteresis curve, which is observed in all sizes, is related to the mechanism involving the reversal of individual nanowires. With an increase in size of the buckyballs, the steps tend to become smoother. It is also observed that the value of both coercivity and remanence decrease with increasing object size. This is a general tendency in micromagnetic, which is related to the reduced impact of exchange effects in larger samples. In this specific case, the size-dependent changes in the hysteresis can partly be related to the evolution of the triple-charge vertices to a three-dimensional vortex-like state, which results in a more efficient flux-closure compared to the triple-headed domain wall structure developing in smaller geometries.

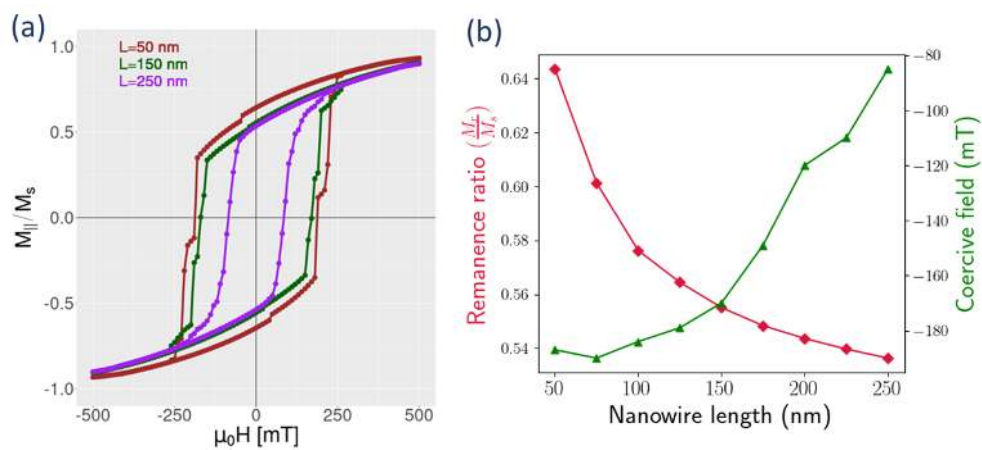

Figure 6.10: (a) Comparison of hysteresis curves of Buckyballs of different sizes. The plots show the reduced magnetization component $M_\parallel/M_\text{s}$ along the field direction. (b) With increasing size, both the remanence and the coercive field strength tend to decrease.

### 6.4.3 Magnonic spectrum of a buckyball: Switchable frequencies

The presence of triple charges has a strong impact on the magnonic spectrum of a magnetic buckyball structure. To demonstrate this, we simulate the small-angle precession modes of the equilibrium zero-field states of the buckyballs by the methods explained in section 6.3. The spectrum of the high-frequency response of a buckyball to a small perturbation at zero field in Fig. 6.11. It shows the case of a buckyball in single-charged state with $L = 100\,\text{nm}$. We can clearly identify five modes, with two lower frequency sharper peaks and three higher frequency broader peaks. These five modes can be classified into two types, based on their geometric origin. The first two peaks modes are oscillations localized at the vertices, in which the intensity of the activity is concentrated at the spheres connecting the nanowires. The remaining higher-frequency modes correspond to the oscillation of standing waves within the nanowires. The results of the discrete Fourier analysis provides information about the spatial profiles of each mode, which are plotted in Figs. 6.12 and 6.13. In these images, the average value of the modulus of the dynamic component of magnetization $\delta\boldsymbol{m}$ at each point over the entire time frame is plotted. The yellow regions indicate areas of high magnonic activity, while the purple regions are the inactive parts, or the nodes.

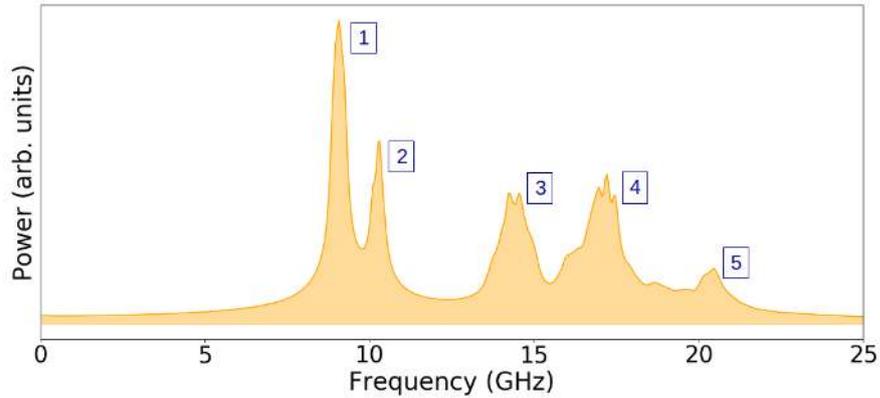

Figure 6.11: Power spectrum of the Fourier-transformed magnetic oscillations in single-charged state at zero field of a buckyball with $L = 100\,\text{nm}$.

The modes 1 and 2 at $9.07\,\text{GHz}$ and $10.31\,\text{GHz}$, respectively, are caused by the activity of the single charged vertices. Almost half of the vertices oscillate at the first frequency and the rest oscillates at the second frequency as shown in figure 6.12. The modes 3, 4 and 5 at $14\,\text{GHz}$, $17\,\text{GHz}$ and $20\,\text{GHz}$, respectively, correspond to the first, second and third order oscillations of the nanowires 6.13.

In the next step, we repeat the same procedure for the triple-charged state and compare the results. A comparison of the frequency response of the triple-charge state with the single-charge state is given in Fig. 6.14. We can see that all the modes we discussed earlier remain essentially the same. However, now the spectrum also contains a pronounced low-frequency mode at $1.4\,\text{GHz}$, which was previously absent. The spatial profile of this mode is displayed in Fig. 6.15. This low-frequency mode is caused by the activity of the $+3/-3$ triple-charge pair. This is an important observation with potential applications: Since the vertices oscillate at considerably lower frequency after changing in the magnetic vertex configuration, a controlled change of the magnetization state opens the possibility to manipulate the sample's magnonic spectrum, and to switch between two distinctly different high-frequency responses. As we have already seen, the triple-charge configurations can be generated and disarrayed in a controlled manner, thereby opening up a new route for reconfigurable magnonic devices.

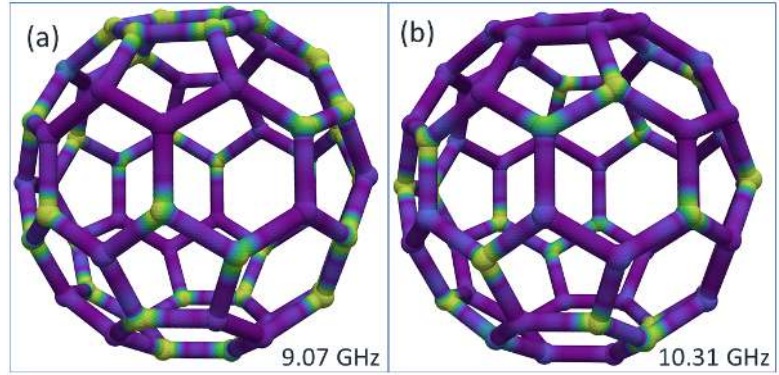

Figure 6.12: Oscillation of the vertices (a) Mode 1 (b) Mode 2 in single-charge state. (light green areas indicate higher magnonic activity).

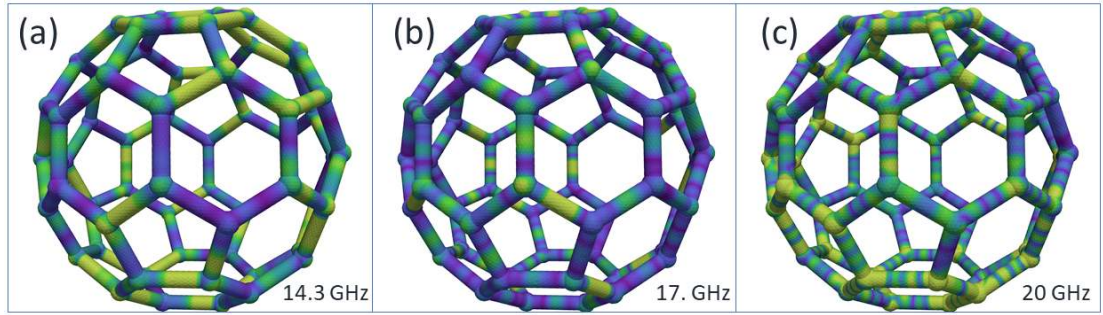

Figure 6.13: Oscillation of the nanowires (a) Mode 3 (b) Mode 4 and (c) Mode 5 in Single-charge states

In addition, the frequency of these triple charges can be controlled by varying the size of the buckyball and by means of an external field. The variation of the frequency of the vertices with changing size of the buckyball is shown in Fig. 6.16(a). Since the frequency of oscillation of the single charges are split into two groups, the frequency of the lowest mode in all sizes is used for the comparison. Both the single charge mode and the triple charge frequencies decreased linearly with an increase in the nanowire length: the frequency of oscillation of the single charged vertex decreased from 11.18 GHz in the smallest sample with nanowire length 50 nm to 1.8 GHz in the largest sample with a nanowire length 250 nm, while the frequency of oscillation of the triple charged vertices decreased from 1.75 GHz in the smallest Buckyball to 0.39 GHz in the largest. This decrease in frequency can attributed to the exchange field contribution which makes the magnetization structure stiffer at smaller length scales.

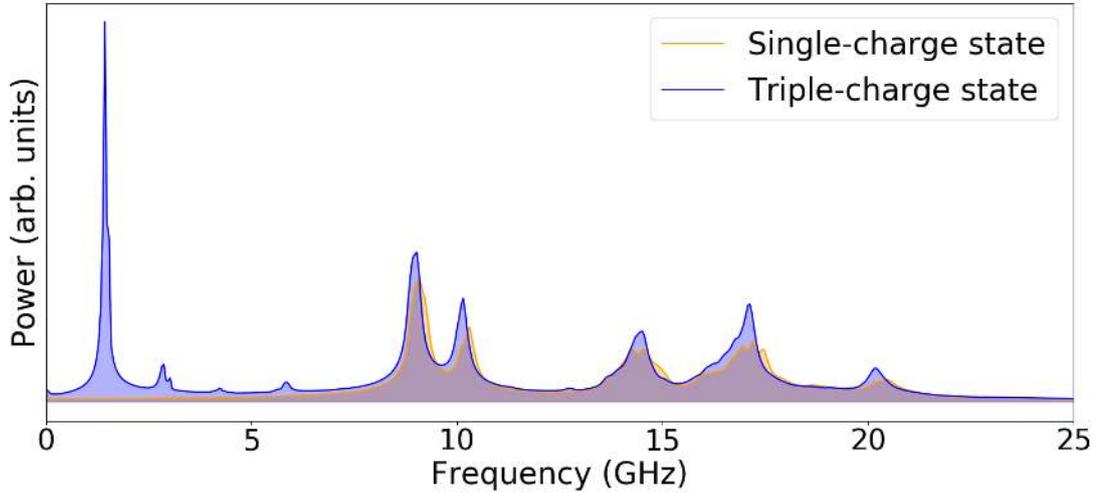

Figure 6.14: Comparison of the frequency response of the triple-charge and single-charge state.

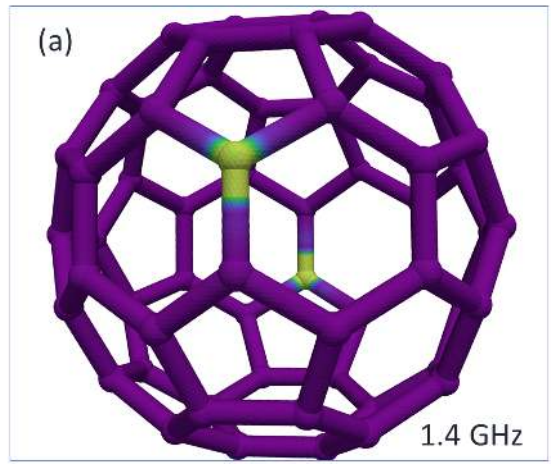

Figure 6.15: Spatial profile of the additional low-frequency mode present in the triple-charge state, corresponding to the magnonic activity of the triple-charge vertices.

In all the cases we simulated - except for the smallest ($L = 25\,\text{nm}$ ), the frequency of oscillation of the single charged vertices always splits into two groups 6.16(b). This splitting into two lower and higher frequency modes is not based on the charge of the vertices or nearest neighbor interactions. We currently don't have a conclusive explanation for this splitting of the ensemble of nominally identical single-charge vertex oscillators into two populations of markedly different frequencies. The selection of the vertices belonging to one mode or the other does not appear to be based on details of the magnetic configuration. It might be the result of complex effects arising from a weak coupling between the oscillators, resulting in the partial synchronization of clusters of vertices. Future studies will be aimed at analyzing this effect in more detail.

The response of the frequency of oscillation of the triple-charged vertices of the buckyball with a side length $L$ of $100\,\text{nm}$ to an external field is shown in Fig. 6.17. The field is applied along the diagonal connecting the two triple charged vertices in the positive direction. We can observe a sharp increase in the frequency of the mode with an increase in the field strength; from $1.4\,\text{GHz}$ at zero field to a frequency of $2.1\,\text{GHz}$ at $70\,\text{mT}$. Beyond this, the oscillation frequency remains constant despite an increase in the field strength. When the field was applied in the opposite direction, the frequency of oscillation was found to be decreasing.

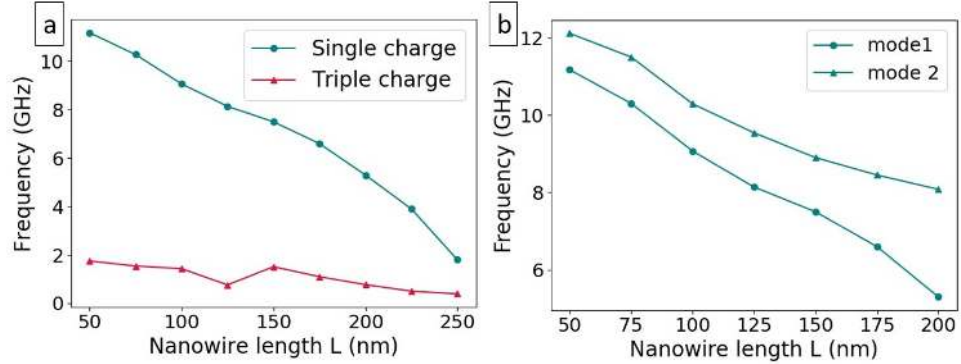

Figure 6.16: (a) Variation of the frequency of the oscillation of the single and triple charge vertices with size of the buckyballs. (b) The oscillation frequency of the single-charged vertices splits into two modes.

## 6.5 Hollow nanoarchitectures

Apart from direct nano-patterning of ferromagnetic materials, three-dimensional structures similar to the buckyball structures discussed in theis chapter can also be fabricated by coating a non-magnetic structure with a ferromagnetic material. Buckyballs fabric-ated through this approach will be effectively hollow inside, at least from a magnetic perspective. They will be made of nanotubes interconnected by spherical shells. The magnetic properties of nanotubes has been extensively investigated in the recent past [212, 213, 54, 53, 214, 215, 216, 217, 218, 219], and it is already known that hollow tubes exhibit distinct magnetic properties that are different from those of solid nanowires. The hollow structure of nanotubes results in magnetic configurations which can close the magnetic flux while avoiding magnetic singularities [212, 215]. It was argued that this may result in faster magnetization switching under external fields and thus offers better possibilities to manipulate magnetic configurations compared to solid cylinders. The presence of an additional *internal* curved surface in these hollow structures is also a point of interest. Hence, it is natural to extend the studies on buckyballs to hollow models, and to compare the results with those obtained for the solid structures.

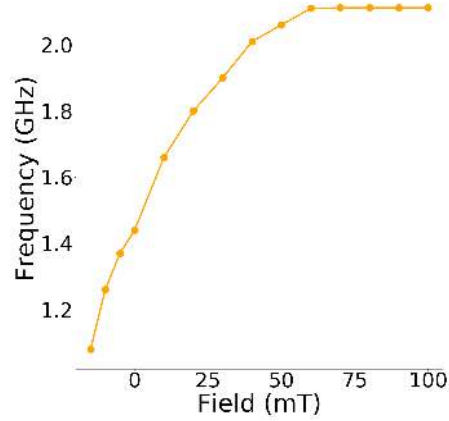

Figure 6.17: Variation of the frequency of the triple charges in the presence of an external field applied along the axis connecting the two triple-charge vertices.

The flexibility of the finite element method permits the accurate modeling of magnetization in such complex geometries. We have simulated a range of hollow buckyballs from side length 100 nm to 200 nm with fixing the geometrical parameters in the following ratio, length of nanotube ($L$), external radius of nanotube ($R_e$), external radius of the spherical shell ($S_e$) internal radius of the nanotube ($R_i$), internal radius of the sphere ($S_i$) 50:6:8:4:6. The mesh length was chosen appropriately such as to ensure an accurate description of the geometry of the thin tubes.

### 6.5.1 Comparison of nanowires and nanotubes

To understand the basic magnetization structure and magnonic response of these two types of buckyballs, we first simulated a miniature versions of their constituent units and compared the results. This involves

1. cylindrical solid nanowire of length $L = 100$ nm, radius $R = 12$ nm

2. cylindrical open tube of length $L = 100$ nm, $R_i = 8$ nm, $R_e = 12$ nm

3. hollow nanotube with closed ends $L = 100$ nm, $R_i = 8$ nm, $R_e = 12$ nm

4. solid nanotube with spheres attached on both ends $L = 100$ nm, $R = 12$ nm, $S = 16$ nm

5. hollow nanotube with hollow spherical shells attached on both sides, $L = 100$ nm, $R_i = 8$ nm, $R_e = 12$ nm

The magnetization structures of these units are shown in Figs. 6.18 and 6.19. We can see that they have almost identical, axial magnetization configurations. There are only minor differences in the ends depending on the presence of spheres or shells.

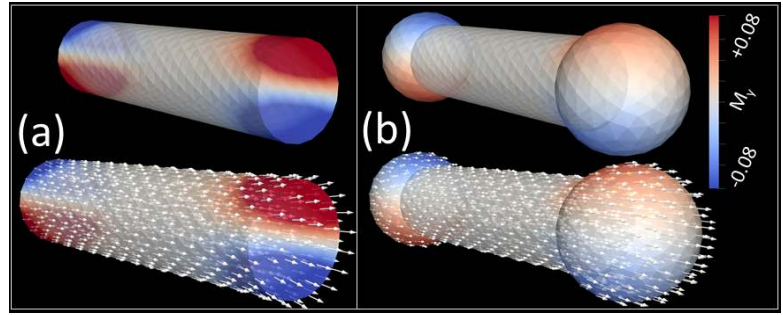

Figure 6.18: Relaxed magnetization structure at zero field of a solid cylindrical nanowire of length $L = 100$ nm, radius $R = 12$ nm (a) and of a solid nanowire with two spheres attached at the end $S = 16$ nm(b).

The relaxed magnetic configurations are excited with a Gaussian field pulse, and the frequency response is analyzed. In all these structures, the primary mode of oscillation is a coherent oscillation of the magnetization component at the two ends. But one can also see a considerable difference in the frequency of oscillation of this mode with a minor change in the geometrical parameters. Specifically, we observe a decrease in the frequency if we add a spherical structure at the end. Comparing the frequency of oscillation of all these structure, we can also see that the hollow tubes with closed ends have a very low frequency of only 610 MHz. Magnetic oscillations at such low frequencies are usually typical for the frequency of gyration of magnetic vortices. But in this case, there are no vortices present in this structure. Here, the low-frequency mode is caused by the slow rotation of the in-plane component of magnetization in the two nanodiscs which form the caps attached at the two ends of the nanotube. In comparison with the frequency of oscillation of other similar structures, this mode in the nanotube with closed end is more intense and sustained for a longer time period. These simple observations indicate that the magnonic properties of hollow structures can be significantly different from those of solid ones, and that they can depend sensitively on details of the geometry of the constituent units.

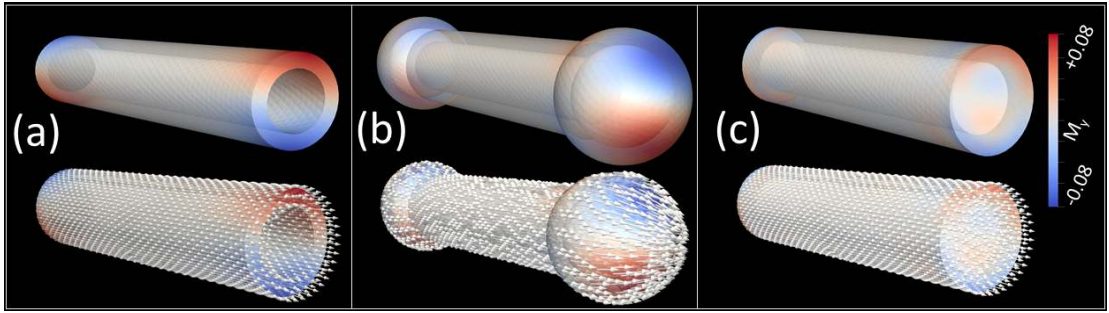

Figure 6.19: (a) Hollow open nanotube ($L = 100\,\mathrm{nm}$, $R_i = 8\,\mathrm{nm}$, $R_e = 12\,\mathrm{nm}$), (b) hollow nanotube with two spherical shells attached at the ends ($L = 100\,\mathrm{nm}$, $R_i = 8\,\mathrm{nm}$, $R_e = 12\,\mathrm{nm}$), and (c) hollow nanotube with closed ends capped by disks ($L = 100\,\mathrm{nm}$, $R_i = 8\,\mathrm{nm}$, $R_e = 12\,\mathrm{nm}$).

| Type of structure | Frequency of primary mode |
|---|---|
| Solid cylindrical nanowire | 6.26 GHz |
| Solid nanowire with two spheres at the ends | 3.74 GHz |
| Hollow nanotube | 9.73 GHz |
| Hollow nanotube with two shells at the ends | 1.85 GHz |
| Cylindrical nanotube with closed ends | **0.610 GHz** |

Figure 6.20: Comparison of the frequency of he primary modes of various hollow and sloid structures with different geometries.

### 6.5.2 Hollow Buckyballs: magnetization structure

In the length ranges concerned, similar to the case of solid nanowires, the nanotubes tend to adopt an axially magnetized state at zero fields, as would be expected from the shape anisotropy effect. As a result, despite differences in the properties of the constituent nanowires and nanotubes, the solid and hollow Buckyballs exhibit similar magnetic structures at the vertices. Based on the direction of magnetization of each nanotube meeting at a vertex we can apply the same classification of vertex configurations as in the case of the solid buckyball be distinguishing between single and triple charges. A detailed visualization of the relaxed magnetic configuration of a hollow buckyball structure with side $L = 100$ nm is displayed in Fig. 6.22. The simulations show that the triple charge vertices have the same triple head-to-head domain wall structure, as seen in the solid model. As the size of the structure is increased, the triple-charge vertex structures evolves towards a three-dimensional vortex-like configuration, in essentially the same way as previously discussed in the case of the solid buckyball.

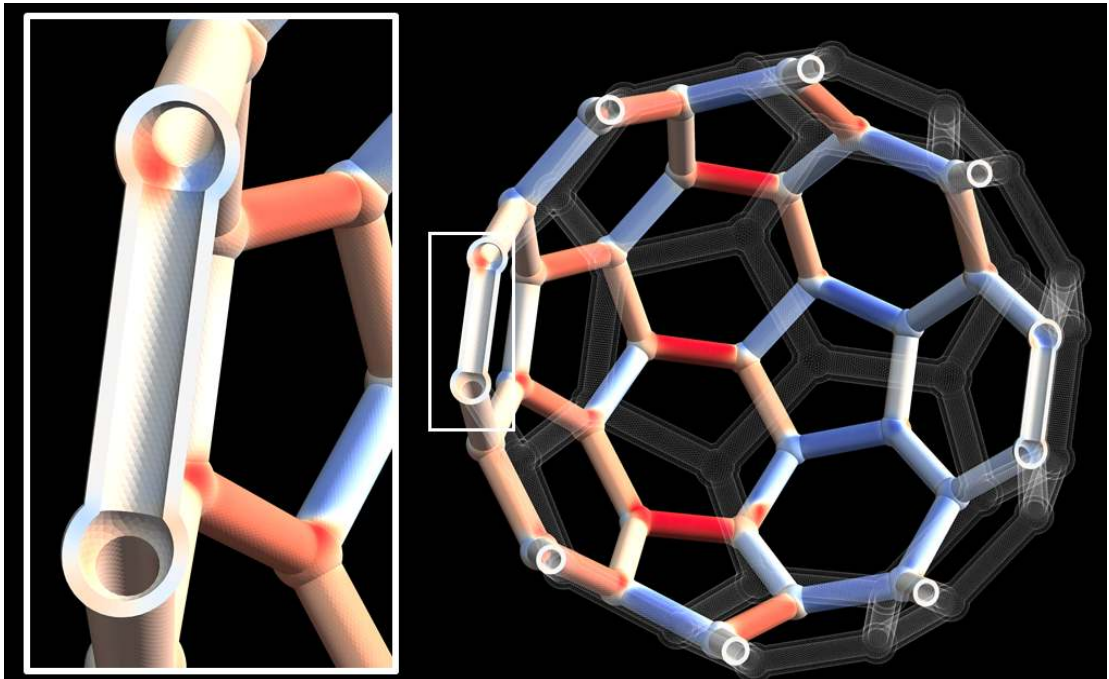

Figure 6.21: Finite-element model of a hollow buckyball with side length $L = 250$ nm. A cross-section of the whole structure is displayed to visualize the hollow tubes. The zoomed-in view on the left highlights one of the hollow nanotubes and shows the nanoshells at the vertices.

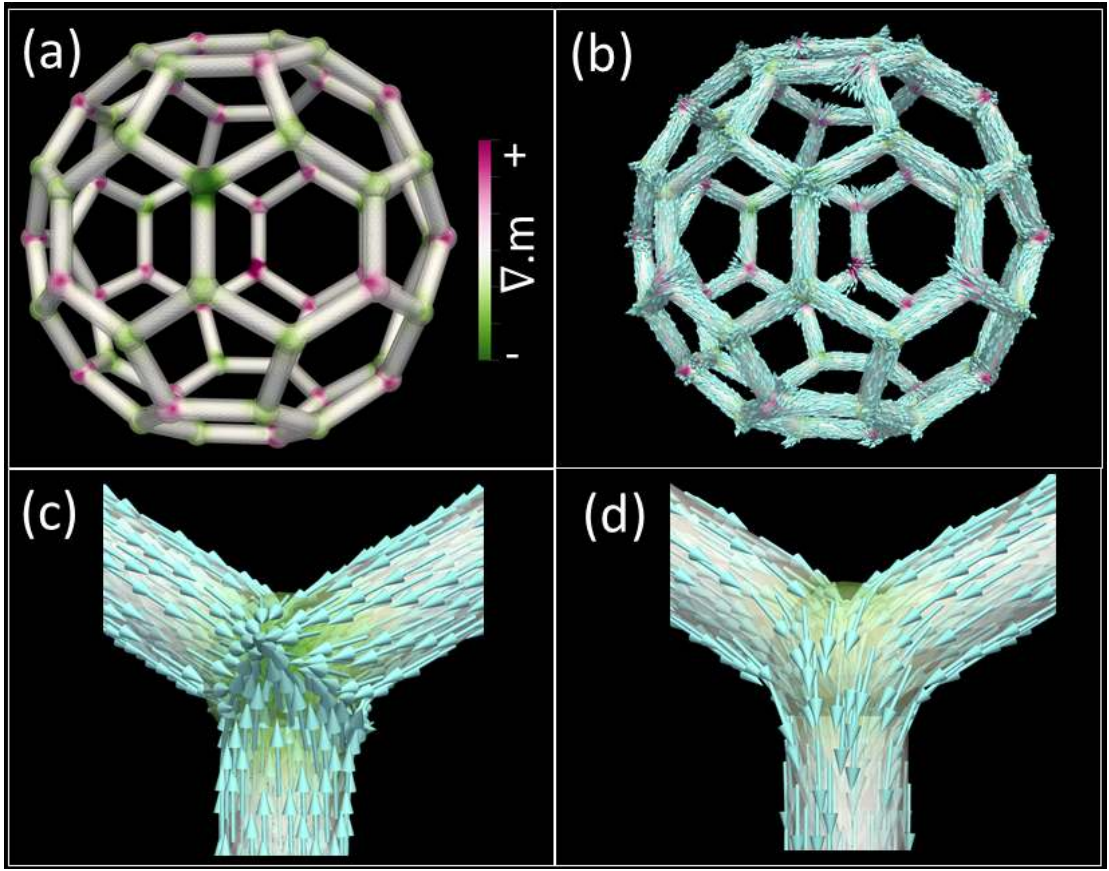

Figure 6.22: Relaxed magnetization structure of a hollow buckyball with side length, $L = 100\,\text{nm}$ (a) the charges at vertices are visualized by plotting the volume charge density $\nabla \cdot \boldsymbol{m}$ (b) The local magnetization is plotted by means of arrows (c) Zoomed-in view of one of the triple charged vertices (d) Zoomed-in view of a single charge vertex structure.

### 6.5.3 Hysteretic properties

The hollow structures exhibit a hysteretic behavior that is quite similar to that of the solid buckyballs. When the structure is saturated by applying a strong magnetic field of $\mu_0 H_{\text{ext}} = 500\,\text{mT}$ along an axis connecting two vertices locate on diametrically opposite positions, and if the field is then gradually reduced to zero, we obtain again a state which contains a pair of triple charges on the two vertices along the axis through which the field was applied. Applying a gradually increasing field in the opposite direction will result in the annihilation of the triple charges by the emission of a domain wall, and a subsequent relaxation back to zero field through a minor loop will again result in a single-charge state which is free of triple charges. The simulated hysteresis curve

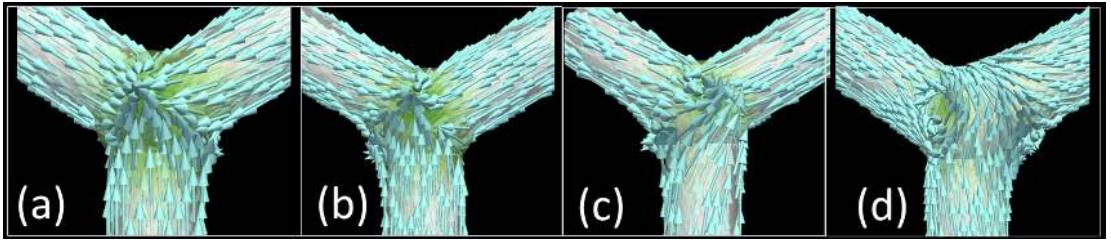

Figure 6.23: Evolution of the triple charge stucture, from a triple-head domain wall structure to a vortex-like state. Stable triple charges at different geometries (a) $L = 100\,\mathrm{nm}$, (b) $L = 125\,\mathrm{nm}$ (c) $L = 150\,\mathrm{nm}$, (d) $L = 175\,\mathrm{nm}$.

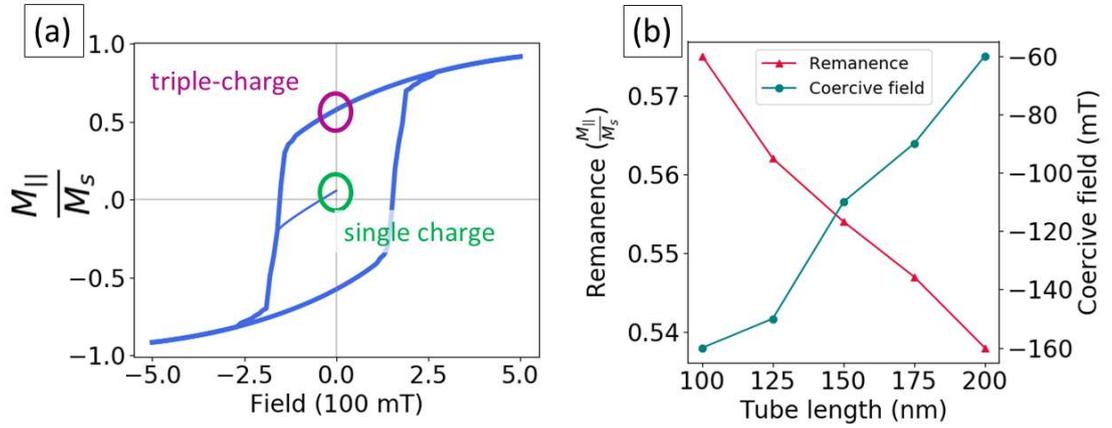

Figure 6.24: (a) Simulated hysteresis curve of the hollow Buckyball with a tube length of $L = 100\,\mathrm{nm}$ (b) Comparison of the variation in remanence ratio and the coercive field strength with size

of the structure with a tube length of $100\,\mathrm{nm}$ and a comparison of the remanence ratio and coercive field strength of the samples with wire length is given in Fig. 6.24. We could observe that both the remanence ratio and the coercive field tends to decrease with size.

### 6.5.4 High-frequency modes

A comparison of the high-frequency response of the magnetization to a short excitation in the case of the triple-charged state of the solid and hollow buckyball is shown in Fig. 6.25. We can identify three major modes which are marked as $\boxed{1}$, $\boxed{2}$ and $\boxed{3}$. As in the case of the solid buckyballs, the first mode at $2.78\,\mathrm{GHz}$ is caused by the activity of the triple charged vertices, the second and third mode at $7.42\,\mathrm{GHz}$ and $9.10\,\mathrm{GHz}$, respectively, are caused by oscillations of the single charged vertices. In comparison to the solid Buckyball, we can see that the characteristic frequency of the triple charges is shifted towards a higher value, while those of the single-charge vertex oscillations are shifted towards lower frequencies. The most striking difference in the magnonic spectrum of the hollow buckyball, compared with the solid case, is the absence of wire modes. The nanotubes connecting the vertices are not magnonically active. This is manifested as the absence of the three higher frequency modes, which we had seen in the previous case.

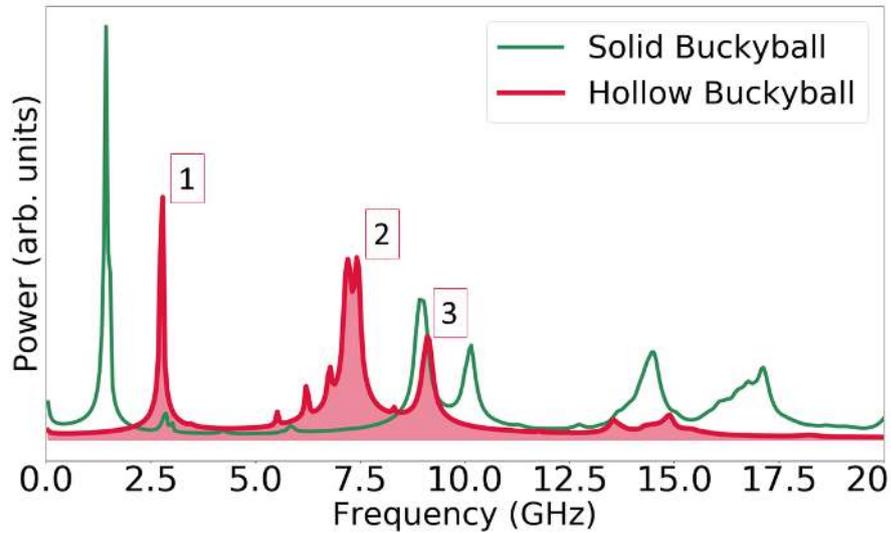

Figure 6.25: Comparison of the frequency response of the triple-charges states of the solid and hollow Buckyballs with side length $L = 100\,\text{nm}$.

## 6.6 Chapter summary

In this chapter, we have investigated the fundamental properties of artificial ferromagnetic buckyball nano-architectures by means of finite-element micromagnetic simulations. The geometric flexibility of the finite element approach allowed as to accurately model the cylindrical nanowires and the spherical vertex regions of these structures. We simulated buckyballs of a wide range of size – from a side length of $25\,\text{nm}$ to $250\,\text{nm}$. In all the sizes we studied, the individual nanowires displayed a single-domain state at remanence, magnetized along the longitudinal direction. Based on the axial direction of the magnetization in each nanowire we could assign an Ising-like dipole moment to the nanowire. According to the relative orientation of these Ising moments, we could identify different magnetic configurations in the vertices with varying degree of magnetic frustration, and we distinguished between the ice-rule obeying single charges and the defect-type triple charges. Due to the three-dimensional spherical structure of the buckyballs, these triple changes can be generated and destroyed in a controlled manner by means of an external field. We studied the magnonic response of these structures and we could identify the vertex modes and the nanowire modes arsing from the corresponding geometric locations. We saw that the presence of triple charged vertices produced a signature with low-frequency peaks in the magnonic spectrum of these buckyballs. The lower frequency of oscillation of the triple charge vertices when compared to the single charged vertices is due to the absence of a strong pinning field in the triple charges due to the symmetrical arrangement of the magnetisation of the nanowires. The frequency of these peaks could be controlled by varying the geometry, and also by applying an external magnetic field. We compared these findings with the results of a hollow buckyball and observed that they exhibit similar behavior to that of the corresponding solid geometries in the concerned length scales.

These structures can be considered as a model system to describe a transition from two to three dimensional artificial spin ice systems. The spherical symmetry of these structure allows the to insert and remove triple-charge defects by means of external magnetic fields. This opens a pathway for magnonic applications as, this control of the magnetic vertex configuration allows to manipulate the high-frequency spectrum of the nanostructures through the occurrence of a pronounced, sharp peak in the magnonic absorption spectrum in the case of triple-charge defects. We hope that the findings of these simulation studies will inspire experimental groups to investigate these properties and to obtain an experimental verification.

# CHAPTER 7

Artificial magnonic crystals

In the previous chapter, we analyzed the micromagnetic properties of the static and dynamic magnetization in buckyball-type geometries. Although these structures are clearly three-dimensional, the nanowire network generating the buckyballs is, in itself, two-dimensional as it covers the surface of a sphere. While this combination of two- and three-dimensional features renders the buckyball geometry interesting to study effects emerging from the transition from two to three dimensions, other structures display a more genuine three-dimensional character. This is the case for periodic nanowire arrays with a periodic geometric arrangement of the vertex positions. The ability to fabricate three-dimensional structures of this type has recently been demonstrated [64, 220, 65]. This is one example showing the important progress in nanofabrication and nanopatterning techniques of three-dimensional patterning techniques, such as FEBID [26, 144], which has made it possible to fabricate magnetic nanostructures of virtually any three-dimensional geometry. Using these techniques, it is possible to fabricate arrays consisting of periodic, three dimensional networks of interconnected nanowires, with vertex positions corresponding to those of atomic positions in a crystal lattice.

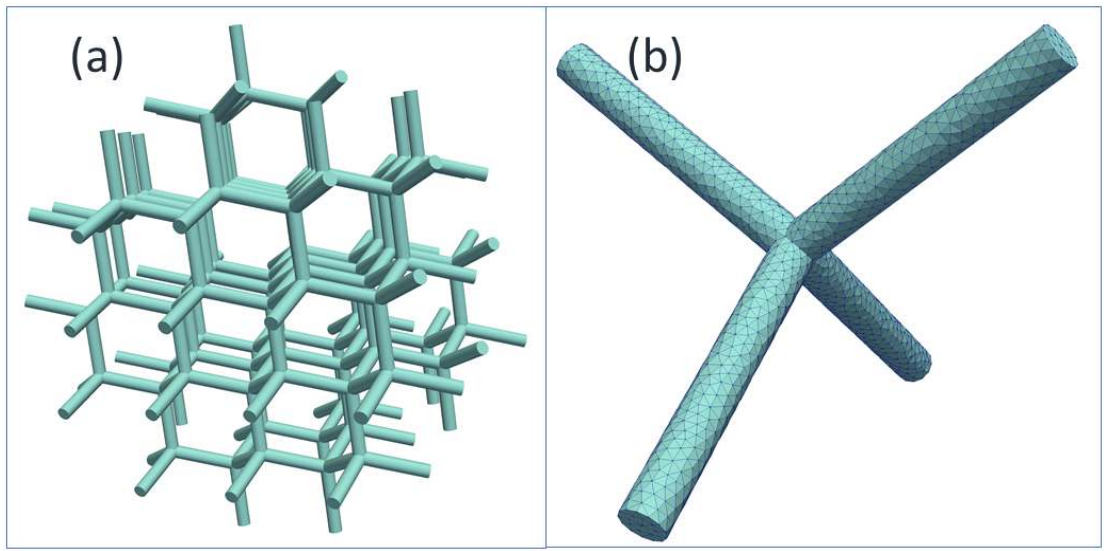

Figure 7.1: (a) Finite element model of a diamond-type structure. (b) Enlarged view of an individual tetrapod unit representing the building block of the artificial diamond structure.

Three-dimensional interconnected nanostructures of this type can be interpreted as an extension from two-dimensional artificial spin-ice systems to three-dimensional ones. On the other hand, their particular geometry is reminiscent of artificial structures which, in the case of optical or acoustic applications, are used for the design of metama-terials [221]. One can therefore also see these three-dimensional nano-architectures as prototypes for a new form of magnetic metamaterials.

Metamaterials are artificially prepared composite materials which exhibit properties that cannot be observed in the bulk state [43, 42, 221], such as negative refractive index [222, 223], unusual acoustic properties [224], exceptionally large optical susceptibilities, non-reciprocal behavior, sign reversal of thermal expansion constant *etc.* The emergence of such exotic properties in metamaterials is a result of their microscopic artificial geometric structure, rather than a consequence of the intrinsic properties of the chemical constituents [42]. This particular property of the metamaterials opens up the potential to manipulate their exotic behaviors by fine-tuning their geometric parameters [225]. An interesting sub-category of such metamaterials are artificial crystals which are formed by a periodic arrangement of building blocks mimicking the behavior of natural crystals [226, 227]. If such structures are fabricated with magnetic materials, they may result in a new type of magnonic materials. Artificial magnonic crystals [228, 229, 230] (AMC) are largely an unexplored sub-category of artificial magnetic nanostructures. In particular, little is known about the magnonic properties of three-dimensional interconnected structures in the context of their potential application as magnetic metamaterials. Analogous to how the propagation of photons and phonons is manipulated in photonic and acoustic crystals, AMC have the potential to be tuned in such a way to control magnons - the fundamental excitation unit in magnetism.

In contrast to photonic an acoustic metamaterials, the ferroic character of AMC's introduces an additional degree of freedom, which can be exploited to generate a varying degree of disorder within the crystal. As discussed already on the example of the buckyball structures, it is possible to change the magnetic configuration at the vertex points, where two or more nanowires meet. Such manipulations of the magnetic structure could be used to modify the magnonic spectrum of the array, since the vertices can act as sources of magnetic frustration due to competing interactions [231, 232, 66], as already known from traditional artificial spin ice (ASI) systems [117]. Because these network of interconnected nanowires can be interpreted as a three dimensional version of the familiar artificial spin ice lattices, their multiple magnetization states could yield different and switchable magnonic properties. Conventionally, most of the research in artificial spin ice structures was concentrated on two-dimensional disconnected grids of patterned single domain elements [117, 119]. However, recent studies have extended the investigation into *interconnected* two-dimensional networks [233, 234, 235] and three-dimensional disconnected structures [41, 220, 236]. In this chapter, using finite-element micromagnetic simulations, we investigate the frustrated states, field dependence and high-frequency properties of two types of artificial magnonic crystals: the diamond lattice [65] structure and the cubic lattice structure.

## Artificial diamond lattice structure

An interesting sub-category of three-dimensional artificial magnonic crystals are diamond-type nano-architectures, which are formed by a network of interconnected nanowires. The nodes where these wires meet in space are located as per the lattice points of a natural Diamond crystal. Four nanowires meet at each node in a tetragonal arrangement and the lattice can be interpreted to be made of individual tetrapod units. When compared to the Buckyball structures discussed in the previous chapter which can be considered as a planar structure wound to form a spherical structure, the Diamond lattice is a full-fledged three-dimensional network.

## 7.1 Methodology

As in the case of the study on the buckyballs, we prepared finite element meshes of these geometries by using Netgen and chose material parameters corresponding to those of FEBID-deposited Cobalt. The diamond-type lattice structure that we simulated had an overall size of $450 \times 450 \times 450$ nm. The array was composed of nanowires with a length of $70.0$ nm and a radius of $7.0$ nm. The entire network contained 202 nanowires interconnected at 83 vertex node points. The finite element model which we used had a maximum cell size of $4.0$ units and contained 704,476 tetrahedral elements.

We performed micromagnetic simulations to investigate the static magnetization structure, different kinds of vertex configurations, their hysteretic behaviour and also their magnetic high-frequency properties.

## 7.2   Static magnetization structure and vertex configurations

Similar to the case of buckyballs, in the geometries which we simulated, the wires are thin enough so that they are axially magnetized at equilibrium state. We can thus interpret the magnetization direction of each wire as that of an Ising-type macroscopic dipole [237] consisting of two equal and opposite magnetic charges ($\pm q$). In such a macroscopic dipole, the $+q$ charge represents the end at the head of the magnetization and $-q$ the tail. In the diamond-type geometry that we studied, each vertex is the intersection point of four nanowires. Accordingly, we can identify different types of vertex configurations based on the relative orientation of the four dipoles meeting at each vertex, and assign a charge value to each configuration [211, 238]. The concept of assuming a charge for each vertex configuration can be justified as follows. Since each nanowire in the relaxed state is magnetized axially, depending on the orientation of magnetization with respect to the vertex, each nanowire either carries magnetic flux towards the vertex or away from it. Since there are four nanowires meeting at each vertex, a balanced state is established when the number of wires with incoming flux and outgoing flux are equal (two/two) and thus cancel out, resulting in a net zero charge. Any imbalance in the number of wires with in- and outflowing flux results in a non-zero net magnetic charge at the corresponding vertex, as in the case of emergent magnetic monopoles in ASIs [239, 240, 241]. Thus, in the diamond lattice or in the case of any other such structure with a coordination number of four, there can exist five different vertex configurations as listed below.

such as,

1. Two-in / two-out (0)

2. Three-in / one-out ($+2q$)

3. One-in / three-out ($-2q$)

4. Four out ($+4q$)

5. Four in ($-4q$)

Out of these five configurations, the first one with a net charge of zero is the ice-rule obeying structure, while the rest can be classified as different types of magnetic defect

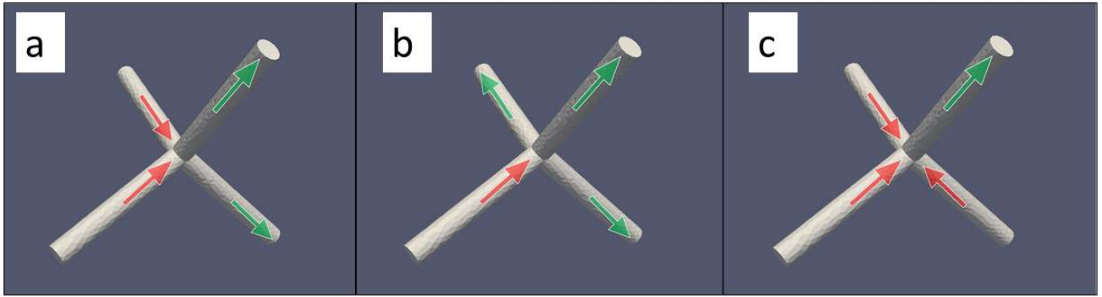

Figure 7.2: (a) Two-in/two-out state (b) One-in/three-out state (c) Three-in/one-out state

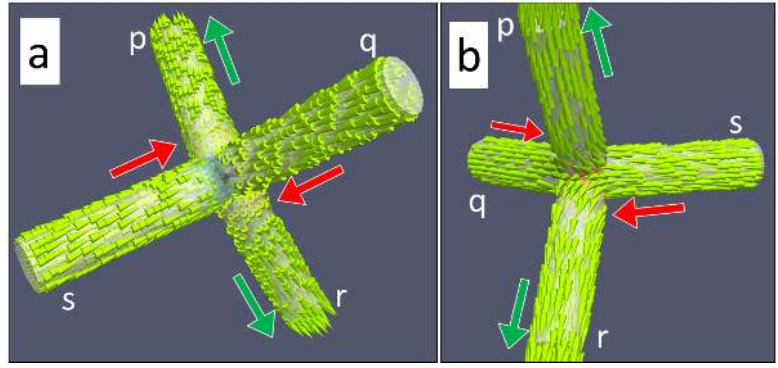

Figure 7.3: (a) The micromagnetic structure of the ice-rule obeying two in - two out state (b) A different orientation of the same structure (the wires are marked for easier identification )

.

charges. As our simulations show, the theoretically proposed configurations number 4 and 5 are energetically unstable and do not develop in the parameter range that we studied. In none of our simulations could we observe these states as stable configurations. However, the states 2 and 3 with a net charge of $+2q$ and $-2q$ respectively are obtained and are found to be stable. For convenience, these states will be referred to as the "double-charge" states henceforth in this thesis. In addition to the five stats listed above, another type of charges can be distinguished if one also considers the magnetization state of the naowires at the surface of the artificial crystal. The dangling free ends of the surface nanowires represent a situation that is different from the vertex structures within the volume. We can assign a charge value of $+1$ or $-1$ to these ends, based on the direction of magnetization at the ends pointing inwards or outwards

Owing to the symmetric, three-dimensional arrangement of the nanowires at each vertex, there is only one ice-rule obeying state. This is in contrast to the two different possible ice-rule obeying state known from two dimensional square ASI, evan though planar square ASIs also have vertices with a coordination number of four [242]. In the case of a 2D square ASI, not all the configuration of three nanowires meeting at a vertex are equivalent with respect to a fourth wire. We can differentiate between the magnetic structure in wire on the opposite side and that in the two orthogonal wires. Because of this difference, there exist two flavors of two-in/two-out state in a 2D square ASI depending on the relative orientation of the incoming/outgoing wires, as indicated in Fig. 7.6. In the diamond lattice, however, all three neighboring wires meeting at a vertex are equivalent with respect to the fourth wire. The angle between any of the wires in the tetragonal arrangement is the same, and therefore the distinction between wires on opposite or neighboring directions does not apply. Consequently only a single flavour of two-in/two-out state exists in this case. As a result of this equivalence of the relative orientation of any pair of nanowires at the vertices, each charge value for the tetrapod-type vertex has only one possible magnetic configuration associated with it.

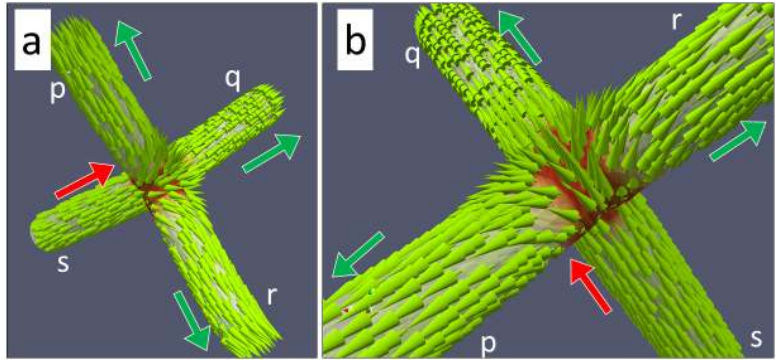

Figure 7.4: (a) The micromagnetic structure of the one-in/three-out state (b) Zoomed-in view highlighting the tail-tail domain wall structure at the vertex

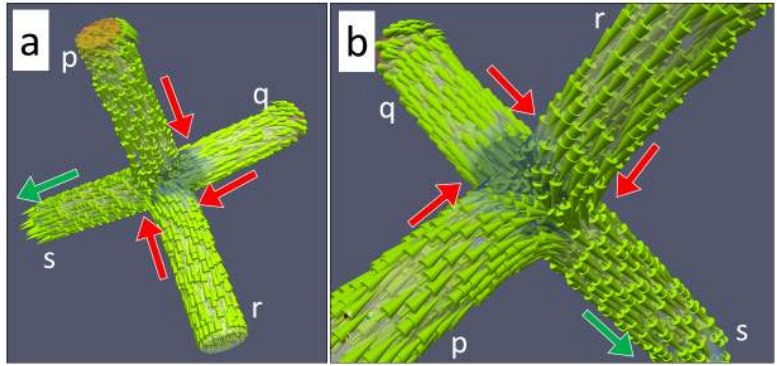

Figure 7.5: (a) Magnetic configuration of the three-in/one-out state (b) Zoomed in view highlighting the three dimensional head-head domain wall

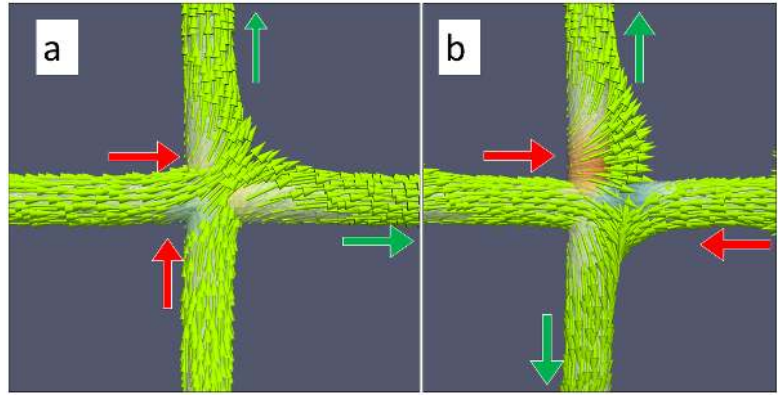

Figure 7.6: Micromagnetic structure of two different flavors of the two-in/two-out state in a 2D square ASI: (a) head-tail state, (b) head-head state

Therefore, the magnetic configuration at each vertex can be uniquely identified based on their charge. To identify the magnetic configurations at each vertex, it is therefore sufficient to calculate the volume charge distribution ($\rho = \nabla \cdot \boldsymbol{m}$) at each vertex, without requiring an individual inspection of each vertex configuration to identify the total number of nanowires magnetized towards or away from the vertex. Since the local divergence can be calculated numerically, this feature is convenient for easily obtaining a statistical analysis of the number and position of each type of vertex configuration in extended lattices. It can also be used to automatize the detection and further analysis of the defect charges, see Fig. 7.7.

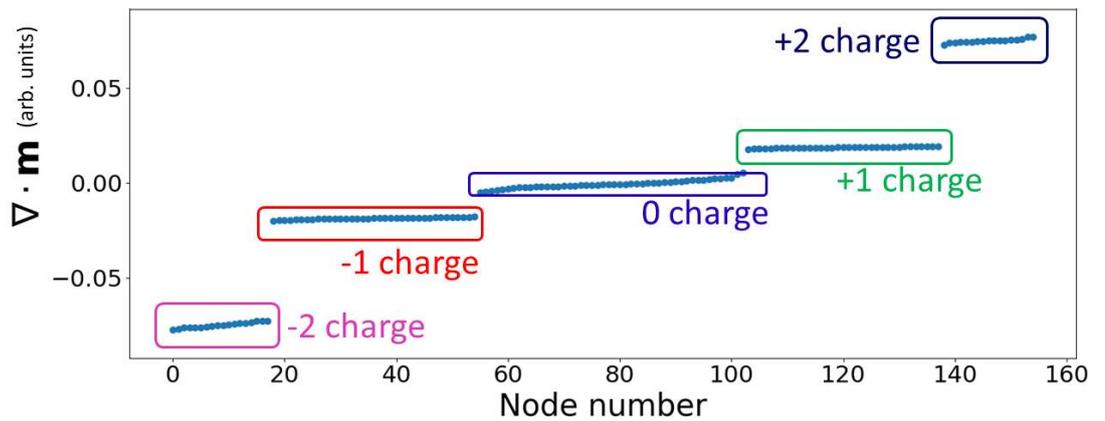

Figure 7.7: Automatized detection of the charge type, number and position of each type of vertex configuration based on the value of $\nabla \cdot \boldsymbol{m}$ at each vertex. Different types of configurations can be uniquely identified by clustering the vertices based on the average value of the local divergence (volume charge density) at each vertex.

The ice-rule obeying zero-charged vertices are energetically optimal states for the lattice. Their configuration results in a minimum possible magnetostatic and exchange energy, while the charged vertex configurations yields both a net magnetic charge, thereby increasing the magnetostatic energy, and a higher degree of inhomogeneity, and thus higher exchange energy. The formation of defect charges within the AMC results in a higher magnetic disorder, which can be quantified by the density of charged vertices within a lattice. Despite their energetically unfavorable structure compared to the ice rue obeying states, the defect charges are stable once they are formed. In general, the removal of a defect charge from the system requires an external field strong enough to magnetically saturate the sample.

| Charge types | Zero-charge / Reference state | +2/-2 state | Disordred state |
|:---:|:---:|:---:|:---:|
| +2 | 0 | 40 | 17 |
| +1 | 36 | 33 | 35 |
| 0 | 83 | 0 | 48 |
| -1 | 36 | 39 | 37 |
| -2 | 0 | 43 | 18 |

Figure 7.8: Summary of the number of the types of vertices with different charge in the case of three different remanent magnetic states of the AMC: The defect-free state obtained after saturation along the $[1, 0, 0]$ direction, the ordered charged state resulting after saturation along $[1, 1, 1]$, and a disordered state obtained from a randomized initial configuration.

Based on the presence or absence of defect charges, we can define two qualitatively different ordered magnetic states within the lattice: A defect-free state, which only contains ice-rule obeying zero-charged vertices, and a defect state which contains several $+2q/-2q$ charged double charge vertices. The two different magnetic states can be simulated in a controlled way by changing the initial configuration of the simulation. Relaxing the structure to a zero-field state when starting from a saturated state along the $[1, 0, 0]$ direction so that none of the wires is aligned parallel the applied field results in a defect-free, zero-charged state. On the other hand, if the initial point is a saturated state with the field applied along the along the $[1, 1, 1]$ direction of the lattice, a subset of the nanowires (one out of six) in the lattice is magnetized parallel the field direction. This initial direction of the magnetization in the subset of nanowires will be preserved if the field is subsequently lowered. Relaxing to zero field from this state results in a state which only contains double charges: All the vertices within the crystal display the magnetic structure of ice-rule violating states, resulting in a periodic distribution of alternate layers of $+2q$ and $-2q$ vertices. Relaxing from a random initial configuration results in a disordered state, which contains both zero charged vertices and double charged vertices distributed randomly within the AMC. The disordered state which we simulated and used for all the upcoming discussions contained 48 zero-charged vertices and 35 double-charged vertices, out of which 17 are of $+2q$ and 18 are of $-2q$ type. A detailed description of the number of each charge type in each lattice type is given in table 7.8. It is to be noted that this disordered state is only one realization out of a quasi-continuum of many possible configurations [243, 244, 245].

## 7.3 Hysteretic properties of the crystal: zero charged state

A simulated hysteresis loop of the network is shown in Fig. 7.9, and a visualization of the evolution of the magnetization of the structure at various stages of the hysteresis is displayed on the next page. For convenience and better visualization, the magnetization of a single tetrapod unit within the bulk of the lattice is visualized, instead of displaying the whole network. The external field is applied along the $x$ direction, so that it is not parallel to any of the nanowires in the network. At the starting field of 800 mT, the network is in an technically saturated state in which the nanowires are magnetized parallel to the field direction rather than along the axial direction. As the field is lowered, the magnetization in the wires rotates smoothly towards the axial direction. On further lowering the field to zero, we obtain a state where all the individual nanowires are magnetized along their axis and the network has a net remanent magnetization along the $x$ direction. This remanent state obtained after saturation along the $x$ direction is highly ordered and contains only zero-charged vertices. Upon increasing the field in the opposite direction, the magnetization along the wire initially remains axial in the case of low fields. However, once the field exceeds about $-240$ mT, the magnetization flips irreversibly towards the field direction resulting, in a sharp jump in the $M$-$H$ curve. After this point, increasing the field further results in the reversible rotation of the magnetization towards the external field. This gradual rotation appears as an almost linear increase of $\boldsymbol{M}_{||}/M_s$ with increasing external field, until saturation is reached.

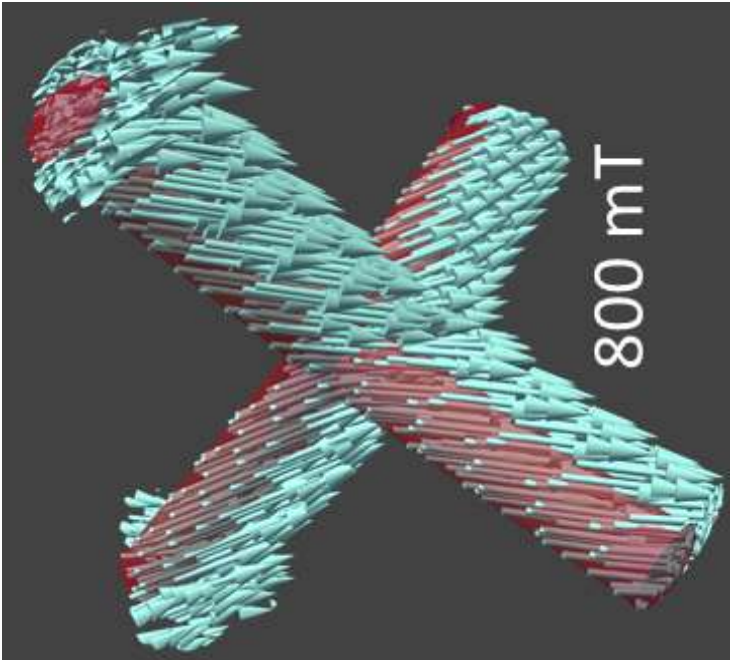
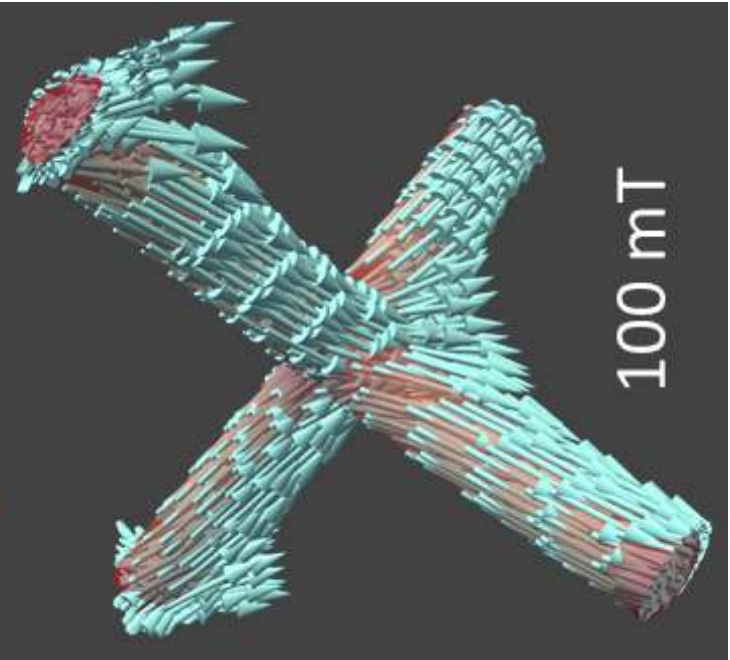
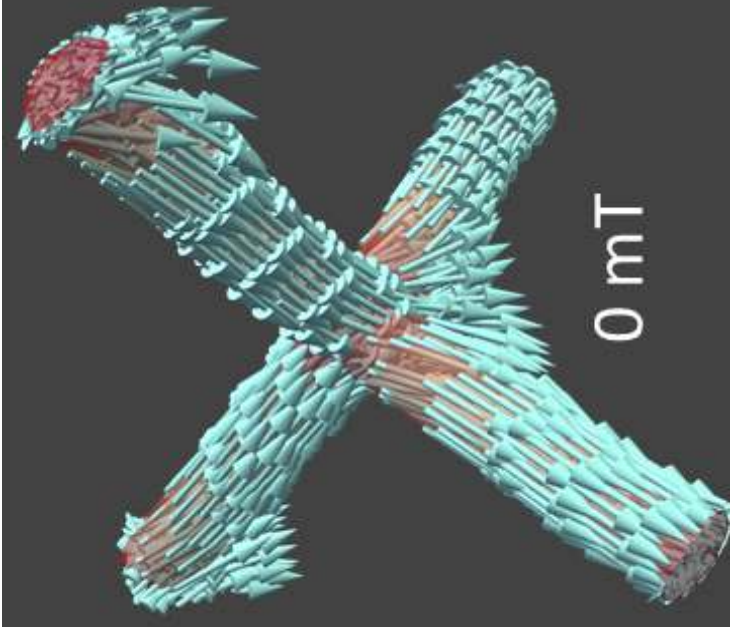
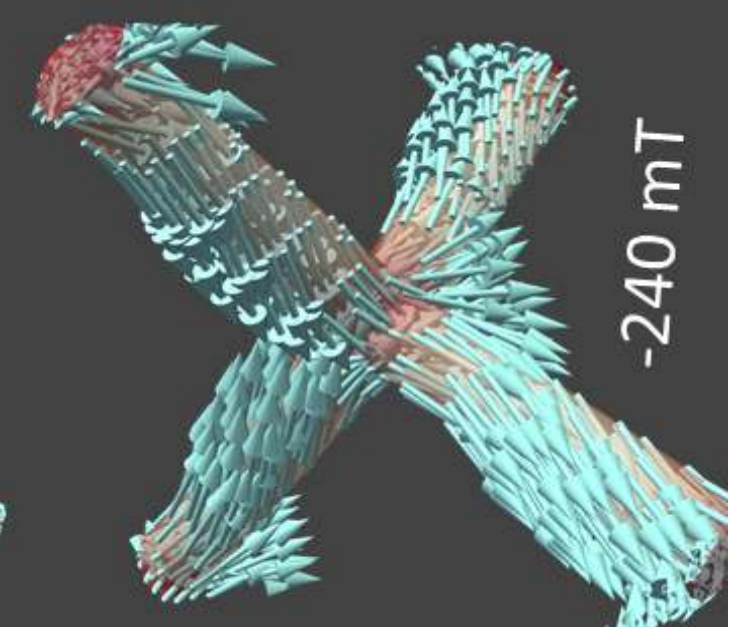
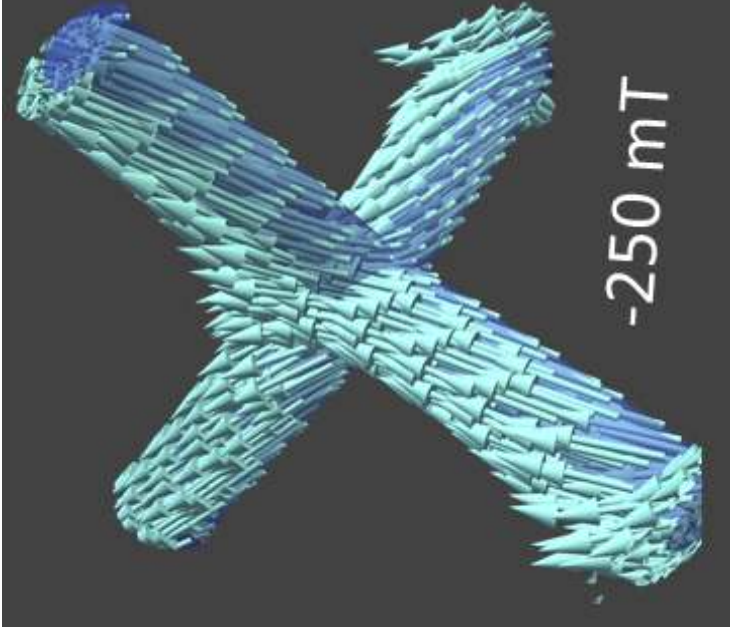
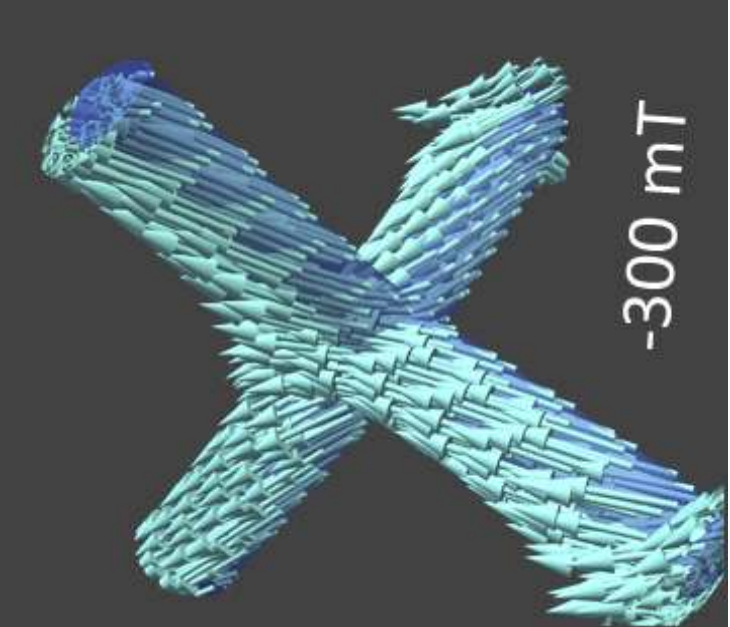

800 mT

100 mT

0 mT

-240 mT

-250 mT

-300 mT

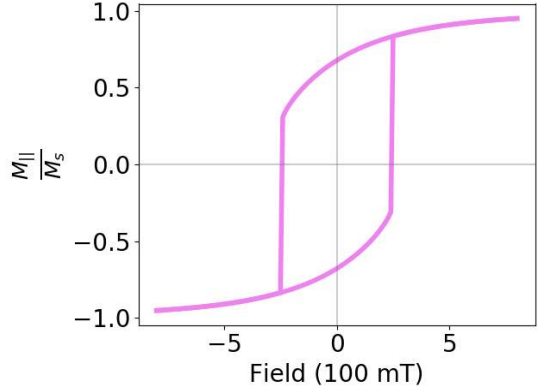

Figure 7.9: Simulated hysteresis loop of the diamond AMC lattice

## 7.4 Magnonic excitations: Zero-charged state

As we have seen already in the case of the buckyball structures, the formation of ice-rule-violating defect charges can have a strong effect on the magnonic response of the artificial spin ices. To demonstrate this, we simulate the small-angle precession modes and extract the frequency modes with a Fourier analysis, as explained in the previous chapters, and compare the magnonic signature of the zero-charge state and the disordered defect state. The magnonic spectrum of the zero-charge state is shown in Fig. 7.10. We can identify five major modes with the frequencies 9.9 GHz, 14.1 GHz, 16.6 GHz, 19.0 GHz, and 28.3 GHz, respectively. Each of these prominent modes can be attributed to distinct magnetic oscillation developing at various geometric locations within the lattice: the dangling wire ends at the surface, the vertices, and the cylindrical nanowires. Each mode is marked with a number from 1 to 5, which will be used in the following discussion to refer to the individual modes.

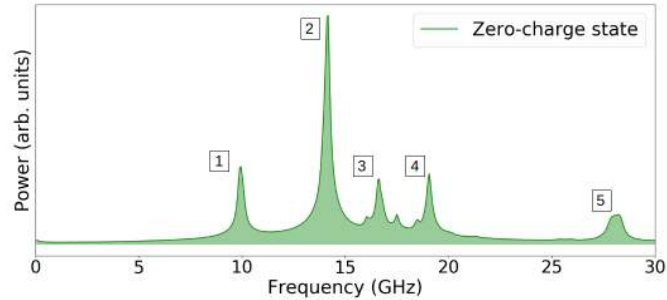

Figure 7.10: Frequency spectrum of the zero-charge state. Different prominent modes are marked from 1 to 5.

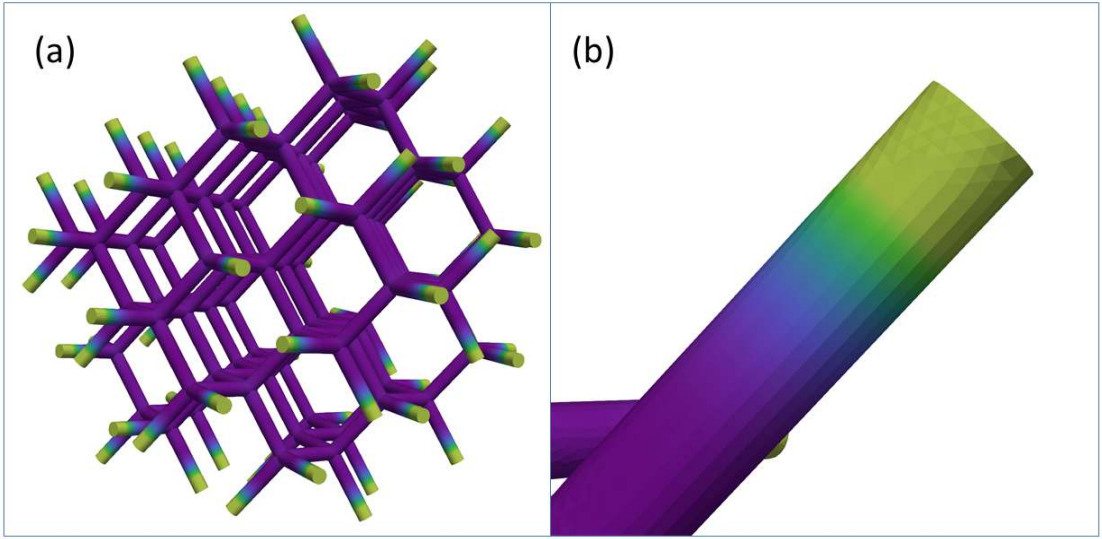

Figure 7.11: (a) Mode 1: 9.9 GHz, Oscillation of the +1 and −1 charges at the free ends of the lattice. (b) Zoomed-in view showing the activity at the ends.

The spatial profile of each mode is extracted using a windowed inverse Fourier transform and illustrated in Figs. 7.11, 7.12, and 7.13. In these visual representations, the time-averaged value of the magnitude of the dynamical component of the magnetization at each discretization point is calculated and displayed. The purple color indicates inactive regions and the yellow regions indicate ares of high magnonic activity. It is to be kept in mind that the displayed plot is a not an instance of time evolution, but an averaged value of the entire time sequence.

We can classify the five modes into two categories: the low frequency modes which are caused by the activity of the ±1 and 0 charged vertices (and free ends) and the higher frequency modes which are caused by different orders of oscillations of the magnetization in the nanowires. The mode $\boxed{1}$ at 9.9 GHz arises from the activity of the ±1 charges at the dangling free ends. The vertices in the bulk and the nanowires are inactive at this frequency (Fig. 7.11). The relative height of the peak $\boxed{1}$ with respect to the other peaks depends on the geometry of the crystal geometry, which determines the surface-to-volume ratio; that is the ratio of ±1 charges to 0 charges. The oscillation of the dangling free ends appears to be decoupled, meaning that we could not observe any phase correlation between the oscillations at different sites. It can be observed that there is no distinction between the oscillation at +1 charges with that of the −1 charges.

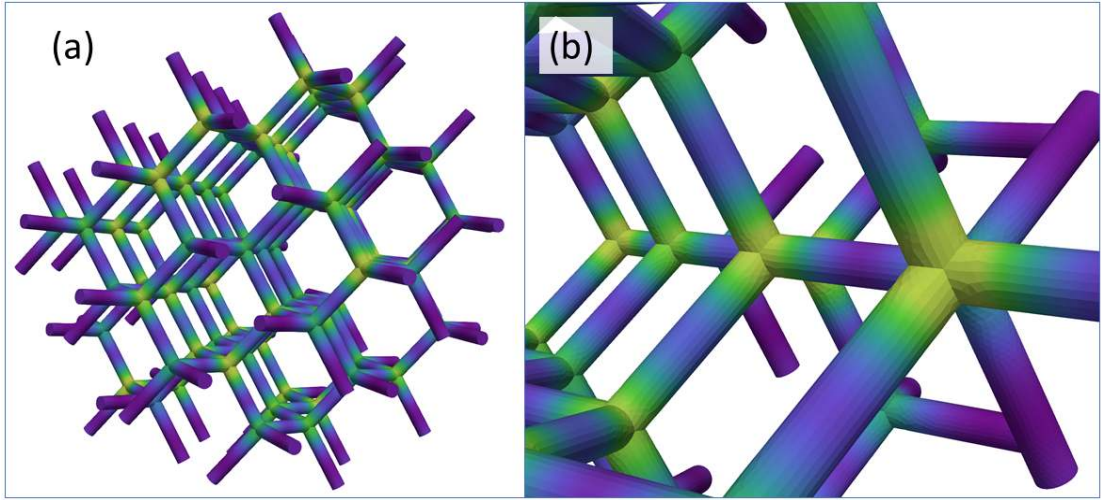

Figure 7.12: (a) Mode 2: 14.1 GHz, Oscillation of the ice-rule obeying zero-charged vertices (b) Zoomed in view showing the activity of the vertices

The most intense peak, marked $\boxed{2}$, develops at 14.1 GHz. The corresponding mode is an oscillation of the magnetization at the zero-charged vertices within the lattice. The spatial profile of the mode is shown in Fig. 7.12. We can observe a clear localization of the mode at the zero-charged vertices. Contrary to the oscillation of the free ends, we can visually identify a phase correlation in the oscillation of the zero charged vertices. This is discussed in more detail in a later subsection.

The modes $\boxed{3}$ $\boxed{4}$ and $\boxed{5}$ at 16 GHz, 19.0 GHz and 28.3 GHz, respectively are the wire modes. These peaks correspond to the development of standing waves within the nanowires. The spatial profile of the three standing waves are displayed in Fig. 7.13. For convenience, the wires are graphically isolated from the rest of the lattice, although in the simulation they are structurally embedded in the lattice. In the case of mode $\boxed{3}$ and $\boxed{5}$ the vertices remain inactive and act as nodes, while for mode $\boxed{4}$ the magnetization also oscillates at the vertices. A clear macroscopic correlation of the oscillations at different sites of the crystal can only be observed for the mode $\boxed{4}$, which is the only one in which the vertices takes part in the activity. This suggests that long-range dynamic correlation results from the dynamic interaction between the vertices. According to the orientation of the nanowires, there are six different distinct directions in the lattice. However, the magnetic modes developed in the nanowires in the defect free state do not appear to be affected by the orientation of the wires.

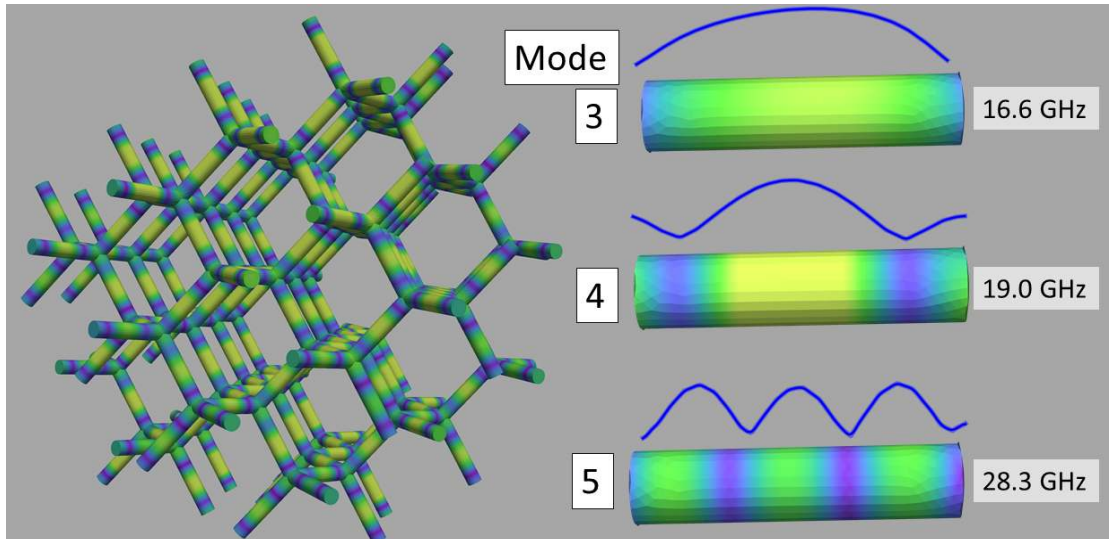

Figure 7.13: High-frequency standing-wave modes in the nanowires.

## 7.5 Magnonic response of defect charges

So far, we have investigated the high-frequency modes arising in the defect-free lattice. We could identify five modes, each corresponding to a specific magnonic activity. In the next step, we repeat the same procedure for the magnetic configuration which contains ice-rule violating defect charges; the disordered state or the double-charge state. A comparison of the magnonic response of this double-charge state with the zero-charge state is shown in Fig. 7.14. On first glance, we can see that the five modes we discussed in the previous section remain almost unchanged. The most striking difference is the appearance of an additional, intense low-frequency mode at 7.9 GHz. This mode describes the oscillation of the double-charged vertices, which were absent in the previous state. The appearance of this peak, caused by the introduction of magnetic defects, is a manifestation of the dependence of the high-frequency properties on the magnetic configuration of the AMC. The zero-charged vertices present in the disordered state oscillate at the same frequency discussed previously, which is quite different from that of the double charges, even though the oscillations occur at the same geometrical location.

The spatial profile of the oscillation of the zero-charged and double charged vertices is shown in figure 7.15. The mode profiles of the zero- and double-charge vertices are complementary, that is, at 7.9 GHz only the double charged vertices are active, and at 14.2 GHz only the zero charged vertices. In spite of this clear separation based on the magnetic configuration, there exists a considerable difference in the intensity of the activity even within the vertices belonging to the same type. We attribute these differences to inhomogeneities in the local magnetic fields arising due to the disordered charge distribution. This difference only concerns the zero charges; we could not observe any difference in the activity between $+2q$ and $-2q$ double charges. Even though the oscillating frequency and the mode profile of the zero-charged vertices remains the same as in the case of the defect free state, a clear phase correlation between different vertices, which was observed in the previous case, was absent in the discorded state. The number of zero-charged vertices in the defect state (48) is almost 40 % less than that of in the defect free state (83). When we compare the intensity of the peaks, we can observe a proportional decrease in the amplitude of the disordered state, which is in accordance with the decrease in the number of zero-charged vertices.

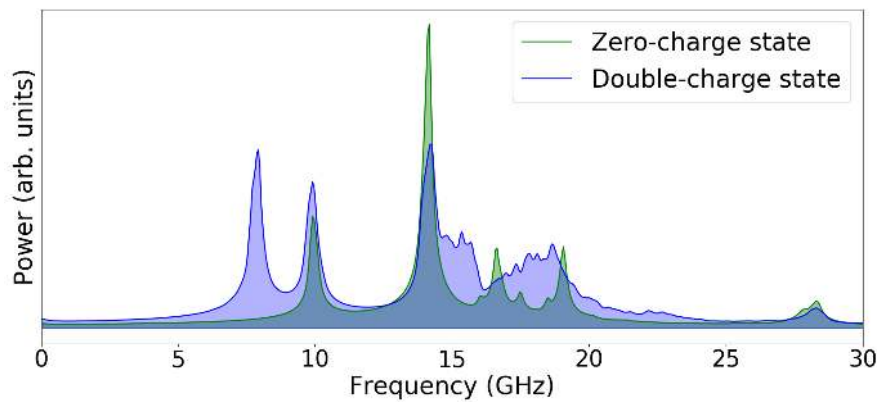

Figure 7.14: Comparison of the magnonic spectrum of the zero-charge and double-charge state

Considering the magnonic modes in the nanowires in the two cases, there is a significant difference in the disordered state. The peaks ③ and ④ describing the high-frequency standing-wave modes in the nanowires are smudged and become almost indistinguishable. Instead of having two clear, well-defined peaks, we now have a broad distribution of overlapping modes from to about 15 GHz to 20 GHz. A detailed isolated Fourier analysis of the individual nanowires shows that the nanowire modes are nevertheless still present in the disordered state. However, their individual frequencies differ from one nanowire to another, depending on several factors, which results in a broad an blurred spectrum, see Fig. 7.6. A more detailed investigation of this effect is discussed in the section 7.6. Compared to the modes ④ and ⑤, the higher-frequency mode $\boxed{5}$ remains largely unaffected by the disorder of the magnetic state. In general, higher-frequency modes are dominated by exchange interactions, which leads to local effective field much larger than those due to dipolar fields. As a result, changes in the dipolar field distribution connected with the disordered magnetic structure can be expected to have a weaker impact on the exchange-dominated higher-frequency modes.

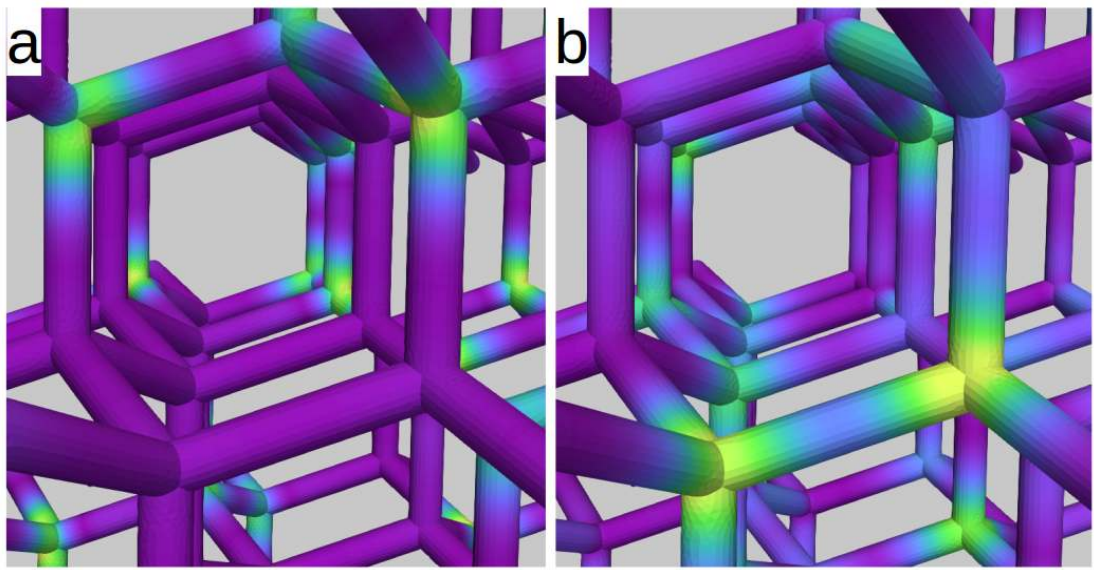

Figure 7.15: Spatial profile of the vertex modes in the disordered state (a) Activity of the zero-charge vertices. (b) Activity of the double-charge vertices. It can be visually observed that the two modes are complementary, meaning that the regions active at one frequency are inactive at the other.

## 7.6 Different types of nanowires: Influence of local fields on the magnonic spectrum

We have classified the vertices into different groups based on their magnetic configuration and we saw how the configuration affects their magnonic response. In a similar approach to analyze the modes in the nanowires, we can classify the nanowires into different groups depending on the charges of the vertices which they connect. Let us consider the case of the zero-charge state, which contains only ice-rule obeying zero-charged vertices and $\pm 1$ charged surface points. In that case, we have three different types of charges if we include the $+1$ and $-1$ charged free ends. Therefore, we can distinguish between three different kinds of nanowires:

1. type 0/0 wires located within the bulk and connecting two zero-charged vertices

2. type +1/0 wires at the surface, connecting a +1 charged end to a zero-charged vertex

3. type −1/0 wires connecting a −1 charged end to a zero-charged vertex

It is to be noted that wires connecting −1 to +1 charges cannot exist, as the ±1 single charges are always formed on the single, dangling free ends. They are always connected to a vertex within the volume, and never with each other. In the ice-rule obeying zero-charge state, there are 83 zero-charge vertices, 36 vertices with +1 charge, and 36 with a charge of −1. As a result, there are 130 wires of 0/0 type, 36 wires of +1/0 type, and 36 wires of −1/0 type. To obtain a deeper understanding about the dependence of the presence different types of nanowire configurations on the magnonic response of the whole structure, we carried out localized Fourier analysis for each nanowires. In this analysis process, each descretization point corresponding to a particular nanowire is identified, isolated and the magnetization dynamics in this region is Fourier-analyzed such as to yield the individual magnonic response of each nanowire.

It is observed that nanowires belonging to the same category display similar magnonic properties. We further found that the peculiar magnonic signature of each type of wire is charge-symmetric: that is, a set of wires connecting charges $x$ and $y$ exhibits a basically identical magnonic spectrum as that of a set of wires connecting $-x$ and $-y$ type charges. This can be explained as follows. In the absence of an external field, the magnetic configuration in a nanowire oscillates under the effect of the dipolar field generated by the magnetic configurations at the vertices. The magnetic volume charge densities developing in a $-2q$ vertex and a $+2q$ vertex act as a sources and sinks for the magnetic field $\boldsymbol{H}$. As a result, a longitudinal magnetic field develops in wires representing a Dirac string [128, 246] connecting $+2q$ a charge to a $-2q$ charge ($-2q \rightarrow +2q$). The presence of such a magnetic field can shift the frequency of the standing wave with respect to the same oscillation at zero field [247]. Thus, all the wires having same kinds of charges attached to their ends are exposed to similar dipolar fields, resulting in similar magnonic properties. This also explains the charge symmetry effect. Due to time-inversion symmetry, a +2 charge and a −2 charge are equivalent, which in turn results in dipolar fields of equal strength.

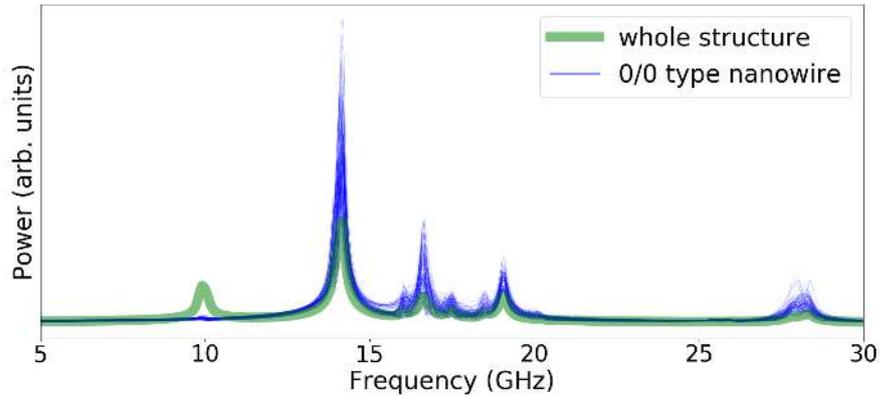

Figure 7.16: comparison of the magnonic response of all the 0/0 type nanowires, with the magnonic response of the whole structure overlayed on top (green) for reference. Each thin, blue line corresponds to a different 0/0 type nanowire.

Since there are three types of wires in the reference state, we can have three types of interactions. A comparison of the magnonic response of all the 0/0 type nanowires is given in Fig. 7.16. One can see that all the individual wires in this group exhibit a similar magnonic spectrum. The minute variations in the power spectrum can be attributed to asymmetries in the next-nearest neighbour interactions, and to variations in the orientation direction with respect to the direction of the field pulse by which the oscillation is triggered. Comparing the magnonic features of the nanowires with that of the whole structure, we can see that the mode $\boxed{1}$ at 9.9 GHz, which is caused by the activity of $\pm 1$ charges, is absent in the 0/0 type. A detailed comparison of the 0/0, +1/0 and −1/0 type nanowire groups is shown in Fig. 7.17. We can clearly see that the +1/0 and −1/0 groups have a similar magnonic response to the field-pulse exitation, and that their spectrum is significantly different from that of the 0/0 group, as expected.

In the case of the disordered double-charge state, which contains $\pm 2$ charged vertices along with zero-charges vertices, more types of nanowires can be observed as explained in the following table. In such a disordered configuration, we can distinguish twelve different types of nanowires, with the 0/0 type being the most common one. A detailed graph depicting the magnonic response of all twelve types of nanowires is presented in Fig. 7.18. We can see that all the nanowires in the same group exhibit a similar magnonic spectrum, especially in the lower-frequency modes below 15 GHz. Above that frequency, the modes tend to broaden and to become less coherent. The charge-symmetric behaviour explained in the previous sub-section can also be observed.

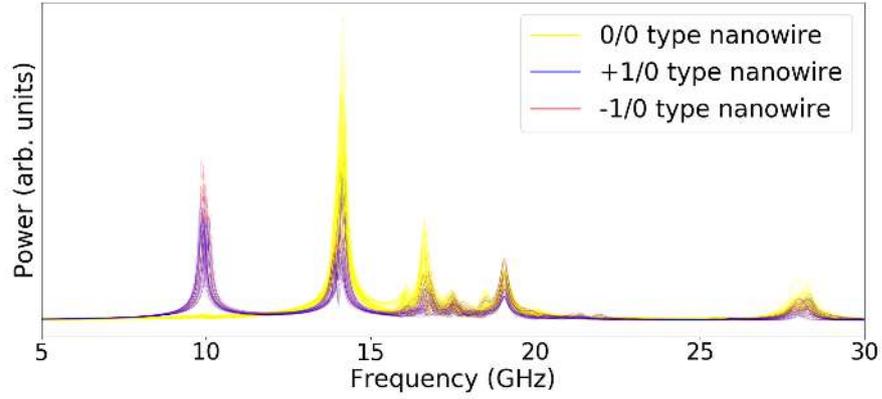

Figure 7.17: Comparison of the magnonic spectrum of 0/0, +1/0 and −1/0 type nanowire groups.

| Types of nanowires | Number |
|---|---|
| +2 / +2 | 15 |
| +2 / +1 | 12 |
| +2 / 0 | 28 |
| +2 / -1 | 5 |
| +2 / -2 | 13 |
| +1 / 0 | 20 |
| +1 / -2 | 3 |
| 0 / 0 | 48 |
| 0 / -1 | 16 |
| 0 / -2 | 32 |
| -1 / -2 | 16 |
| -2 / -2 | 4 |

## 7.7 Tuning the mode frequencies with an external field

To investigate the field dependence of the frequency of the modes, small-angle precession modes of the lattice are simulated in the presence of an external field. The resulting changes of the frequency can be displayed in a heatmap representation of the modes, as shown in Fig. 7.19. Since the equilibrium magnetization state was obtained by relaxing to zero field from saturation along a particular direction, the structure retains an effective magnetization along that direction of initial saturation. In the case of the spectrum shown in Fig. 7.19(a) the field is applied parallel to the initial saturation direction. The result indicates that the frequency of the modes can be linearly controlled by varying the strength of the external field. Applying the field in the opposite direction results in a corresponding reduction of the frequency. Reducing the field below $-200\,\mathrm{mT}$ would lead to a switching of the magnetization and a thus a departure from the ASI behavior typical of that specific magnetic configuration. The heatmap shown in Fig. 7.19(b) refers to the case where the external field is applied perpendicular to the direction of the remanent magnetization. Comparing this field-dependent evolution of the spectrum to the previous one, one can clearly identify a strong influence of the direction of the applied field on the resulting frequency response. In the case of a field applied perpendicular to the remanent magnetization direction, we observe a splitting of the mode lines into several branches as the field strength is increased. This lift of degeneracy can be explained based on the dependence of the frequency of the nanowires on the external field, analogous to the discussin on the previous section in the case of local dipolar fields. The oscillation frequency of the nanowires splits into four branches depending on the relative orientation with respect to the external field.

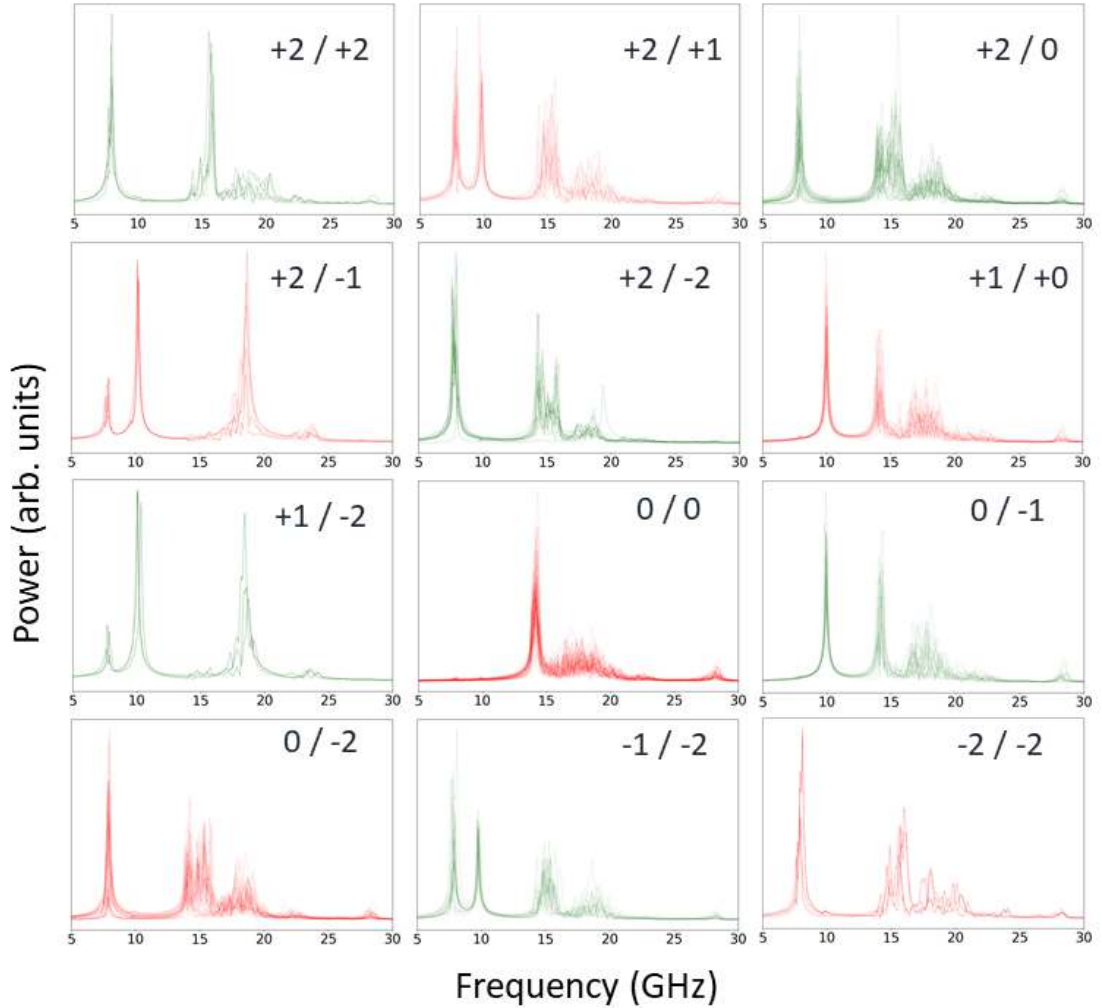

Figure 7.18: Comparison of the magnonic response of all twelve types of wires

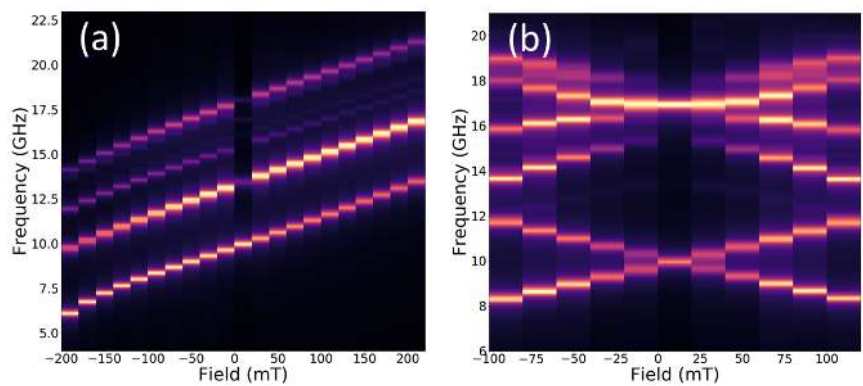

Figure 7.19: Field response of the frequency of the modes in the case of (a): an external field applied parallel to the direction of effective magnetization of the lattice and (b) a field applied perpendicular to the direction of effective magnetization.

## 7.8 Effect of a geometric defect in the crystal lattice

The strong impact on the magnonic spectrum of the diamond-type artificial lattice by the formation of defect charges is comparable with that of Nitrogen vacancy centers (NV centers) on the optical properties in the case of a Nitrogen-doped real diamond crystal. To further investigate the potential for manipulating the properties by defects in the artificial lattice, we conducted the same studies as described before, but on a structure with geometrical defects. The individual defects in the AMC are introduced by removing single nanowires from the bulk of the lattice. Such a removal results in the creation of two vertices with odd coordination in the bulk of the lattice, namely with a coordination number 3. Similar to how we assigned a charge value to each vertex configuration in section 7.2, we can now assign a charge value to these structural defect vertices as well. Since an odd number of wires is meeting at these kinds of vertices, they cannot adopt a zero-charge configuration. Similar to the case of the ice-rule obeying states in the buckyball structure, only two types of configurations can be found in these vertices:

1. Two-in/one-out state ($+1.5q$)

2. One-in/two-out state ($-1.5q$)

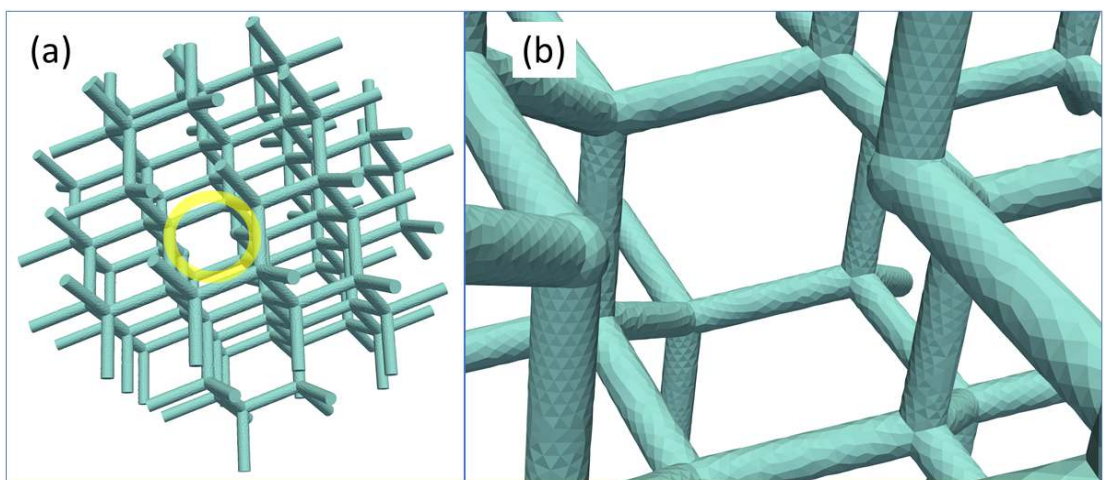

Figure 7.20: (a) Diamond lattice structure with a geometrical defect (yellow circle). (b) Zoomed-in view highlighting the defect vertices with coordination number 3.

Theoretically, one could imagine also to find three-in and three-out configurations at these vertices. But in our simulations we never observed a situation like this, in which all the wires of a defect vertex are magnetized towards or away from the vertex. To be consistent with the previous discussion of the charged vertices in the buckyball lattice, one could have assigned charges of $+1$ and $-1$ to these two magnetic states forming at the tripods of the defect vertices, respectively. But it is necessary to differentiate between these charges and those from the $\pm 1$ charges occurring at the surface. The distinction is not only required because the magnetic configuration is entirely different in those two case, but also because the numerically computed average value of $\nabla \cdot \boldsymbol{m}$ at the geometrical defect points is significantly different from that at the free ends. It lies between that of the single charges at the surface and the value of the double-charged vertices. We therefore attribute a value of $\pm 1.5q$ to the charges at these structural defects; more for practical than for formal reasons.

The magnonic spectrum of the ice-rule obeying zero-charged state of the structure with the geometrical defect is shown in Fig. 7.21. We can identify two strong peaks at 9.9 GHz and 13.7 GHz, respectively. As already seen, these peaks correspond to the activity of the $\pm 1$ charged vertices at the free ends and to the zero-charged ice-rule obeying vertices, respectively. In between these peaks we can observe a low-intensity mode (labelled "2") at 11.7 GHz. The profile of this mode is localized at the $\pm 1.5$ charged defect vertices. It has a low intensity because the signal originates from just two defect vertices out of a total of more than 100 surface and volume vertices in the AMC. Besides the appearance of a this mode, the introduction of geometrical defects results in the shifting of the wire modes $\boxed{3}$, $\boxed{4}$ and $\boxed{5}$ towards lower frequencies. Adding more structural defects exacerbates this effect and leads to further changes in the spectrum.

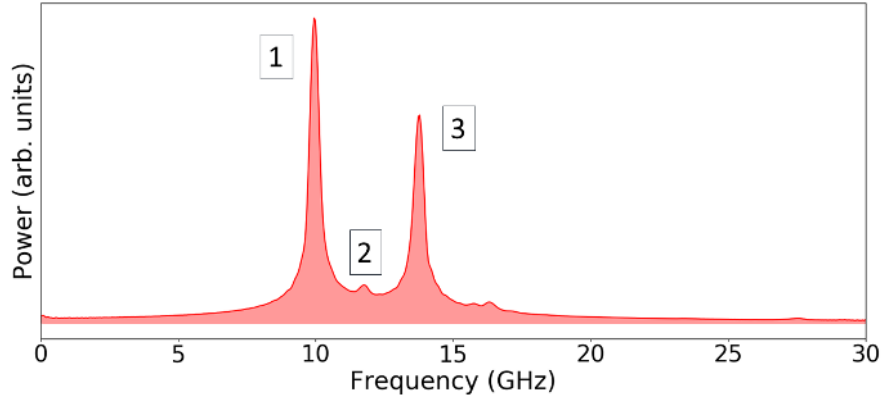

Figure 7.21: Magnonic spectrum of a diamond-type AMC lattice with two geometrical defects. The weakly developed peak $\boxed{2}$ can be directly ascribed to the defect in the lattice.

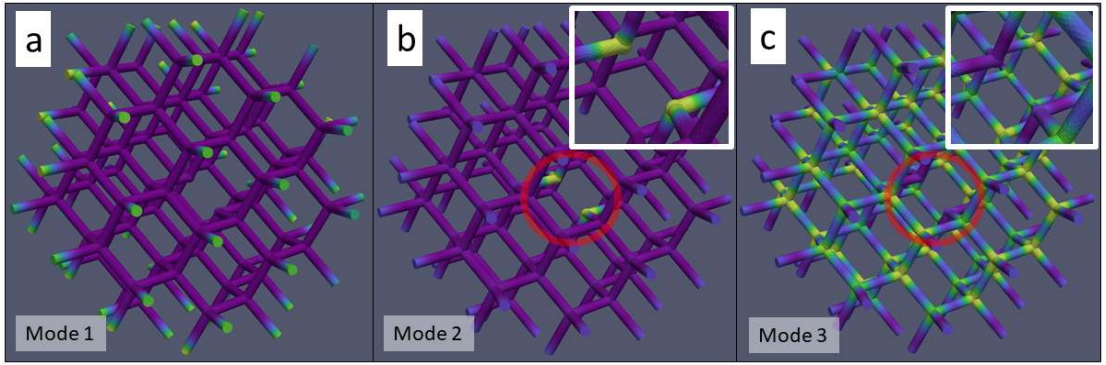

Figure 7.22: Spatial profile of various modes in the AMC lattice with structural defect. (a) Oscillation of the $\pm 1$ charges at the ends, (b) oscillation of the $\pm 1.5$ charges at the defect vertices (the inset shows a zoomed view on the defect vertices), and (c) oscillation of the zero-charged regular vertices, with the inset showing a zoomed view on the defect vertices.

Upon increasing the density of structural defects, the extent to which the spectrum changes quickly reaches a degree at which it becomes impossible to consider the defects as mere perturbations or modifications of the defect-free state. An illustration of the spatial profile of the mode at the structural defect is shown in Fig. 7.22.

## 7.9 Macroscopic spin wave in a diamond-type AMC

As indicated in the previous section, we could observe a clear long-range phase correlation in the oscillation of the zero-charged vertices in the defect free state. The oscillation appears to propagate from the core of the lattice towards the surface in the form of a macroscopic spin wave with a wavelength of approximately twice the wire length. To verify this observation, we modelled an elongated AMC network with a length of 20 times the estimated wavelength of the macroscopic spin wave. The relaxed magnetization structure was obtained by saturating the sample along the longitudinal direction and then relaxing the system to a zero field state. This resulted in a well-defined defect-free magnetization state in which the whole structure has a net magnetization along the longitudinal direction and where all vertices within the crystal are in a two-in/two-out zero-charged configuration. To investigate the possible existence and propagation of macroscopic spin waves in the form of correlated oscillations of the zero-charged vertices within the artificial lattice, the elongated structure was excited with a spatially homogeneous Gaussian pulse. The magnetic field pulse was applied perpendicular to the longitudinal direction. The subsequent global response of the vertex modes is analyzed with the same methods we used before. In order to verify whether such macroscopic vertex modes can propagate through the lattice, we also studied the case in which the elongated structure was excited only at one of its ends by means of a spatially inhomogeneous pulse. In both cases, the magnetic ring-down dynamics triggered by the field pulse was analyzed.

The high-frequency spectrum of the elongated structure, obtained from a homogeneous field pulse excitation, is shown in Fig. 7.24. Unsurprisingly, the structure exhibits high-frequency response that is similar to the one of the other, smaller diamond-type lattice structure. The extracted oscillations of the vertex modes are displayed in Fig. 7.25. Contrary to what we expected, we could not observe any longitudinal variation in the phase of the oscillation. All the vertices appear to be oscillating at the same phase along the entire lattice. However, we could detect a propagation along the radial direction of the lattice. We can visually identify a phase delay in the oscillation of the vertices near the surface from that of the vertices in the bulk, in particular thos close to the central axis. This effect is probably caused by small differences in the properties of the vertices depending on their nearest-neighbor interactions. The vertices lying closer to the surface of the lattice are connected to two other bulk vertices while the two other branches are dangling free ends. Contrary to this, in the case of vertices within the bulk, all the four nanowires originating from them are connected to four other vertices within the volume. This difference in nearest-neighbor interaction could be responsible for the appearance of variations of the properties of the vertices within the lattice depending on their location, which in turn could explain a phase delay between their oscillation.

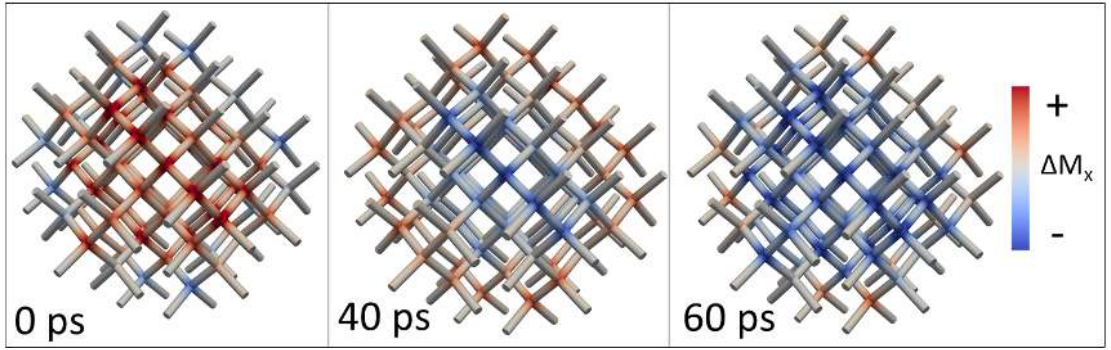

Figure 7.23: Time evolution of the vertex mode in an elongated diamond-type AMC. The oscillation appears to be propagating from the core towards the surface.

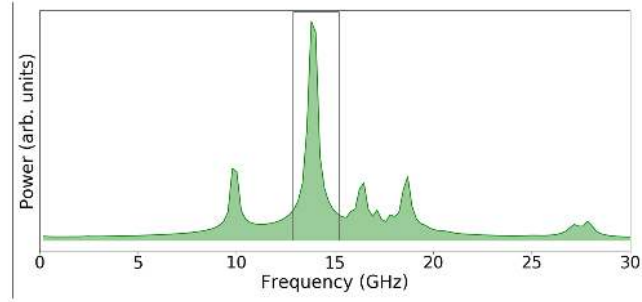

Figure 7.24: Frequency response of the elongated lattice upon excitation by a field pulse. The vertex modes which are of interest are marked.

To test whether a localized perturbation can propagate along the lattice in the form of a macroscopic spin wave, the elongated lattice was excited at only one of its ends with a position-dependent Gaussian field pulse. As usual, the following dynamical response was recorded and the oscillation of the vertices was extracted. The time evolution of the oscillation of the vertices is displayed in Fig. 7.26. We can see that the oscillations of the vertices do not propagate through the lattice. This suggests that the magneto-dipolar coupling between the vertices is not strong enough for an efficient energy transfer and for a related spin-wave propagation.

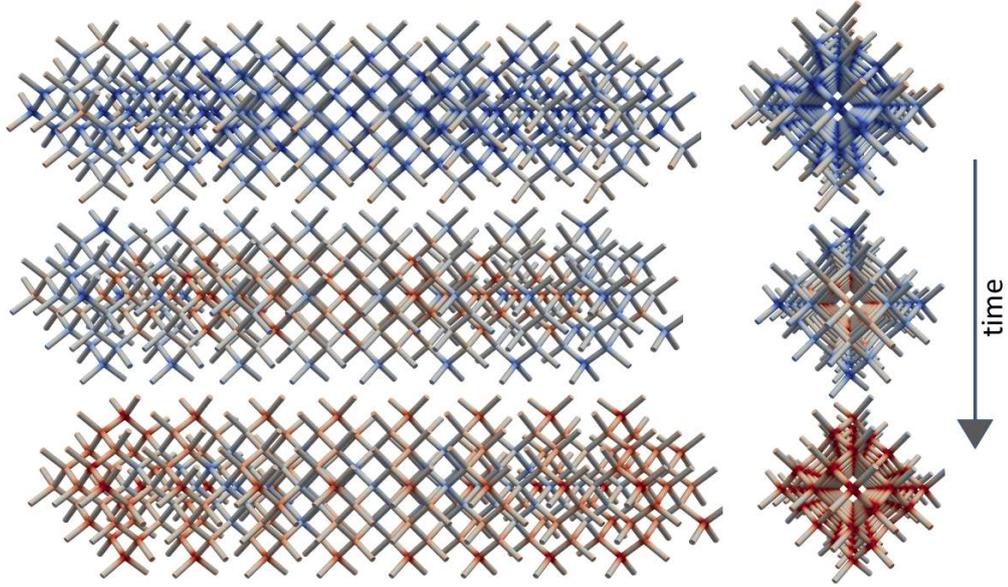

Figure 7.25: Time evolution of the vertex modes of the elongated lattice with the corresponding lateral view. Even though there is no propagation along the length of the lattice, we can see a radially propagating behaviour from the centre of the lattice towards the exterior region.

## 7.10 Cubic lattice structure

In the previous section, we investigated the formation of different vertex configurations in a diamond AMC and their effect on the magnonic signature of the network. To demonstrate the robustness of these observations, we expand our investigation into a different kind of AMC: the cubic lattice. The cubic lattice structure is formed by the interconnection of nanowires in a simple cubic arrangement, so that six wires meet at each vertex. Our model structure was made of cylindrical nanowires of radius $7.0\,\mathrm{nm}$ and length $70.0\,\mathrm{nm}$, arranged in a $7 \times 6 \times 5$ lattice, see Fig. 7.27.

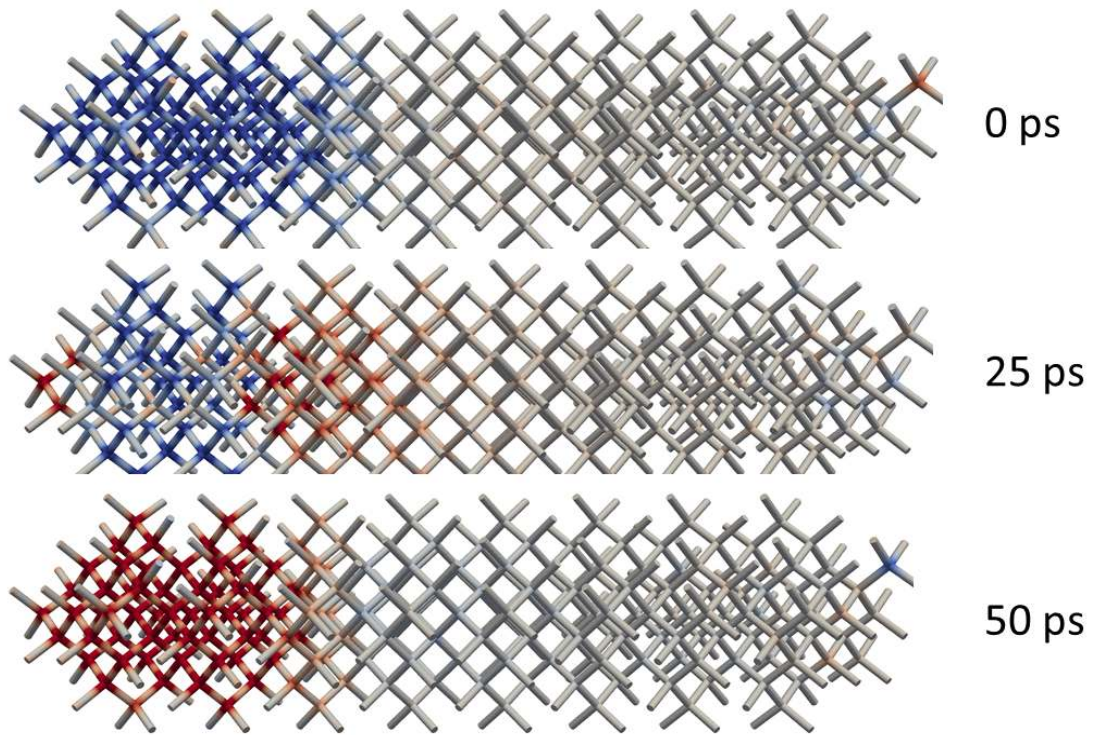

Figure 7.26: Oscillation of the vertices after being excited locally. We can see the development of a standing wave, but the wave does not propagate through the lattice.

### 7.10.1 Vertex configuration

In contrast to the four nanowires meeting at the vertices of the diamond structure, there are six nanowires at each vertex in the case of a cubic lattice. Adopting the same principles as before, we assign an Ising type magnetization to each nanowire. Based on the total number of Ising type moments coming in or leaving out at each vertex site, we can have seven different types of vertex configurations in the cubic lattice:

1. Six out (+6)

2. Five-out / one-in (+4)

3. Four-out / two-in (+2)

4. Three-out / three-in (0)

5. Two-out / four-in (−2)

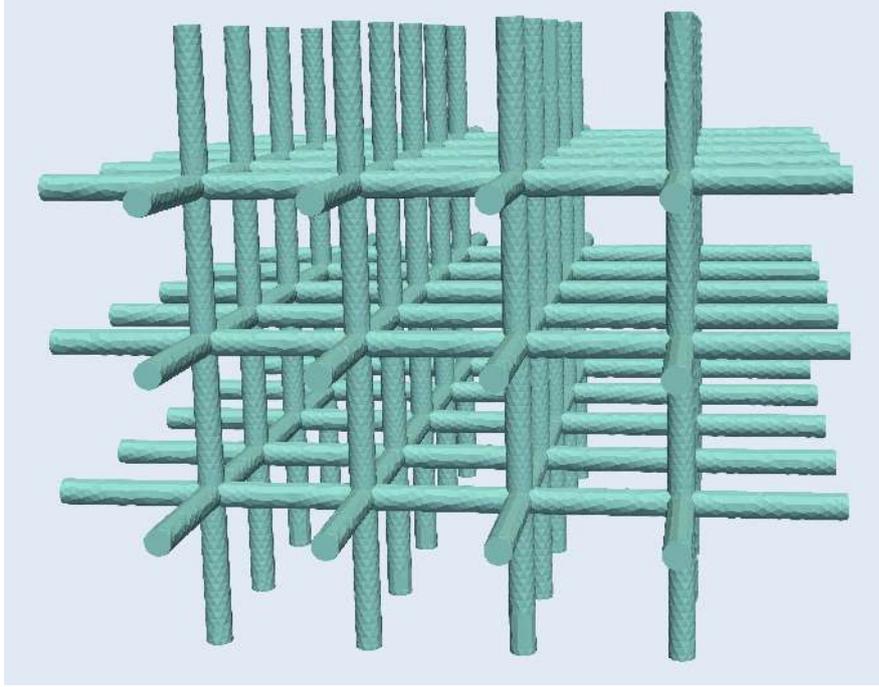

Figure 7.27: Geometry and finite element mesh of a cubic lattice structure

6. One-out / five-in ($-4$)

7. Six in ($-6$)

Out of the seven theoretically possible vertex configurations listed above, the first and the last states with sixfold charges could not be observed in any of the simulations we carried out, while all the other vertex states can be stable at zero field.

In addition to these vertex configurations, the dangling free ends at the surface of the network can also be assigned a charge of $\pm 1$, where the sign depends on the direction of magnetization. We note that the total charge of the system, given by the sum of volume charges and surface charges is zero. Moreover, the charge distribution at the surface is related to the charge distribution in the volume. According to Gauss's divergence theorem, the total magnetostatic volume charge density can be expressed as $\int_D \nabla \cdot \boldsymbol{M} \, dV = \int_S \boldsymbol{M} \cdot \boldsymbol{dA}$. Accordingly, one can conclude that the total volume charge within the bulk depends on total number of surface vertices with magnetization pointing inwards or outwards.

There are a few qualitative differences between the diamond AMC and the cubic version of an artificial crystal. In addition to having different types of vertex charges, there can also be different flavors of the same charge type, depending on the relative orientation of the incoming and outgoing wires, which results in different micromagnetic structures and properties. This is due to the special geometrical arrangement of the nanowires at the vertices. In comparison to the diamond structure, where all the three neighboring wires are equivalent with respect to the each other, in the cubic lattice there exist two different types of neighboring wires: four orthogonal wires, which are mutually equivalent, and one pair of opposite wires, which is qualitatively different from the others. Based on this there can exist two different types of zero-charged vertices, as shown in Fig. (7.28). The initial head-head type zero charge is characterized by the presence of a head-to-head and tail-to-tail domain wall at the vertex as the direction of magnetization in two opposite wires are antiparallel to each others. In contrast, in the second head-tail type zero charge, the opposite wires are magnetized in the same direction. Even though these two versions of zero-charge vertices ave completely different magnetization structure, they have the same value of magnetic volume charge density $\boldsymbol{\nabla} \cdot \boldsymbol{M}$ and, as will be seen later, different magnonic signatures.

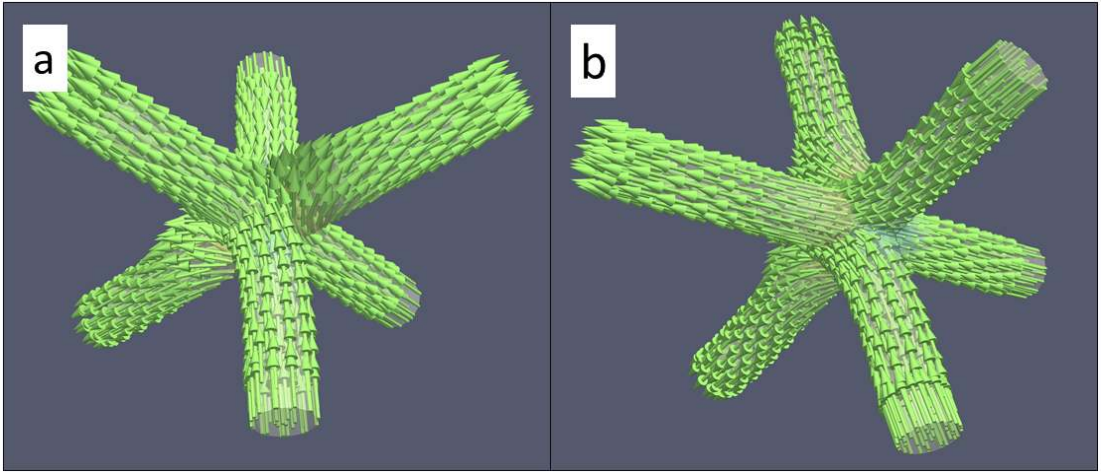

Figure 7.28: Two different types of zero state configuration (a) Head-head state containing a pair of head-head/tail-tail domain wall, (b) Head-tail state with a magnetization structure describing a continuous flow across all three branches.

Even though there can exist two different flavours of $\pm 2q$ double charges differing by the relative orientation of the two incoming / outgoing wires, in our simulation we could only observe one type, in which the two incoming wires are orthogonal to each other. The other flavor would be the case where two wires which are magnetized towards the vertex are arranged opposite to each other along one axis, resulting in the formation of three head-to-head domain-walls in the vertex. In contrast to the double and zero charges, there exists only one type of $\pm 4q$ quadruple charges as there are five wires magnetized towards the vertex and one away from it. The states that can be obtained by moving this single wire to another position can be mapped onto the initial state by means of rotational transformations.

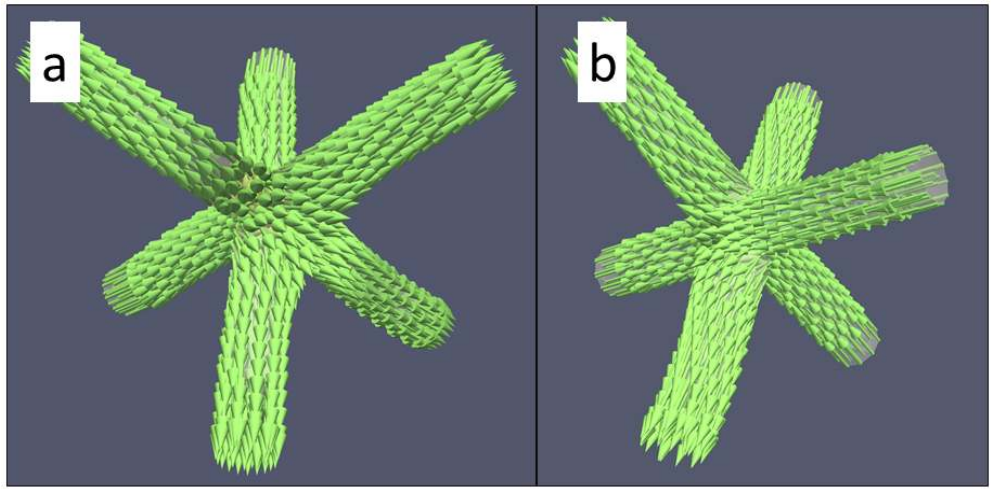

Figure 7.29: Defect-type double-charge vertex configurations: (a) $+2q$ charge formed by a two-in/four-out configuration. The two incoming wires are orthogonal to each other, the other possible state in which the two incoming wires are mutually opposite is not observed in the simulations. (b) $-2q$ charge formed by the inverse, four-in/two-out configuration.

Out of all these possible vertex configuration,s the zero-charged vertices are the only ice-rule obeying states. Both, the $\pm 2q$ and $\pm 4q$ type vertices represent defect charges. By saturating the AMC in different directions by means of external fields and then relaxing to a zero field state, we could generate three qualitatively different ground states.

- Reference state: Contains only head-tail type zero charge vertices. This state can be produced by saturating the entire structure by applying a strong external field along the $[1, 1, 1]$ direction and then relaxing to zero field.

- $\pm 2$q state: contains head-head type zero charged vertices and $\pm 2q$ charged vertices. Can be prepared by saturating the structure along [1,0,0] direction and relaxing to zero.

- $\pm 4$q state: contains $\pm 4q$ charged vertices along with other types listed above.

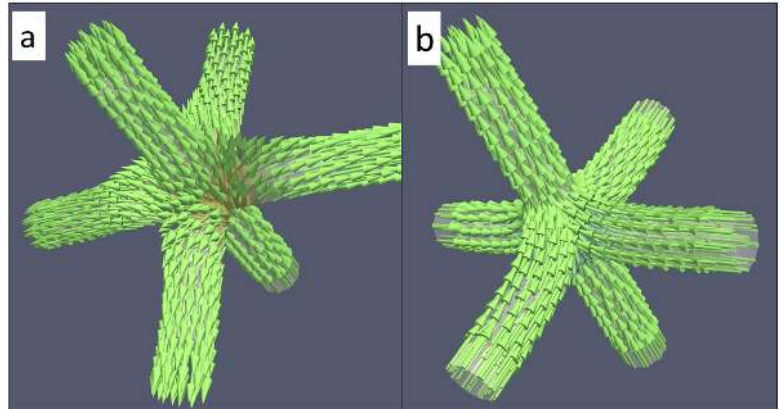

Figure 7.30: Defect quadruple charge configurations: (a) $+4q$ formed by the five-out/one-in state (b) $-4q$ charge pattern formed by the five-in/one-out state. There is no multiplicity for quadruple charges.

Relaxing to zero field from a random-initial configuration results in this magnetic state.

A statistics of the number of different types of vertex configurations in each type is given in table 7.31. The existence of more number of charge types and multiple flavors of each charge type in the cubic lattice opens a wider potential for manipulation. To investigate these effects, we simulated these three states and compared their high-frequency properties.

| Charge types | State 1 | State2 (+2q state) | State 3 (+4q state) |
|---|---|---|---|
| +4q | 0 | 0 | 1 |
| +2q | 0 | 7 | 20 |
| +1q | 47 | 47 | 54 |
| 0 Head-tail type | 60 | 40 | 13 |
| 0 Head-head type | 0 | 6 | 12 |
| -1q | 47 | 47 | 40 |
| -2q | 0 | 7 | 13 |
| -4q | 0 | 0 | 1 |

Figure 7.31: Overview of the number of different types of charged vertices in cubic lattices with different macroscopic magnetization state, as described in the main text.

### 7.10.2 Magnonic excitations

High-frequency small-angle precession modes of the magnetization are simulated by exciting the various relaxed states with a small perturbation and recording the ring-down process of the magnetization. The frequency response to this perturbation is studied for different zero-field remanent states and the effect of the various vertex configuration on the magnonic properties are compared. We can identify three different types of modes in the lattice differing by their geometric localization: vertex modes -in which the activity is concentrated at the vertex points, end modes - which are caused due to the oscillation of the dangling free ends, and wire-modes which are caused by the formation of standing modes in the nanowires.

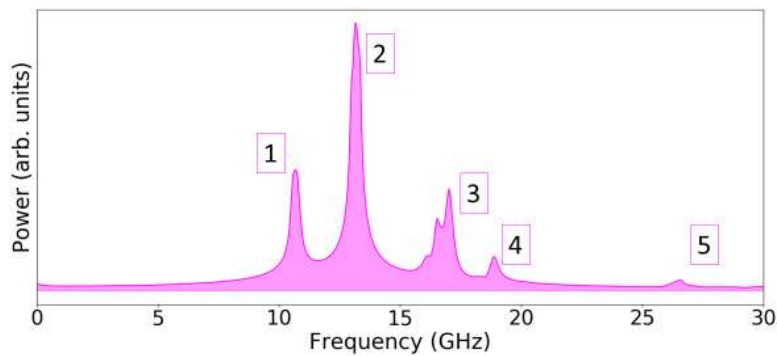

Figure 7.32: Frequency response of the reference state. Modes are marked from $\boxed{1}$ to $\boxed{5}$.

The high-frequency spectrum of the reference state which only contains ice-rule obeying, head-tail type zero-charged vertices is shown in Fig. 7.32. We can already observe a similarity in the behavior with the case of the diamond AMC. The lowest frequency mode – denoted with $\boxed{1}$ at 10.67 GHz – is caused by to the oscillation of the $\pm$1q charge free ends. In contrast to the case of the diamond AMC, where all the free ends oscillate with similar intensity, irrespective of their geometric location on the surface, in the case of the cubic lattice we can see a clear difference in the intensity in the activity of the surface ends depending on their location. The ends of the wires lying parallel to the $x$ axis are oscillating at a much weaker intensity compared to those of the other two axes. This is because the pulse used to excite the system was applied along the $x$ direction, thus it generated a relatively weak torque in regions where the magnetization is nominally parallel to the $x$ axis, while the nanowires oriented along the $z$ and $y$ axes experienced a comparatively much stronger torque. Mode $\boxed{2}$ at 13.6 GHz is the vertex mode, where the activity is concentrated at the zero-charged vertices. Even though all the zero-charged vertices are actively participating in the oscillation, we can see an inhomogeneity in their intensities due to the variations in the local fields. The higher frequency wire modes connected with standing waves formed in the a nanowires are also present in the cubic lattice. Modes $\boxed{3}$, $\boxed{4}$ and $\boxed{5}$ are such wire modes in which the activity is caused by the formation of standing spin waves of different order within the nanowires. The spatial profiles of these modes are shown in Fig. 7.33. For mode $\boxed{3}$ and $\boxed{4}$ the ends of the wires - the vertex points - act as nodes, while for the highest frequency mode $\boxed{5}$ the vertex points take part in the oscillation and the nanowire contains two node planes.

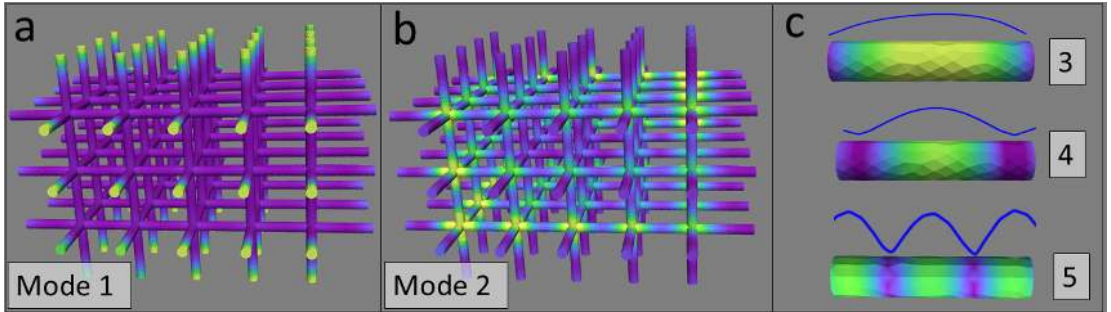

Figure 7.33: Mode profiles of the reference state. (a) Edge modes corresponding to mode $\boxed{1}$ (b) Vertex modes corresponding to mode $\boxed{2}$ (c) Wire modes corresponding to modes $\boxed{3}$, $\boxed{4}$ and $\boxed{5}$

As already seen in the case of buckyballs and in the diamond AMC, the introduction of defect charges in the crystal will activate additional frequency modes which are absent in the defect-free reference state. This is demonstrated by comparing the frequency response to a short-pulse excitation of the two defect states with that of the reference state. The "+2q state" contains in total three charge types, compared to just one type in the reference state: (i) head-head type zero-charged vertices and (ii) +2q/ − 2q defect charges along with the (iii) head-tail type zero-charged vertices already seen in the reference state. The presence of these two results in two additional peaks: mode $\boxed{6}$ at 8.8 GHz corresponds to the activity at head-head type zero charges and mode $\boxed{7}$ at 9.4 GHz corresponds to the activity of ±2q-charged vertices. It is to be noted that despite having same magnetic charge densities, the head-tail and head-head type zero charges oscillate at significantly different frequencies, because of their differences in the micromagnetic structure and consequently the difference in the effective exchange and demagnetizing fields. Inspecting the intensities of the peaks of the modes in the two states, there is a decrease in the intensity of the mode [2] in accordance with the decrease in the number of head-tail type zero charged vertices. The introduction of different charge types results in the broadening and smudging of the wire modes, especially of the mode [3]. As discussed previously in the case of the diamond AMC lattice with disordered magnetic structure, this effect can be attributed to the influence of local dipolar fields arising from the charges at the vertices.

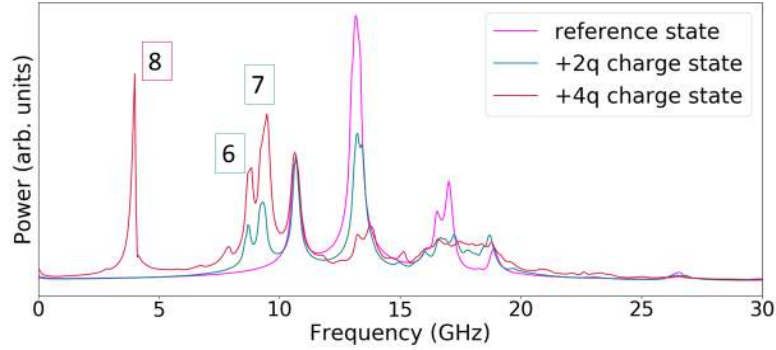

Figure 7.34: Comparison of the frequency response of the three different magnetic states. The additional peaks observed in the $+2q$ charged state ([6], [7]) and $+4q$ charged state ([8]) are marked .

The third magnetic state, the "$+4q$ state" contains a pair of $+4q/-4q$ charges in addition to the second state. Mode [8], which is present only in the third state, is generated by these quadruple charges. Despite there being only two quadruple charges, mode [8] is as intense as the contribution arising from the 60 zero-charged vertices in the first state. This difference in intensity is a result of a significantly higher magnonic activity of the quadruple charges. In the third state, the intensity corresponding to the oscillation of the $+2q$ and the head-head zero-charges is increased in proportion to an increase in their number in the AMC consequently resulting in a drop of the intensity of the mode corresponding to the head-tail zero charged to almost $1/7^{th}$. The introduction of the quadruple charges and an increase in the number of the double charges in the cubic lattice combined has a strong impact on the wire modes. In this case, the modes are further broadened and intertwined, becoming basically indistinguishable from one another.

Comparing the frequency of oscillation of different types of charged vertices, we can see that, in general, all vertices with the same charge type oscillate at the same frequency. Moreover, there is a consistent tendency that the oscillation frequency decreases with increasing magnitude of the charge, while the relative oscillation strength increases with increasing magnitude. The different charge-dependent frequencies are summarized in Fig. 7.35. These general observations apply to all cases, except for the zero-charged vertices, which in the case of the cubic AMC exist in two different flavors oscillating at two different frequencies. The appearance of different frequencies for vertices of equal charge reflects the multiplicity of the possible flavors of these magnetic configurations.

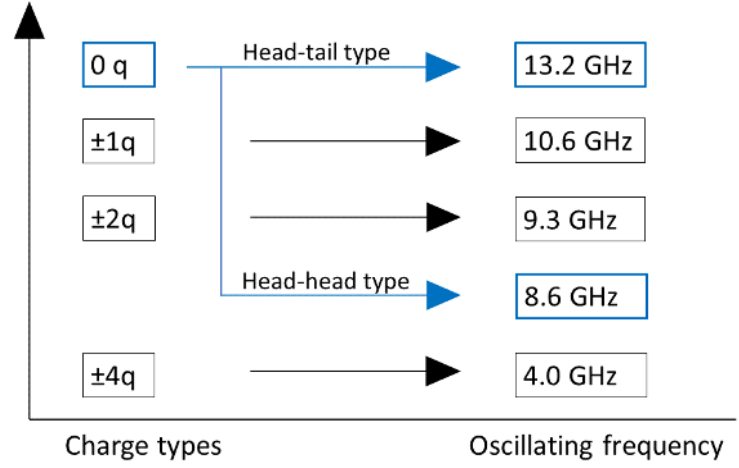

Figure 7.35: Charge-dependent frequency of different vertex types in a cubic AMC lattice.

## 7.11   Chapter summary and outlook

To summarize, we have conducted detailed finite-element micromagnetic simulation studies on three-dimensional diamond and cubic-type artificial magnonic crystals. We looked into the details of the magnetization structures developing at the vertices of these structures. All our simulations yielded axially magnetized nanowires at equilibrium states, resulting in a spin-ice behavior in these three-dimensional arrays. Based on this observation and a corresponding Ising-type spin assignment to the indivdual nanowires, we could classify different vertex configurations with varying levels of magnetic frustration and energetics. We could identify and automatize the recognition of these vertex configurations based on the magnetostatic volume charge densities they produce. The quasistatic magnetic properties of the diamond AMC in an external magnetic field were studied by simulating hysteresis loops. We found that the magnetic switching in these structures involved the abrupt reversal of individual nanowires, resulting in Barkhausen-type jumps in the simulated hysteresis curves. With such hysteretic studies, we also demonstrated how different magnetic configurations – including in particular magnetic defect structures – can be selectively generated in the lattice by saturating the structure at various crystallographic directions and then relaxing to a zero field state. In addition to creating magnetic defects by external fields, we could also inject frustrated states by manipulating the geometry of the lattice through the insertion of a structural defect. We simulated the small-angle precessional modes within these AMC and, by means of inverse Fourier transform, extracted and analyzed their magnonic origins. In general, the modes can be classified into two types: the fundamental modes including the surface, vertex, and wire modes which are related to corresponding geometric constituents, and the defect modes which originate from the magnetic defects within the AMC. We could see that the formation of magnetic defects gives rise to distinct additional modes and yields unmistakable fingerprints in the magnonic spectrum. We verified the robustness of these results by extending a similar investigation into the cubic AMC and found that the general observations are valid there as well. The existence of a larger number of vertex configurations and even multiple flavors of the same charge type in the cubic lattice offers additional methods to control the frequency properties.

The phenomena of magnetic defects creating additional frequency peaks in the spectra of these AMC are analogous to the changes in the optical spectra of natural diamonds by the presence of chemical defects [248, 249]. The ability to control the formation of these defects by field and geometry manipulation opens up an entire new field for spin manipulation using these artificial magnonic crystals. The re-configurability offered by the artificial spin ices, when combined with the advantages of magnonic crystals, makes these new architectures an promising tool for magnonic applications. Moreover, this behavior of the magnetic charges can be exploited as an indirect means to detect the presence or absence of these defects in an AMC network. This type of investigations using finite-element simulations on the high-frequency properties of interconnected three dimensional networks has not been reported extensively in the literature; and though these results are rather convincing and promising, they require experimental confirmation. We anticipate that our findings, which demonstrate that high-frequency modes can be controlled through both the 3D geometry and the magnetic structure of an AMC, will inspire experimental groups to study the magnonic properties of these structures and will further promote three-dimensional interconnected nanowire arrays as a new category of magnonic metamaterials with promising potential for applications in devices for storage, magnonic operations [250] [251], and neuromorphic computations [207, 250].

# CHAPTER 8

Conclusion

In this thesis, we conducted extensive micromagnetic finite-element simulations to study the magnetization structures and the resulting high-frequency properties of different types of three-dimensional nano-architectures. The initial part of this research was concerned with the development of a python-based postprocessing software; a Fourier analysis tool which was necessary to carry out the frequency analysis of these studies. The working principles and the usage of this tool is explained in Chapter 2. This Fourier analysis software was helpful in identifying and isolating the high-frequency modes developed in the structures which we simulated. The flexibility of python permitted to modify the code effortlessly based on the concerned problem and to carry out advanced analysis on the results.

We investigated primarily four different types of geometries,

1. Three-dimensional fractal Sierpinski structures

2. Three-dimensional buckyballs

3. Three-dimensional diamond AMC

4. Three-dimensional cubic AMC

## 8.1 Three-dimensional Sierpinski structures

A Sierpinski tetrahedron is formed by four tetrahedrons arranged on the vertices of a larger tetrahedron. Higher iterations of the fractals are generated by dividing the individual tetrahedrons into smaller units. We kept the overall size of the fractal constant at 512 nm and simulated five generations of the fractal. The individual tetrahedron units in the first two stages of the fractal were large enough so that their equilibrium magnetization structure at zero field consisted of the formation of a three-dimensional vortex structures while for the higher stages the individual units existed in a single domain state and the flux closure was achieved by the formation of of vortex-like structure were the magnetization of the individual units curl together in a ring. The magnetization reversal of these structures was investigated by simulating their hysteresis loops. The high-frequency magnonic response of the relaxed magnetic configurations of these fractals to a small perturbation was simulated by exciting these states with a Gaussian pulse, recording the magnetic ring-down for an extended period of time, followed by a discretized Fourier analysis. We could see that the fractal nature of these geometries had a considerable impact on their magnonic spectrum. Especially, in the third and fourth stages we could see the appearance of a wide-band frequency response stretching across several GHz instead of individual isolated peaks. This property of these fractals has potential applications in a wide range of fields, such as microwave absorption for telecommunication antennas and for the development of radar absorbing materials for military applications.

## 8.2 Buckyball nano-architectures

The Buckyball nano-architectures are made of 90 cylindrical nanowires, interconnected at 60 spherical vertices so that three nanowires meet at each vertex to form a spherical network of interconnected hexagons and pentagons. The Buckyball can be regarded as a model system which depicts a transition from two to three dimensional spin-ice systems. To restrict the parameter space, we fixed the ratio between the length of the nanowire $L$, the radius of the nanowire $R$ and the radius of the spheres $S$ at the vertices to ratio 25:3:4 and simulated a wide range of Buckyballs by varying the side length $L$ from 25 nm to 250 nm. We investigated their relaxed magnetic states, different types of vertex configurations, their hysteretic properties and high-frequency magnonic excitation modes. In the size ranges we studied, the individual nanowires were magnetized in single domain state, i.e., they are magnetized along the axis. A situation of interest, from a micromagnetic perspective, arises at the vertex points where three such nanowires meet. Depending on the direction of magnetization of each nanowire and the Ising-like dipole moment of each nanowire, we could identify different vertex configurations. We noted that these structures exhibit artificial spin ice behavior, including the existence of different degenerate vertex configurations, the appearance of monopole-like magnetic defects, and a quasi-continuum of nearly degenerate states with different spatial distributions of defect structures. Based on the direction of the magnetization of the three nanowires meeting at a vertex, we distinguished two different types of vertex configurations: the ice-rule obeying single-charged vertices and the defect-type triple charges with monopole-like properties. Owing to the three-dimensional nature of the buckyballs structures, the defect-type triple charges could be generated and removed in a controlled way by means of a suitable sequence of external magnetic fields. We could thus selectively generate two qualitatively different magnetic states in such a buckyball: a defect-free state which only contained ice-rule obeying single charged vertices and a defect-state which contained a pair of triple charge vertices along with single-charge vertices.

We simulated the small-angle precession modes of the relaxed magnetic configurations and compared the results. There exists two different magnetic modes in these buckyballs based on their geometric origin: the vertex modes which are localized at the vertices and the higher-frequency wire modes which can be attributed to the formation of standing waves in the nanowires. We could see that the presence of triple-charged vertices results in the appearance of a characteristic, intense low-frequency mode in the magnonic spectrum. This is a potentially important observation, as this feature in combination with the possibility to insert and remove the triple charges reposnible for this peak through external fields opens up a pathway for re-configurable magnonic applications. This effect can also be exploited as an indirect tool to probe the magnetic configurations in such nanoarchitectures. The robustness of the results was verified by simulating hollow buckyballs made of nanotubes and spherical shells. In the simulation of these geometries we observed that, in the concerned length-scales, the hollow structures exhibited a quite similar behavior to that of the solid structures.

## 8.3 Three-dimensional diamond AMC

The diamond lattice was formed by a three-dimensional network of cylindrical nanowires so that four nanowires meet at each vertex point in a tetragonal orientation. The vertex points where four such wires meet are distributed in space corresponding to the atomic locations of a natural diamond lattice - hence the name. The network we simulated was made of nanowires of length 70 nm and thickness 14 nm. The extended network contained 202 such nanowires, interconnected at 83 vertex points. The whole structure has a dimension of $450 \times 450 \times 450$ nm Similar to the case of buckyballs, the individual nanowires in the network were magnetized axially at zero fields and their interaction at the vertex points leads to frustrated vertex configurations. We could identify two types of vertex configurations: the ice-rule obeying two-in/two-out zero charged vertices and the defect charges with three in / one out arrangement. Based on the distribution of these different types of vertices we could develop three qualitatively different states in such a diamond lattice: a reference state which only contained ice-rule obeying zero charged vertices a defect state which contained only defect vertices which were periodically distributed in the lattice and a disordered state which contained both types of vertices which were all randomly distributed.

Investigation into the magnonic excitations of the lattice revealed the existence of surface modes, vertex modes and wire modes based on the geometric origin of the active regions. The presence of defect charges created additional low frequency modes which were absent in the reference state. We could also create additional modes in the spectrum by introducing geometric defects in the lattice. Since the defects can be used to manipulate the magnonic spectrum and their formation could be controlled, these types of periodic AMC networks offers immense potential for magnonic applications.

## 8.4 Cubic AMC

The cubic lattice is formed by the interconnected network of cylindrical nanowires so that six wires meet at each vertex in an orthogonal way. The wires had a similar dimension to that of the diamond AMC and they exhibited similar single domain behavior. The increased coordination number in the cubic structure (6) compared to the diamond (4) and the orthogonal arrangement of wires results in the emergence of additional possible vertex configurations and even different flavors of the same type of configuration. Along with the vertex configurations seen in the diamond lattice, there exist two additional vertex types in the cubic lattice; two flavors of the zero charged vertices and $\pm 4q$ quadruple charges. As expected, the presence of double charges and quadruple charges introduces characteristic fingerprint modes in the magnonic spectrum of the lattice. Despite having same number of moments coming in and leaving out, and having moreover almost identical magnetic volume charge density, the two flavors of two-in/two-out vertex configurations exhibit completely different dynamic properties as a consequence of the difference in their local micromagnetic structure. The availability of a larger number of defect-type vertex configurations in the cubic lattice offers additional pathways for manipulating the magnonic properties of the lattice.

## 8.5 Summary

To summarize, we carried out fundamental investigations on the static and magnonic properties of several three-dimensional nanoarchitectures. The geometric flexibility of our finite-element approach allowed an accurate modelling of their complex geometries, especially the curved surfaces and the interior surfaces of the hollow structures. We demonstrated how the magnonic properties of artificial spin ice lattices can be manipulated and re-configured by means of defect charges. Investigations on three-dimensional magnetic nanostructures and nanoarchitectures is still at an early stage, especially as far as their high-frequency dynamics is concerned. The studies conducted in this thesis have been performed with the goal of achieving a better understanding of the properties of these novel three-dimensional structures. One can assume that many of their properties and potential applications are yet to be discovered. We hope the findings of this thesis will aide the experimentalists to gain a deeper insight into the physics of these structures and also inspire future studies to follow.

## Funding

This work was funded by the LabEx NIE (ANR-11-LABX-0058NIE) in the framework of the Inter-disciplinary Thematic Institute QMat (ANR-17-EURE-0024), as part of the ITI 2021-2028 program supportedby the IdEx Unistra (ANR-10-IDEX-0002-002) and SFRISTRATUS (ANR-20-SFRI-0012) through the FrenchProgramme d'Investissement d'Avenir. The authors ac-knowledge the High Performance Computing center ofthe University of Strasbourg for supporting this work byproviding access to computing resources.

UNIVERSITÉ DE STRASBOURG

# Simulations micromagnétiques de nano-architectures tridimensionnelles

# Micromagnetic simulation of three-dimensional nano-architectures

## Résumé

Cette thèse traite de simulations micromagnétiques de la dynamique hyperfréquence de l'aimantation dans des nanoarchitectures tridimensionnelles (3D) constituées de réseaux de nanofils interconnectés. Les propriétés magnétiques de tels nanomatériaux artificiels sont fortement influencées par leur géométrie. En étudiant la structure magnétique statique de ces systèmes, nous montrons des configurations correspondant à des réseaux de glace de spin artificiels 3D, avec des interactions frustrées et des structures de défaut monopolaires aux points d'intersection. Nos simulations révèlent une activité élevée de ces sites dans les excitations magnétiques de haute fréquence. Nous étudions diverses nanoarchitectures 3D et montrons que leur géométrie et leur structure magnétique donnent des signatures hyperfréquences caractéristiques. Le contrôle de ces caractéristiques pourrait ouvrir de nouvelles voies pour la recherche magnonique et dans le développement de métamatériaux magnétiques reprogrammables.

Mots clé : Simulations micromagnétiques, nanostructures tridimensionnelles, dynamique de l'aimantation, hyperfréquences

## Résumé en anglais

The thesis discusses micromagnetic simulation studies on the high-frequency magnetic dynamics in three-dimensional (3D) nanoarchitectures made of interconnected magnetic nanowire networks. Such artificial magnetic materials with nanoscale features have recently emerged as a vivid topic of research, as their geometry has a decisive impact on their magnetic properties. By studying their static magnetization structure, we find that these systems display a behavior analogous to that of 3D artificial spin ice lattices, with frustrated interactions and the emergence of monopole-like defect structure at the wires' intersection points. Our simulations reveal a high activity of these defect sites in the magnonic high-frequency spectrum. We study various 3D nanoarchitectures and show that their geometry and magnetization state results in characteristic high-frequency signatures. Controlling these features could open new pathways for magnonics research and reprogrammable magnetic metamaterials.

Keywords: Micromagnetic simulations, three-dimensional magnetic nanostructures, high-frequency magnetization dynamics